\documentclass{article}
\usepackage[utf8]{inputenc}

\usepackage{natbib}
\usepackage{graphicx}
\usepackage{graphicx,color}
\usepackage{amsmath}
\usepackage{amssymb}
\usepackage{graphicx}
\usepackage{epstopdf}
\usepackage{inputenc}
\usepackage{amsfonts}
\usepackage[table]{xcolor}
\usepackage{arydshln}
\usepackage{mathtools}
\usepackage{geometry}
\usepackage{float}
\usepackage{ulem}
\usepackage{longtable}
\usepackage[colorlinks=true,linkcolor=blue,citecolor=blue]{hyperref}
\usepackage{fancyhdr}
\usepackage{pgfplots}
\usepackage{booktabs}
\usepackage{natbib}
\usepackage{color}
\usepackage{subfigure}
\usepackage{multirow}
\usepackage{threeparttable}
\usepackage{enumitem}
\usepackage{colortbl,multirow}
\usepackage[linesnumbered,ruled,vlined]{algorithm2e}% http://ctan.org/pkg/algorithm2e
\DontPrintSemicolon
\usepackage{pdflscape}
\usepackage{adjustbox}
\usepackage{pgfplots}
\pgfplotsset{compat=1.7}

\title{A Bayesian Skew-heavy-tailed modelling for loss reserving}%{A Bayesian heavy-tailed and asymmetric modelling for loss reserving} 

\author{William L. Leão$^{1}$ \hspace{0.1cm} and \hspace{0.1cm} Viviana G. R. Lobo$^{1}$\footnote{{{\it Corresponding author}: Viviana G R Lobo, Departamento de M\'etodos Estat\'{\i}sticos, Instituto de Matem\'atica, Universidade Fe\-de\-ral do Rio de Janeiro, Av. Athos da Silveira Ramos, Centro de Tecnologia, Bloco C, CEP 21941-909. \newline {\it E-mail}: {\tt viviana@dme.ufrj.br}. {\it Homepage}: https://sites.google.com/site/dme/viviana}}   \\
\textit{$^{1}$Instituto de Matem\'atica, Departamento de Métodos Estatísticos} \\ 
\textit{Universidade Federal do Rio de Janeiro, Brazil }
}
\date{}

\begin{document}

\maketitle

\begin{abstract}
This paper focuses on modelling loss reserving to pay outstanding claims. As the amount liable on any given claim is not known until settlement, we propose a flexible model via heavy-tailed and skewed distributions to deal with outstanding liabilities.  The inference relies on Markov chain Monte Carlo via Gibbs sampler with adaptive Metropolis algorithm steps allowing for fast computations and providing efficient algorithms. An illustrative example emulates a typical dataset based on a runoff triangle and investigates the properties of the proposed models. Also, a case study is considered and shows that the proposed model outperforms the usual loss reserving models well established in the literature in the presence of skewness and heavy tails.

\end{abstract}


\section{Introduction}\label{sec1}

\subsection{Background}

%%% motivation
An insurance policy guarantees the payment provided by an insurer to the policyholder if a loss occurs conditionally on a premium payment. However, in some cases, insurance companies often do not settle outstanding claims immediately because some claims are reported with some delay, which means that claims should be paid with a time delay of months, years, or decades. In this context, is crucial for the insurance company be the ability to predict as accurately as possible the amount necessary to compose the reserve for claims that incurred but not yet reported (IBNR, which means the liabilities that have arisen but are reported after the accounting date) or settled to maintain their solvency. 

Traditionally, some of the existing statistical approaches for dealing with the problem, such as the chain ladder method (\cite{Taylor1984}), Bornhuetter–Ferguson algorithms (\cite{BFerguson1972}) and Mack method (\cite{Mack1993}) to compute loss reserves are not able to provide satisfactory performance, because is based on deterministic methods that return a pointwise estimate and do not take into account stochastic uncertainties involved. Also, the usual techniques are not flexible to capture irregular and extreme claims in the dataset.

%%% revisao de literatura sobre loss reserving mais de prposta de modelagem (modelos usuais for loss reserving)
%{\color{red} USUAL MODELS WITH GAUSSIANTY:} 
Several papers have proposed models based on  stochastic techniques or modifications of the usual chain-ladder method for loss reserving. For example, \cite{Verrall2000} investigates the relationship between chain-ladder and some stochastic models. Other authors, \cite{Kremer1982}, \cite{RenshawVerrall1998}, \cite{Verrall1996}, \cite{HabermanRenshay1996} consider the use of log-linear models, such as the analysis of variance via ANOVA-type and ANCOVA-type modelling. \cite{Verrall1989}, \cite{Jong1983},  \cite{Verrall1990}, and \cite{Ntzoufras02} consider flexibility of the models based on state-space. For more details about loss reserving stochastic methods, see \cite{Ntzoufras02} and \cite{EnglandVerrall2002}.
%%% falar de modelos que fazem transformacao na variavel resposta e nao e legal as vzs

%%% modelos de mistura de escalas para ibnr (RELAXING GAUSSIANTY ASSUMPTION)
A usual assumption to models loss reserving takes into account that reserves have Gaussian errors (see \cite{Verrall1989}, \cite{Ntzoufras02}, and among others shown previously). However, these processes often have common characteristics such as non-Gaussianity, outlying claims, and non-constant variance. In this way, if these behaviour are not incorporated into the model assumption, the predictions for loss reserving might be poor. Relaxing the assumption, some proposals consider modelling reserves via scale mixture forms. \cite{Choy2008} proposed a generalized-{\it t} error distribution via a scale mixture of uniforms (SMN) to take into account the presence of potential outliers. \cite{Jennifer16} consider scale mixtures (Gaussian and Uniform) based on heavy-tailed distributions including Cauchy, Student-t, EP, logistic, and Laplace in a univariate loss reserving data modelling. They consider modelling the amount of claims paid to the insureds of an insured product during the 18 delayed years and conclude that the log-ANCOVA-t model affords better performance in the heavy-tailed class proposed. \cite{Shi2012} and \cite{Goudarzi2018} consider the same approach by \cite{Verrall1989} and  \cite{Jennifer16} however in a multivariate loss reserve setting. The authors include an additional parameter associated with the calendar year. The calendar effects are captured  by a random effect term, via a random-walk model or an autoregressive model. {Recently,  \cite{ChanJennifer2022} propose adopting the  conditional autoregressive range model to the mean function of the loss reserve distribution to explain the persistence of the development year effect.}

%a random walk model and also a first-order auto-regressive model for calendar effects to incorporate the association between accident and development periods in the state-space model. %(os dados que que goudarzi utiklizou  ele utilizou nao sao legais para capturar caudas pesadas talvez colocar isso na hora de analisar os dados... na parte de discussao.)

%% pegar gancho com assimetria
%%% revisao sobre modelos assimetricos e da caudas pesadas e nossa proposta
Although these models allow fatter tails than Normal, they are not able to capture skewed claims. In this manner, \cite{Azzalini1985} and \cite{AZZALINI96} propose a class of distributions based on skew-normal (SNLM). Other authors, such as \cite{Arellano2005} and \cite{Arellano2007} propose skew-normal linear mixed models (SNLMM). \cite{Pigeon2013} consider a family of multivariate skew-symmetric (MSS) based on the skew-normal distribution to model individual reserve claims. Although this model allows skewness, they do not take into account heavy-tail distributions. To allow simultaneously both skewness and thick tails in the modelling, \cite{Steel1998} propose a family of univariate skew-t distribution. Another family of skew-t distribution is proposed by \cite{Jung2001} extending the idea in \cite{Azzalini1985} where the skew-normal is a special case.  Furthermore, \cite{genton2006} introduce the multivariate skew-slash model as an alternative to skew-t and skew-normal models. 

%% Nossa proposta aqui
\cite{Jennifer16} address the importance of improving their proposed model considering modelling skewness. However, few propose to accommodate both skewness and high kurtosis have been proposed yet in the loss reserving literature. \cite{Julia13} performs the estimation of the log-ANOVA and dynamic ANOVA models using the normal, Student's t, Skewed Normal and Skewed Student's t distributions. \cite{Generoso19} considers a class of loss-reserving models
based on the quantile regression approach developed for modelling the tails of the claims considering the asymmetric Laplace distributions and \cite{ChanJennifer2022} consider the Generalized beta type II distribution to capture the characteristics of loss reserving distribution including skewness, peakness and long right tails. We propose flexible models based on a family of Skew-heavy-tailed distributions that deal with outstanding liabilities. Our class of models include Student-t, Slash and Variance-Gamma (also known as generalized Laplace, \cite{Dilip1990}) based on a scale mixture of skew-Normal distribution. This proposal gives a suitable characterization for reserve claim data. The inference procedure follows the Bayesian framework that has the advantage of taking into account the uncertainty associated with the estimates through the posterior distribution. Also, we can access the loss reserving of the predictive distribution for outstanding liabilities and the prior knowledge might be elicited over past experience from an insurance company. 

From a full Bayesian view, the posterior distribution is commonly not available from their analytical form, and numerical integration and simulation are considered, in particular, Markov chain Monte Carlo (MCMC) methods \citep{gamerman} are used in this work. Several actuarial papers for loss reserving such as, \cite{Jennifer16}, \cite{Goudarzi2018}, and \cite{ChanJennifer2022} obtain the posterior distribution via MCMC through open sources software (\texttt{OpenBUGS}, \texttt{WinBUGS} and \texttt{RStan} \cite{Winbugs}, \cite{lun09bug}, \cite{rstan}). Although \texttt{BUGS} and \texttt{Stan} are user-friendly Bayesian software, it could be very restrictive if considering additional complexity in the modelling process and/or computationally exhaustive. For this, we address the posterior distribution via adaptive MCMC (\cite{Andrieu2008}) jointly with the Gibbs sampler techniques, allowing for fast and smart feasible computations. For the Skew-Variance-Gamma distribution, we regard the technique proposed by \cite{Philippe1997} to generate truncated Gamma distributions.

Briefly, our proposed method for loss reserving allows accommodating skewness, heavy tails, and outlying claims behaviour in the dataset considering Skew-heavy-tailed models and taking into account Bayesian inference procedure through
Gibbs sampler with adaptive Metropolis algorithm steps MCMC techniques with computational efficiency that we can obtain through the development of the proposed method in the free software \texttt{R}, with the use of the \texttt{Rcpp} library \citep{Rcpp13,Rcpp18} that provides easy integration between the programming language \texttt{C++} and \texttt{R}, and through the use of the linear algebra library \texttt{Armadillo} for language \texttt{C++} \citep{Sanderson2016,Sanderson2018} and loops implemented efficiently, we were able to save considerable computational time.

%\subsection{An illustrative example from \cite{Jennifer16}}

\subsection{Outline of the paper}
The remaining of the paper is organised as follows. Section \ref{sec2} describes the class of Skew-heavy-tailed distributions. In subsection \ref{sec2.1}, the Bayesian procedure for loss reserving is introduced and algorithms are presented (see subsection \ref{sec2.2}). Section \ref{sec3} includes some simulation studies and examples that have already been presented in \cite{Jennifer16} in order to illustrate the performance of our proposed models. %In subsection \ref{sec3.2} a real dataset for an insurance company based in Brazil is presented with some discussion. 
Concluding remarks are given in Section \ref{sec4}. Some aspects related to properties of the proposed model, the posterior distribution, model comparison, and real datasets, are presented in Appendices \ref{apA}, \ref{apB}, \ref{apC} and \ref{apD}, respectively.

\section{ Skew-heavy-tailed modelling}\label{sec2}
 In this section, we propose new approaches modelling based on a family of Skew-heavy-tailed models for outstanding claims predictions. We define a scale mixture of the skew-normal family and study some of its properties. 
 
 The insurance data follows a structure typically presented by a run-off triangle (see Table \ref{tab1}), showing the development of the claims over time for each period. Let a random variable $Y_{ij}$ the amount that the insurance company paid with a delay of $j-1$ years (development year) for accidents that occurred in year $i$ (policy year), $i=1, \ldots, n$, $j= 1, \ldots, n-i+1$. After the business has been running for $t$ years, the data available has the form $Y_{\Delta} :=\left\{ Y_{ij}: j \leq  n-i+1, i \leq n \right\}$ and $Y_{ij} >0$, $\forall i$ and $\forall j$ (\cite{Verrall1989}, \cite{Kremer1982}), usually called run-off triangle. For example, $y_{2,2}$ means that the event occurred in year 2 and was reported in the same year, whereas $y_{2,6}$ means that the event was settled after four years before the policy year.  The main interest lies on predicting the unobserved claim amounts, which means the lower triangle (light grey cells, $Y_{\nabla}:= \left\{Y_{ij}: j > n-i +1 \right\}$) should be estimated. 

%to the problem of estimating the expected value of IBNR-c1aims. 
\begin{table}[H]
\begin{center}
\caption{Run-off triangle: structure of claim amount dataset.}\label{tab1}
    \begin{tabular}{c|ccccc}
        \hline
          {\it policy year}  & \multicolumn{5}{c}{{\it lag year} $j$}\\
        $i$ & 1 & 2 & $\ldots$ & $n-1$ & $n$ \\
        \hline
        1 & $y_{1,1}$ & $y_{1,2}$ & $\ldots$ & $y_{1,n-1}$&$y_{1,n}$  \\       
        2 & $y_{2,1}$ & $y_{2,2}$ & $\ldots$ & $y_{2,n-1}$&    \multicolumn{1}{c}{\cellcolor{lightgray}} \\       
        $\vdots$ & $\vdots$ & $\vdots$ & $\vdots$ &  \multicolumn{2}{c}{\cellcolor{lightgray}}  \\  $n-1$ & $y_{n-1,1}$ & $y_{n-1,2}$ &  \multicolumn{3}{c}{\cellcolor{lightgray}} \\ %   & \cellcolor{lightgray}  &  \cellcolor{lightgray}  \\       
        $n$ & $y_{n,1}$ &  \multicolumn{4}{c}{\cellcolor{lightgray}} \\ %   &  \multirow{-3}{*}{\cellcolor{lightgray}}   &   \cellcolor{lightgray}  & \cellcolor{lightgray}  \\
        \hline
\end{tabular}
\end{center}
\end{table}

\subsection{A class of Skew-heavy-tailed distributions }\label{sec2:sec2.1}

Let $Z_{ij}=log(Y_{ij})$ be the log claim amount paid to the policyholder from the policy-year $i$ with $n-1$ years of delay, and $j \leq n-i +1$, $i \leq n$.  The stochastic representation of the variable for loss reserving is given by the formulation as a mixture of scales as follows

\begin{equation}\label{sec2:eq1}
     Z_{ij} = \mu_{ij} + \sigma\frac{ \varepsilon_{ij}}{\sqrt{\lambda_{ij}}}, 
\end{equation}

\noindent where $\mu_{ij}$ represents the mean of the log claim amount, originated in year $i$ and was paid with a delay of $n-1$ years. The variable $\varepsilon_{ij}$ is a zero mean Skew-Normal distribution, with $\sigma^2$ as a scale parameter, and $\kappa$ skewness parameter. The component $\lambda_{ij}$ is a positive mixing variable and it is responsible for controlling the tail of the distribution and is independent of $\varepsilon_{ij}$. The p.d.f. of $Z_{ij}$ is said to have a Skew-Mixture-Normal representation with $\mu_{ij} \in \Re$, $\sigma^2 \in \Re^{+}$ and $\kappa \in \Re$ if it can be expressed as
\begin{eqnarray}\label{sec2:eq2}\nonumber 
    f_{Z}(z \mid \mu, \sigma, \kappa) &= &\int_{0}^{\infty} \mathcal{S}\mathcal{N}(z ; \mu, \sigma^2,  \lambda, \kappa) p(\lambda) d \lambda, \\
     &=& 2 \int_{0}^{\infty} \phi(z \mid \mu, \lambda^{-1}\sigma^2) \Phi\left(\kappa \frac{z-\mu}{\sigma}\right)p(\lambda) d\lambda, 
    \end{eqnarray}

\noindent where $\phi$ follows a Gaussian distribution and $\Phi$ follows a cumulative distribution of the standard Normal distribution, respectively. 
Conditional on $\lambda_{ij}$, the variable $Z_{ij}$ has to mean function $\mu_{ij}$ that is specified for parameters that depend on policy-year and lag-year effects, variance $\lambda_{ij}^{-1}\sigma^2$ and skewness $\lambda_{ij}^{-1/2}\kappa$. Gaussian behaviour is only assumed given the mixing variable $\lambda$. The representation is given by $Z_{ij} \mid \mu_{ij}, \sigma^2, \lambda_{ij}, \kappa \sim \mathcal{S}\mathcal{N}(\mu_{ij}, \lambda^{-1}_{ij}\sigma^2, \lambda_{ij}^{-1/2}\kappa)$.  Notice that when $\lambda_{ij} \neq 1$, $Z_{ij}$ has a heterogeneous variance, and if $\lambda_{ij}$ is integrated out the resulting is non-Gaussian distribution. On the other hand, if $\lambda_{ij} =1$, the resulting model is the usual Skew-Normal (\cite{AZZALINI96}). For $\kappa = 0$ and conditional on $\lambda_{ij}$, the Skew-Normal distribution reduces to the Normal distribution, that is, the loss reserving follows $Z_{ij} \mid \mu_{ij}, \sigma^2, \lambda_{ij} \sim \mathcal{N}(\mu_{ij}, \lambda_{ij}^{-1}\sigma^2)$.  

To keep the model feasible for the inference procedure, we define the loss reserving modelling through a hierarchical representation of distributions according to the stochastic representation proposed by \cite{Henze1986}.
%\begin{proposition}\label{sec2:prop1}
    Let $T_1$ and $T_2$ standard Normal variables. If $\varepsilon$ follows a standard Skew-Normal distribution with density given by equation (\ref{sec2:eq2}) with $\lambda=1$, then $\varepsilon$ is able to be written as a location-scale mixture of Gaussian distribution given by
     \begin{equation}
        \varepsilon=  \rho |T_{1}| +  \sqrt{(1- \rho^2)} T_{2},
     \end{equation}
     \noindent where $\rho= \frac{\kappa}{\sqrt{1+\kappa^2}}$, $|T_{1}|$ denotes the absolute value of $T_{1}$, and $T_{1}, T_{2}$ are independent $\mathcal{N}(0,1)$. As we can see, $\varepsilon$ is written as a mixture of variables from $T$ that follows a standard Gaussian distribution. 

    % \noindent Proof. {\color{red}escrever aqui o caminho... e talvez citar um apendice}
%\end{proposition}
%{\color{red} For more details about the setting see Garay (2010), Branco e Dey (2001), Henze, 1986 e Azzalini (1985).} 
Thus, equation (\ref{sec2:eq1}) can be written as
\begin{equation}\label{sec2:eq3}
 %Z_{ij} = \mu_{ij} + \sigma\rho \lambda_{ij}^{-1/2}|T_1| + \lambda_{ij}^{-1/2} %\sqrt{\sigma^2(1- \rho^2)} T_2,
  Z_{ij} = \mu_{ij} + \sigma \lambda_{ij}^{-1/2} \left( \rho |T_{1;i,j}| +  \sqrt{(1- \rho^2)} T_{2;i,j} \right).
\end{equation}
Therefore,
\begin{eqnarray}\nonumber
    Z_{ij} \mid T_{ij}, \lambda_{ij} &\sim& \mathcal{N}\left( \mu_{ij} + \rho T_{ij}; \sigma^2\lambda_{ij}^{-1} (1-\rho^2)\right) \\
     T_{ij} \mid \lambda_{ij} &\sim& \mathcal{N}\mathcal{T}\left(0, \sigma^2 \lambda_{ij}^{-1}; (0, \infty) \right) \\ \nonumber
      \lambda_{ij} &\sim& p(\cdot \mid \nu), 
\end{eqnarray}

\noindent where $T_{ij}= \lambda_{ij}^{-1/2}|T_{1;i,j}|$ and $\nu$ controls the tails of the distribution. Then, to generate a Gaussian distribution, we proceed through three steps, that is, we generate first from the mixture distribution $\lambda_{ij}$, after from the conditional $T_{ij} \mid \lambda_{ij}$ generate a Truncate Gaussian distribution and next from the conditional distribution $Z_{ij} \mid T_{ij}, \lambda_{ij}$ using the stochastic representation given in equation (\ref{sec2:eq3}).
The p.d.f. of $Z_{ij}$ is rewritten as
\begin{align} \nonumber
		f_Z(z \mid \cdot)&=\int_{-\infty}^{\infty}\int_{0}^{\infty}f(z,\lambda,t \mid \cdot)dt d\lambda\\ \nonumber
		    &=\int_{-\infty}^{\infty}\int_{0}^{\infty}f(z|\lambda,t, \cdot)f(t \mid \lambda, \cdot)p(\lambda)dt d\lambda\\ 
		    &=\int_{-\infty}^{\infty}\int_{0}^{\infty}\mathcal{N}\left(z\mid  \mu+ t\rho,\sigma^2\lambda^{-1}(1-\rho^2)\right)\mathcal{NT}\left(z \mid 0, \sigma^2\lambda^{-1}; (0,\infty)\right)p(\lambda \mid \nu)dt d\lambda.
	\end{align}
\vspace{0.5cm}

\noindent \textit{Some properties of the Skew-Heavy-Tailed class} The resulting mean and variance of the distribution $Z_{ij}$, defined in equation \eqref{sec2:eq1} is given by%obtained by integrating out the mixing distribution $lambda_{ij}$.

\begin{equation}\label{sec2:eq6}
 %E(Z_{ij})= \mu_{ij} + \sqrt{\frac{2}{\pi}} \sigma \gamma E\left[ \frac{1/\lambda}{\sqrt{1+\gamma^2(1/\lambda)}} \right]
 E(Z_{ij}\mid \cdot)= \mu_{ij} + \sigma \rho\sqrt{\frac{2}{\pi}} E\left(\lambda_{ij}^{-1/2}\right)
\end{equation}

\begin{equation}\label{sec2:eq7}
 %Var(Z_{ij})= \sigma^2 \left[ E(1/\lambda) - \frac{2}{\pi} \gamma^2 E^2\left( \frac{1/\lambda}{\sqrt{1+\gamma^2(1/\lambda)}} \right)\right].
Var(Z_{ij} \mid \cdot)= \sigma^2  \left( E\left(\lambda_{ij}^{-1}\right) - \rho^2\left(\frac{2}{\pi}\right)E^2\left(\lambda_{ij}^{-1/2}\right)\right).
\end{equation}

\noindent Notice that if $\lambda_{ij}=1$, then $E(Z_{ij}) = \mu_{ij} + \sqrt{\frac{2}{\pi}}\sigma \rho$ and $Var(Z_{ij})=\sigma^2\left( 1 - \frac{2}{\pi} \rho^2\right)$ follows an usual Skew-Normal distribution, it implies that the distribution does not have ticker tails than Skew-Normal. Also, if $\lambda_{ij}=1$ and $\kappa=0$, then $\rho=0$ and we return a Gaussian distribution. %For more details about the properties of the Skew-Normal mixture models see \cite{Ferreira2011}. 
As follows, we addressed the prior specification of the mixing variable $\lambda_{ij}$ and the parameter associated with skewness $\kappa$. 
\vspace{0.5cm}

\noindent \textit{Modelling the mixing $\lambda_{ij}$} A first and natural approach would be to assume that $\lambda_{ij} \sim \mathcal{G}\left(\frac{\nu}{2}, \frac{\nu}{2} \right)$. This implies that $Z_{ij}$ is a Skew-Student-t (or just Skew-t) distribution with $\nu>0$ degrees of freedom, denoted by $Z_{ij} \mid \mu_{ij}, \sigma^2,\nu, \kappa \sim \mathcal{S}\mathcal{S}\mathcal{T}(\mu_{ij}, \sigma^2,\nu,\kappa)$ and the Gamma distribution ($\mathcal{G}(\cdot, \cdot)$). Therefore, the expected value and variance given in equations  (\ref{sec2:eq6}) and (\ref{sec2:eq7}), unconditional on the mixing parameter are obtained by computing $E(\lambda_{ij}^{-1/2}) = \left[{\Gamma(\frac{\nu-1}{2})}/{\Gamma(\frac{\nu}{2})}\right]\left(\frac{\nu}{2}\right)^{1/2}$ and $E\left(\lambda_{ij}^{-1}\right) = \frac{\nu}{2}/(\frac{\nu}{2}-1)$, with $\frac{1}{\lambda_{ij}}\sim \mathcal{I}\mathcal{G}(\frac{\nu}{2},\frac{\nu}{2})$.

Alternatively, if we assume a Skew-Slash model, then $\lambda_{ij}$ has Beta distribution ($\mathcal{B}e(\cdot,\cdot)$), with mean is $\frac{\nu}{\nu+1}$, and variance is $\frac{\nu}{(\nu+1)^2(\nu+2)}$, hence $\lambda_{ij} \sim \mathcal{B}{e}(\nu, 1)$, and $Z_{ij} \mid \mu_{ij}, \sigma^2,\nu, \kappa \sim \mathcal{S}\mathcal{S}(\mu_{ij}, \sigma^2,\nu,\kappa)$. This implies that the prior marginal distribution of $\lambda_{ij}$ is concentrated around one for bigger values of $\nu$, and as $\nu$ decreases, the distribution becomes more spread out. Analogously, $E(\lambda_{ij}^{-1/2}) = \frac{\nu}{\nu-1/2} $ and $E\left(\lambda_{ij}^{-1}\right) = \frac{\nu}{\nu-1}$, respectively. 

Another distribution that belongs to the skew-heavy-tailed family assumes that $\lambda_{ij} \sim \mathcal{I}\mathcal{G}(\frac{\nu}{2},\frac{\nu}{2})$, where $\nu$ is responsible to capture ticker tails, then $Z_{ij} \mid \mu_{ij}, \sigma^2, \nu, \kappa \sim  \mathcal{S}\mathcal{V}\mathcal{G}(\mu_{ij}, \sigma^2,\nu,\kappa) $. The mean and variance of Skew-Variance-Gamma distribution follow the equations that have been mentioned previously considering 
 $E(\lambda_{ij}^{-1/2}) = \left[{\Gamma(\frac{\nu+1}{2})}/{\Gamma(\frac{\nu}{2})}\right]\left(\frac{\nu}{2}\right)^{-1/2}$ and $E\left(\lambda_{ij}^{-1}\right) =1$. For more details about the distributions and their properties see Appendix \ref{apA}. %{\color{red}vamos colocar algo de propriedade dos modelos??? tipo curtose e assimetria de forma geral}

\begin{figure}[H]
\begin{center}
\begin{tabular}{cc}
\includegraphics[width=5cm]{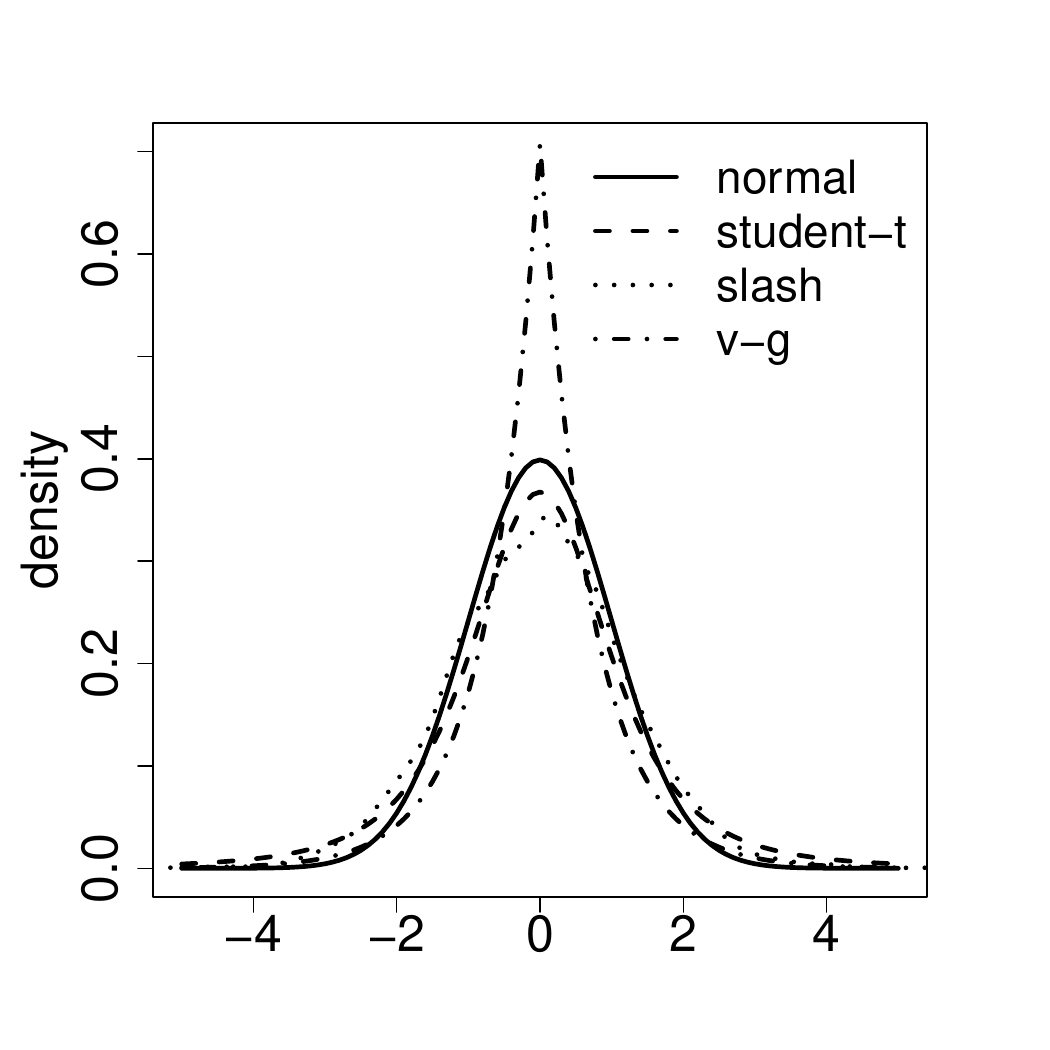} &  \includegraphics[width=5cm]{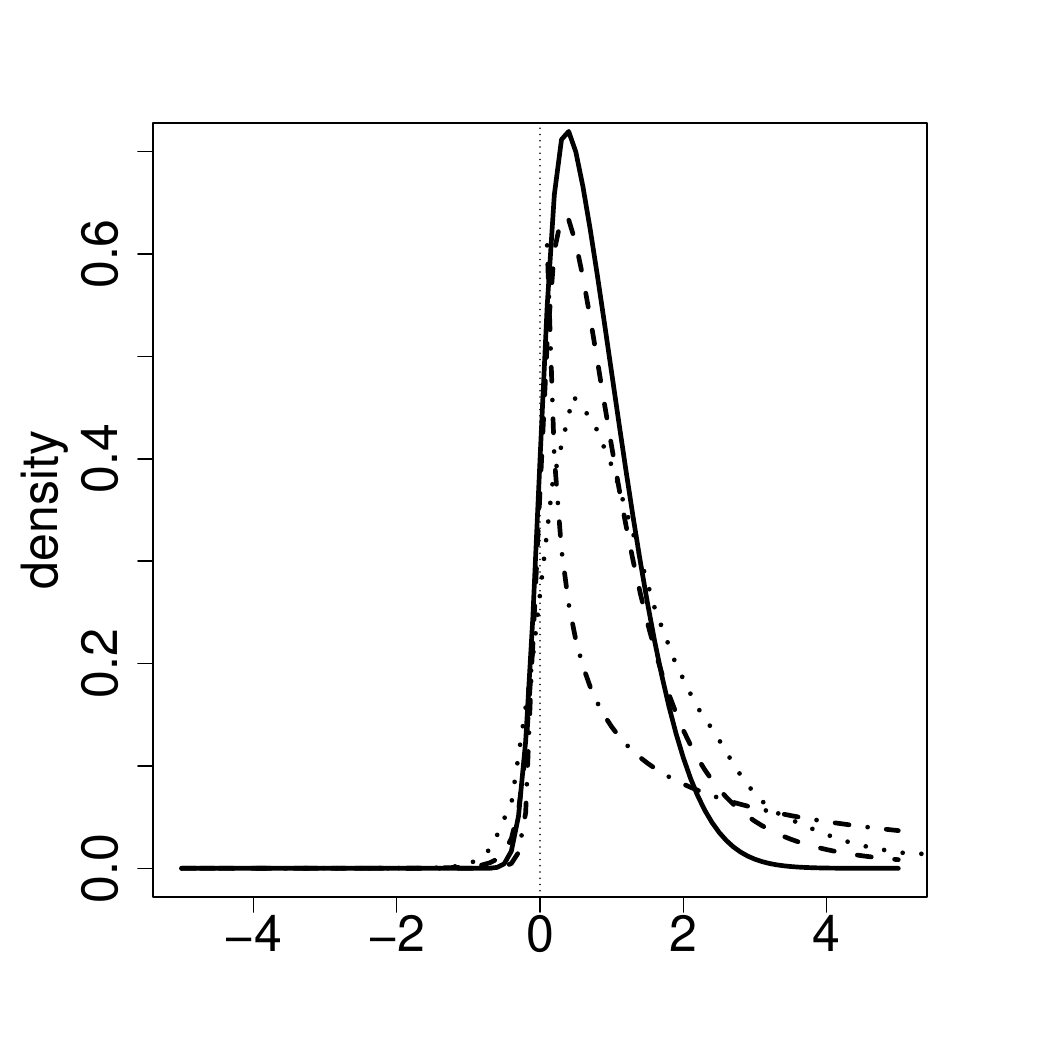} \\
 (a) symmetric-heavy-tail distributions  & (b) skew-heavy-tail distributions \\   
\end{tabular}
\end{center}
\caption{Illustration for the class of heavy-tailed distribution compared with Normal distribution (solid line): (a) symmetric distributions and (b) asymmetric distributions,  Student-t (dashed line), Slash (dotted line), and Variance-Gamma (dotdash line). } \label{sec2:fig1}
\end{figure}
Figure \ref{sec2:fig1} illustrates the behaviour of heavy tails distribution when compared to the Gaussian ($\mu=0$, $\sigma=1$).  As we can see, Panel (a) presents the symmetric distributions: the Student-t ($\mu=0$, $\sigma=1$, $\nu=3$), the Slash ($\mu=0$, $\sigma=1$, $\nu=3$) and the Variance-Gamma ($\mu= 0$, $\sigma=1$, $\nu=1$). In particular, for the Student-t distribution and the Slash distribution, as $\nu \rightarrow \infty$ we retrieve Normal tails, while smaller values of $\nu$ will induce thicker tails. Moreover, for the Variance-Gamma, parameters $\mu$, $\sigma$, and $\nu=3$ are responsible for, location, spread, and excess kurtosis (peakedness), respectively. Notice that, the tails of the Variance-Gamma distribution decrease slower than the Normal distribution and for $\frac{1}{\nu} \rightarrow 0$ implies Normal tails because kurtosis is $ 3(1+ \frac{1}{\nu}) \approx 3$. Otherwise, larger values of $\nu$ suggest peakedness. Panel (b) of Figure \ref{sec2:fig1} shows an example of skewness in the distributions with $\kappa=5$. 

%\noindent \textit{Modelling the mean $\mu_{ij}$} %The simpler way to model reserves is by considering a two-way analysis of variance, the ANOVA model given by 
 %\begin{equation}\label{sec2:eqmean1}
 % \mu_{ij} = \mu + \alpha_{i} + \beta_{j} + \gamma_{t=i+j-1},
 %\end{equation}
 %\noindent for $i=1, \ldots, I$ and $j=1, \ldots, J-i+1$, where $\mu$ is a level, $\alpha_i$, $\beta_{j}$ and $\gamma_{t}$ represent the accident year, development year, and calendar year effects, respectively. To avoid parameter identification problems, define $\sum_i \alpha_i = \sum_j \beta_j=0$.
 
 %If we assume that the accident year and calendar year effects are linear, the ANCOVA model is considered and follows
 % \begin{equation}\label{sec2:eqmean2}
 %     \mu_{ij} = \mu + i \alpha + \beta_{j} + (i+j-1)\gamma,
 % \end{equation}
%\noindent for $i=1, \ldots, I$ and $j=1, \ldots, J-i+1$. To avoid parameter identification problems, define $ \sum_j \beta_j=0$.
\vspace{0.5cm}
\noindent \textit{Modelling the mean $\mu_{ij}$} The mean function can be modelled assuming a dynamic linear structure where the parameters depend on each other in a time-recursive way. The idea is to take into account the interaction between accident and development year in the mean. The dynamic linear mean can be written as
\begin{equation}\label{sec2:eqmean3}
    \mu_{ij}= \mu + \alpha_i + \beta_{ij} +  \gamma_{t=i+j-1},
\end{equation}
\noindent for $i=2, \ldots, n$,  $j=2, \ldots, n-i+1$ and $\mu$ is a general mean between $i$ and $j$, $\alpha_i$, $\beta_{ij}$ and $\gamma_{t}$ represent a factor which reflects expected changes because of the accident year, a factor the interaction among accident year and development year, and calendar year effects, respectively. Notice that the dynamic variable $\beta_{ij}$ allows the development year pattern to change with accident years, which means, they capture the accident-development year interaction effects with hardly any parameters. Following a Bayesian formulation, $\alpha_i= \alpha_{i-1} + u_i$, $u_i \sim N(0, \sigma^{2}_{\alpha})$, $\beta_{ij}= \beta_{i-1, j} + v_i$, $v_i \sim N(0, \sigma^2_{\beta})$ and $\gamma_t$ as a random walk, $\gamma_t= \gamma_{t-1} + \omega_t$, $\omega_t \sim N(0, \sigma^2_{\gamma})$. Also, for parameter identification set the constraints as $ \alpha_1= \beta_{1j}=\gamma_1=0$. More details see \cite{Ntzoufras02}, \cite{Goudarzi2018}, \cite{Shi2012} and \cite{ChanJennifer2022}. %{\color{red}talvez fosse interessante falar um pouco mais do papel de cada parametro aqui ... qual objetivo de cada um e tal.. pensar aqui}

\subsection{Posterior distribution and inference procedure}\label{sec2.1}

Assume that for each period where $i \leq n$ and $j \leq n -i +1$ the loss reserving observations are available. Let $\mathbf{z}=(z_1', \ldots, z_n')$ the observational vector, $\mathbf{T}= (T_1', \ldots, T_n')$, and mixing vector $\boldsymbol{\lambda}=(\lambda_1', \ldots, \lambda_n')$, the vectors that contains the information about $z_{ij}$, $T_{ij}$, and $\lambda_{ij}$, respectively. The distribution obtained from equation (\ref{sec2:eq1}) is given by hierarchical structure as follows
%{\scriptsize
\begin{align}\label{sec2:eq10} \nonumber
	Z_{ij} \mid T_{ij}, \lambda_{ij}&\sim\mathcal{N}\left(\mu+\alpha_i+\beta_{ij}+\gamma_{t}+\rho T_{ij},\lambda_{ij}^{-1}\sigma^2\left(1-\rho^2\right)\right)\\ \nonumber
	\alpha_i&\sim\mathcal{N}\left(\alpha_{i-1},\sigma_{\alpha}^2\right)\\ \nonumber
	\beta_{ij}&\sim\mathcal{N}\left(\beta_{i-1,j},\sigma_{\beta}^2\right)\\ \nonumber
	\gamma_{t}&\sim\mathcal{N}\left(\gamma_{t-1},\sigma_{\gamma}^2\right)\\ \nonumber
	T_{ij} \mid \lambda_{ij}&\sim\mathcal{NT}_{[0,\infty)}\left(0,\sigma^2 \lambda_{ij}^{-1}\right)\\ 
	\lambda_{ij}&\sim p(\lambda_{ij} \mid \nu),
\end{align}
\noindent where $z_i'=(z_{i1}, \ldots, z_{in})$, $T_i'=(T_{i1}, \ldots, T_{in})$ and $\lambda_i'=(\lambda_{i1}, \ldots, \lambda_{in})$. 
As mentioned previously, we follow the Bayesian approach to make inference, predictions, access the goodness-of-fit, and model comparison that are obtained from the joint posterior distribution of the parameters. In particular, we address the hierarchical structure of our proposed model in equation (\ref{sec2:eq10}) in the proposed estimation algorithm to generate from the joint posterior and to make predictions.

The model description is complete after assigning a prior distribution for the static parameter vector $\boldsymbol{\theta}=\left(\mu,\rho,\sigma^2,\sigma_{\alpha}^2,\sigma_{\beta}^2,\sigma_{\gamma}^2,\nu\right)$. We assign {vague} independent priors to the parameters in $\boldsymbol{\theta}$.  A sensitivity analysis of priors is considered for these choices (see Section \ref{sec3.2}). In particular, we assume  $\mu\sim\mathcal{N}(0,s_{\mu}^2=100)$, $(1+\rho)/2\sim \mathcal{B}e(c=1,d=1)$, $\sigma^2\sim\mathcal{G}(a_{\sigma},b_{\sigma})$, $\sigma_{\alpha}^2\sim\mathcal{G}(a_{\alpha},b_{\alpha})$, $\sigma_{\beta}^2\sim\mathcal{G}(a_{\beta},b_{\beta})$ e $\sigma_{\gamma}^2\sim\mathcal{G}(a_{\gamma},b_{\gamma})$, with $a=0.001$ and $b=0.001$. For mixing parameter $\nu$ associated with the Student-t, the Slash, and the Variance-Gamma distribution, we assign a prior $\mathcal{G}(a_t=12,b_t=0.8)$, $\mathcal{G}(a_s=0.2,b_s=0.05)$, $\mathcal{G}(a_{vg}=12,b_{vg}=0.8)$, respectively. Notice that larger values of $\nu$ lead to approximate skew-Normal, while smaller values of $\nu$
suggest very thick tails, with $\kappa >0$ and hence $-1 \leq \rho \leq 1$. %We assume $a_t =$ and $b_t = $, favouring larger values of $\nu$, which represents the Skew-Gaussian model.
%
%and for proposed by \cite{Fonseca08}, $p(\nu)\propto\left(\frac{\nu}{\nu+3}\right)^{\frac{1}{2}}\left[\psi'\left(\frac{\nu}{2}\right)-\psi'\left(\frac{\nu+1}{2}\right)-\frac{2(\nu+3)}{\nu(\nu+1)^2}\right]^{\frac{1}{2}}$, and for the Slash and the Variance-Gamma models, we assign $\nu \sim \mathcal{U}(0,100)$ and $\nu \sim \mathcal{U}(1, 100)$, respectively.

Let $\boldsymbol{\Theta}= (\boldsymbol{\alpha}, \boldsymbol{\beta}, \boldsymbol{\gamma}, \boldsymbol{\theta})$ be the parameter vector associated with the skew-heavy-tailed model such that, $\boldsymbol{\alpha}= (\alpha_2, \ldots, \alpha_n)'$, $\boldsymbol{\beta}=(\beta_2', \ldots, \beta_n')$, with $\beta_i'=(\beta_{i1},  \ldots, \beta_{in})$, and $\boldsymbol{\gamma}= (\gamma_2, \ldots, \gamma_{n})$ the dynamical parameters for $i=2, \ldots, n$ and $j=2, \ldots, n$ with $\alpha_1= \beta'_1=\gamma_1=0$. Following Bayes’ theorem, the posterior distribution of $\boldsymbol{\Theta}$, $\boldsymbol{\lambda}$, ${\bf T}$, given the observed data, $\mathbf{z}$, is proportional to 
{\small
\begin{eqnarray}\label{eq11}\nonumber
    p(\boldsymbol{\alpha}, \boldsymbol{\beta},\boldsymbol{\gamma}, \boldsymbol{\theta} ,\boldsymbol{\lambda}, {\bf T} \mid  \mathbf{z}) &\propto& \left(\prod_{i=1}^{n}\prod_{j=1}^{n-i+1} f_{\mathcal{N}}(z_{ij} \mid \boldsymbol{\theta}, \lambda_{ij}, T_{ij}) f_{\mathcal{N}\mathcal{T}}(T_{ij} \mid \lambda_{ij}) p(\lambda_{ij} \mid \nu) \right)   \\ \nonumber 
     &\times& \left(\prod_{i=2}^{n} p(\alpha_i \mid \alpha_{i-1}) \prod_{i=2}^{n} \prod_{j=2}^{n-i+1} p(\beta_{ij} \mid \beta_{i-1,j}) \prod_{t=2}^{n} p(\gamma_t  \mid \gamma_{t-1}) \right)  \times \pi(\boldsymbol{\theta})\\ \nonumber
  %   &\times& \pi(\boldsymbol{\theta}),  \\ \nonumber
 &\propto& \prod_{i=1}^{n}\prod_{j=1}^{n-i+1}\left\{\left[\frac{\lambda_{ij}}{\sigma^2\left(1-\rho^2\right)}\right]^{1/2}\exp\left\{-\frac{\lambda_{ij}}{2}\left[\frac{\left(z_{ij}-\mu-\alpha_i-\beta_{ij}-\gamma_{t}-\rho T_{ij}\right)^2}{\sigma^2\left(1-\rho^2\right)}\right]\right\}\right\}\\
  &\times& \prod_{i=1}^{n}\prod_{j=1}^{n-i+1}\left\{\left[\frac{\lambda_{ij}}{\sigma^2}\right]^{1/2}\exp\left\{-\frac{\lambda_{ij}}{2\sigma^2}T_{ij}^2\right\}\boldsymbol{I}_{(0,\infty)}(T_{ij})p\left(\lambda_{ij}|\nu\right)\right\}\nonumber\\
  &\times& \left(\sigma_{\alpha}^2\right)^{-(n-1)/2}\exp\left\{-\frac{1}{2\sigma_{\alpha}^2}\sum_{i=2}^{n}\left(\alpha_{i}-\alpha_{i-1}\right)^2\right\}\nonumber\\
  &\times& \left(\sigma_{\beta}^2\right)^{-(n-2)(n-1)/4}\exp\left\{-\frac{1}{2\sigma_{\beta}^2}\sum_{i=2}^{n-1}\sum_{j=2}^{n-i+1}\left(\beta_{ij}-\beta_{i-1,j}\right)^2\right\}\nonumber \\
  &\times& \left(\sigma_{\gamma}^2\right)^{-(n-1)/2}\exp\left\{-\frac{1}{2\sigma_{\gamma}^2}\sum_{t=2}^{n}\left(\gamma_{t}-\gamma_{t-1}\right)^2\right\}\nonumber\\
  &\times& \exp\left\{-\frac{\mu^2}{2s_{\mu}^2}\right\}\left(\sigma^2\right)^{-\left(a_{\sigma}+1\right)}\exp\left\{-\frac{b_{\sigma}}{\sigma^2}\right\}(1+\rho)^{c-1}(1-\rho)^{d-1}\boldsymbol{I}_{(-1,1)}(\rho)\nu^{a_{l}-1}\exp\left\{-b_{l}\nu\right\}\nonumber\\
  &\times& \left(\sigma_{\alpha}^2\right)^{-(a_{\alpha}+1)}\exp\left\{-\frac{b_{\alpha}}{\sigma_{\alpha}^2}\right\}\left(\sigma_{\beta}^2\right)^{-(a_{\beta}+1)}\exp\left\{-\frac{b_{\beta}}{\sigma_{\beta}^2}\right\}\left(\sigma_{\gamma}^2\right)^{-\left(a_{\gamma}+1\right)}\exp\left\{-\frac{b_{\gamma}}{\sigma_{\gamma}^2}\right\},%\nonumber
\end{eqnarray}
}

\noindent where $f_\mathcal{N}( \cdot)$ denotes the density of the Gaussian distribution, $f_{\mathcal{N}\mathcal{T}}(\cdot)$ the density of the Truncated Gaussian distribution, $\pi(\cdot)$ stands for the prior distribution of the static parameters in $\boldsymbol{\theta}$, and $\boldsymbol{I}_{A}(\cdot)$ is a characteristic function of the subset $A$ and $l\in\{t,s,vg\}$. See that the resulting posterior distribution does not have an analytical solution, and we appeal to Markov chain Monte Carlo techniques {(MCMC)} \citep{gamerman} to obtain samples from this distribution. %In particular, posterior samples are obtained through a Gibbs sampler algorithm with steps of the Metropolis-Hastings algorithm. 
The inference procedure is detailed in Appendix \ref{apB}.

\subsection{Computational issues}\label{sec2.2}

Conditional on the latent variables $\mathbf{T}$ and $\boldsymbol{\lambda}$ the Gaussianity is preserved and hence simple to build an algorithm to estimate the model in Section \ref{sec2:sec2.1}. Below, we describe the sampling scheme from the posterior full conditional distributions through the Gibbs algorithm of the proposed model.

\begin{algorithm}[hbpt]
	\SetAlgoLined
	%\KwResult{Write here the result }
	Initialize $\boldsymbol{\theta}^{(0)}$, $\boldsymbol{\alpha}^{(0)}$, $\boldsymbol{\beta}^{(0)}$, $\boldsymbol{\gamma}^{(0)}$, $\mathbf{T}^{(0)}$ and $\boldsymbol{\lambda}^{(0)}$\;
	\For{$l=1,\dots,M$}{
		$\boldsymbol{\alpha}^{(l)}\sim p\left(\boldsymbol{\alpha} \mid \boldsymbol{\alpha}^{(l-1)},\boldsymbol{\beta}^{(l-1)},\boldsymbol{\gamma}^{(l-1)},\boldsymbol{\theta}^{(l-1)},\mathbf{T}^{(l-1)},\boldsymbol{\lambda}^{(l-1)},\mathbf{z}\right)$\;
        $\boldsymbol{\beta}^{(l)}\sim p\left(\boldsymbol{\beta} \mid \boldsymbol{\alpha}^{(l)},\boldsymbol{\beta}^{(l-1)},\boldsymbol{\gamma}^{(l-1)},\boldsymbol{\theta}^{(l-1)},\mathbf{T}^{(l-1)},\boldsymbol{\lambda}^{(l-1)},\mathbf{z}\right)$\;
        $\boldsymbol{\gamma}^{(l)}\sim p\left(\boldsymbol{\gamma} \mid \boldsymbol{\alpha}^{(l)},\boldsymbol{\beta}^{(l)},\boldsymbol{\gamma}^{(l-1)},\boldsymbol{\theta}^{(l-1)},\mathbf{T}^{(l-1)},\boldsymbol{\lambda}^{(l-1)},\mathbf{z}\right)$\;
        $\boldsymbol{\theta}^{(l)}\sim p\left(\boldsymbol{\theta} \mid \boldsymbol{\alpha}^{(l)},\boldsymbol{\beta}^{(l)},\boldsymbol{\gamma}^{(l)},\boldsymbol{\theta}^{(l-1)},\mathbf{T}^{(l-1)},\boldsymbol{\lambda}^{(l-1)},\mathbf{z}\right)$\;
        $\mathbf{T}^{(l)}\sim p\left(\mathbf{T} \mid \boldsymbol{\alpha}^{(l)},\boldsymbol{\beta}^{(l)},\boldsymbol{\gamma}^{(l)},\boldsymbol{\theta}^{(l)},\mathbf{T}^{(l-1)},\boldsymbol{\lambda}^{(l-1)},\mathbf{z}\right)$\;
        $\boldsymbol{\lambda}^{(l)}\sim p\left(\boldsymbol{\lambda} \mid \boldsymbol{\alpha}^{(l)},\boldsymbol{\beta}^{(l)},\boldsymbol{\gamma}^{(l)},\boldsymbol{\theta}^{(l)},\mathbf{T}^{(l)},\boldsymbol{\lambda}^{(l-1)},\mathbf{z}\right)$\;
	}
	\caption{Gibbs algorithm with adaptive Metropolis step.}
\end{algorithm}

For the parameter vector $\boldsymbol{\theta}$ whose complete conditional distribution is unknown ($\nu$ and $\rho$), we consider an additional step based on the adaptive Metropolis-Hastings algorithm (\cite{Andrieu2008}). We employ the same procedure for the mixing parameter $\boldsymbol{\lambda}$. 

More details about the complete conditional distributions of the parameters, latent effects and mixing variables are described in Appendix \ref{apB}. The algorithm was coded in \texttt{R} using \texttt{RStudio Version 2022.12.0-353} (\cite{RCoreTeam}). Due to the high autocorrelation of the samples generated from the latent effects and the unknown parameters obtained through the adaptive method, the algorithm will require many iterations to generate samples from the posterior distribution $p(\boldsymbol{\alpha}, \boldsymbol{\beta},\boldsymbol{\gamma}, \boldsymbol{\theta}, \boldsymbol{\lambda}, {\bf T} \mid  \mathbf{z})$.  For this, we use the \texttt{Rcpp} library to build improved loops of complete conditional distributions by performing part of the code implementation with the \texttt{C++} programming language and with the open source \texttt{C++} library for statistical computing, \texttt{armadillo-11.4.3}, available at the website \url{https://arma.sourceforge.net/}. %The source code is available at \url{https://github.com/williamleao/SHTLR}. 

\subsection{Predictions and Model Comparison}

In order to check the predictive accuracy of the proposed model, consider a vector $(\mathbf{z}^{obs}, \mathbf{z}^{pred})$, with $(n+h)$ dimensional where $\mathbf{z}^{obs}$ and $\mathbf{z}^{pred}$ representing, respectively, the $n$-th dimensional observed and the $h$-th dimensional out-of-sample values of $\mathbf{z}$. Here, the vector $\mathbf{z}^{obs}$ represents the information available in the upper triangle unless the observations are left out in the inference procedure, represented by $\mathbf{z}^{pred}$. The idea is to verify if the proposed model is able to provide good performance to predict the lower triangle. 

To obtain samples from the posterior predictive distribution $p(\mathbf{z}^{pred} \mid \mathbf{z}^{obs})$, we appeal to composition sampling where assuming that $\boldsymbol{\theta}^{(k)}, \boldsymbol{\alpha}^{(k)}, \boldsymbol{\beta}^{(k)}, \boldsymbol{\gamma}^{(k)}, \boldsymbol{\lambda}^{(k)}, \mathbf{T}^{(k)}$ represent the $k$-th sample from the joint posterior distribution $p(\boldsymbol{\theta}, \boldsymbol{\alpha}, \boldsymbol{\beta}, \boldsymbol{\gamma}, \boldsymbol{\lambda}^{obs}, \mathbf{T}^{obs} \mid \mathbf{z}^{obs})$. Thus, samples from $p(\mathbf{z}^{pred} \mid \mathbf{z}^{obs})$ are obtained by computing from %the integral as
%\begin{eqnarray}\nonumber
%    p(\mathbf{z}^{pred} \mid \mathbf{z}^{obs}) &=& \int \int \int p(\mathbf{z}^{pred} \mid \mathbf{T}^{pred}, \boldsymbol{\lambda}^{pred}, \boldsymbol{\Theta}) p(\mathbf{T}^{pred} \mid \boldsymbol{\lambda}^{pred}) p(\boldsymbol{\lambda}^{pred} \mid \nu) p(\boldsymbol{\Theta} \mid \mathbf{z}^{obs}, \boldsymbol{\lambda}^{obs}, \mathbf{T}^{obs}) d\boldsymbol{\Theta}d\boldsymbol{\lambda}^{pred} d\mathbf{T}^{pred} \\ 
 %    &=& 
%\end{eqnarray}
\begin{itemize}
    \item [(i)] $\boldsymbol{\lambda}^{pred} \mid \boldsymbol{\lambda}^{(k)}, \nu^{(k)}$ and
    \item [(ii)] $\mathbf{T}^{pred} \mid \mathbf{T}^{(k)}, \boldsymbol{\lambda}^{pred}$ and
    \item [(iii)] $\mathbf{z}^{pred} \mid \mathbf{z}^{obs}, \boldsymbol{\lambda}^{pred},\mathbf{T}^{pred}, \boldsymbol{\theta}^{(k)}, \boldsymbol{\alpha}^{(k)}, \boldsymbol{\beta}^{(k)}, \boldsymbol{\gamma}^{(k)}$.
\end{itemize}
\noindent Notice that (i) represents the distribution responsible for accommodating the tails of the distribution as seen previously in Section \ref{sec2:sec2.1}. The distributions in (ii) and (iii) are the Truncated Gaussian and the Gaussian, respectively. The same procedure is considered to obtain the lower triangle.

To model comparison, we consider measures based on scoring rules and the root mean square prediction error (RMSPE) assessing the predictive performance of the competing models.  Scoring rules provide summary measures for the evaluation
of probabilistic forecasts, by assigning a numerical score based
on the predictive distribution and on the event or value that materializes \citep{GneitRaf07}. In this case, we consider the Interval Score (IS), the Width of Credible Interval (CWI), and the Continuous Ranked Probability Score (CRPS). A brief description is presented in Appendix \ref{apC}.

\subsection{Bayesian residual analysis}\label{sec2.5}
Consider $Z_{ij}$ observations at policy year $i$ with development $j$ as defined in equation (\ref{sec2:eq10}) such that $Z_{ij} \mid T_{ij}, \lambda_{ij}, \boldsymbol{\Theta} \sim \mathcal{N}\left(\mu_{ij}+\rho T_{ij},\lambda_{ij}^{-1}\sigma^2\left(1-\rho^2\right)\right)$. Then the standardized Bayesian residual for the mixture model is

\begin{equation}
    r_{ij}= \frac{Z_{ij}- E\left( Z_{ij} \mid \boldsymbol{\Theta}, \lambda_{ij}, T_{ij} \right) }{\sqrt{Var\left( Z_{ij} \mid \boldsymbol{\Theta}, \lambda_{ij}, T_{ij} \right)}},
\end{equation}
with $i\leq n$ and $j\leq n-i+1$ . If the errors have Gaussian distribution, then approximately 95\% of the individual residuals are expected to be in the interval $[-2,2]$. Otherwise, there is some evidence that this observation could be a potential outlier.

\newpage
\section{Applications}\label{sec3}

This section presents two applications to the discussion about loss reserving in insurance. A realistic simulated dataset that emulates a real behaviour present in loss-reserving data to evaluate the performance of our proposed methodology; and a case study that considers the claim amounts paid to the insureds of an insurance company from 1978 to 1995. These data have been previously analysed by \cite{Choy2008} and \cite{Jennifer16}. The idea of the case study is to illustrate the flexibility of our proposed model which is able to account jointly skewness and heavy tails when compared to well-known models used in the literature of loss reserving. Codes with the simulated study and inference procedure are available at \url{https://github.com/williamleao/SHTLR}.

%Our proposal is able to account jointly skewness and heavy-tails instead of only a heavy-tailed model in \cite{Jennifer16}.  %The second application focuses on a real dataset from a life insurance company based in Brazil from XXX and XXX. ....bla bla bla {\color{red} usar os dados aqui do soa igual la na mariana}

We fit the models described in Table \ref{tabmodels}, which are particular cases of the general model proposed in Section \ref{sec2}. To verify how our proposed models are able to predict throughout the lower triangle in {the case study}, we consider a cross-validation predictive evaluation where a part of the upper triangle was used for training, and the remaining was left out of the inference procedure for validation. 
\begin{table}[H]
  \caption{Competing models fitted to data applications: Normal ($\mathcal{N}$), Student-t ($\mathcal{ST}$), Slash ($\mathcal{S}$), Variance-Gamma ($\mathcal{VG}$), Skew-Normal ($\mathcal{SN}$), Skew-Student-t ($\mathcal{SST}$), Skew-Slash ($\mathcal{SS}$), Skew-Variance-Gamma ($\mathcal{SVG}$).   }\label{tabmodels}
\centering
  \begin{tabular}{|lccc|}
    \hline
\textbf{class}  &  \textbf{model}  & $\lambda$ & $T$\\
    \hline
&$\mathcal{N}$ & $1$ & -  \\
non-Skew&$\mathcal{ST}$ & $ \mathcal{G}(\nu/2,\nu/2)$  & - \\
&$\mathcal{S}$ &      $\mathcal{B}e(\nu,1)$ & - \\ 
&$\mathcal{VG}$ & $ \mathcal{I}\mathcal{G}(\nu/2,\nu/2)$   & - \\
\hline
&$\mathcal{SN}$ & 1& $\mathcal{N}\mathcal{T}_{[0,\infty)}(0, \sigma^2)$ \\
Skew &$\mathcal{SST}$ & $ \mathcal{G}(\nu/2,\nu/2)$  & $\mathcal{N}\mathcal{T}_{[0,\infty)}(0, \sigma^2\lambda^{-1})$\\
&$\mathcal{SS}$ & $\mathcal{B}e(\nu,1)$ & $\mathcal{N}\mathcal{T}_{[0,\infty)}(0, \sigma^2\lambda^{-1})$\\ 
&$\mathcal{SVG}$ &  $ \mathcal{I}\mathcal{G}(\nu/2,\nu/2)$  & $\mathcal{N}\mathcal{T}_{[0,\infty)}(0, \sigma^2\lambda^{-1})$\\
\hline
    \end{tabular}
\end{table}

\subsection{A simulated Skew-Heavy-tailed dataset}\label{sec3.1}

%mu.t=8.8501
%nu.t=2.3393
%sig2.t=0.1536
%rho.t=-0.9513
%salp2.t=0.1088
%sbeta2.t=0.0447
%sgama2.t=0.1084

This section presents a simulated study to investigate the convergence of the parameters and predictive performance of the competing models in identifying skewness and heavy tails. For this, we consider simulating an artificial run-off triangle with $16 \times 16$ (totalling $n=136$ observations in the upper triangle)  from a Skew Student-t distribution, assuming that the mean function $\mu_{ij}= \mu + \alpha_{i} + \beta_{ij} + \gamma_t$, with $\mu=9$, $\alpha_{i} \mid \alpha_{i-1} \sim \mathcal{N}(0, \sigma^2_{\alpha}= 0.13 )$, $\beta_{ij} \mid \beta_{i-1 j} \sim \mathcal{N}(0, \sigma^2_{\beta}= 0.05 )$, $\gamma_{t} \mid \gamma_{t-1} \sim \mathcal{N}(0, \sigma^2_{\gamma}= 0.13 )$,  $i=2, \ldots, n$, $j=2, \ldots, n-i+1$, $t= i+j-1$, and parameters related on variability, kurtosis and asymmetry as {$\nu=3 $,  $\rho=-0.89$ and $\sigma^2= 0.14 $}, respectively, which were defined based on the posterior point estimates obtained for the parameters in the dataset from \cite{Jennifer16} . We are interested in predicting the total reserve (light grey cells as shown in Table \ref{tab1}, which means $Y_{\nabla}$). The true total loss reserving given by the lower triangle is 33,047.39 currency units.  Priors on parameters are the same as shown in Section \ref{sec2.1}. The burn-in and lag for spacing of the chain were selected so that the effective sample size was around 5,000 samples. To model comparison, we consider the use of scoring rules and evaluate parameters responsible for kurtosis and asymmetry and compare four space-state competing models: $\mathcal{N}$, $\mathcal{ST}$, $\mathcal{SN}$ and $\mathcal{SST}$, that last representing the data-generating model. Also, the usual chain ladder method was considered to project the loss reserving as a naive technique. %Codes for the simulated dataset are available at \url{https://github.com/williamleao/SHTLR}, which includes the chain ladder method, the Skew-Normal, and the Skew-t algorithms.

\begin{figure}[H]
    \centering
    \begin{tabular}{ccc}
    \includegraphics[width=4.5cm]{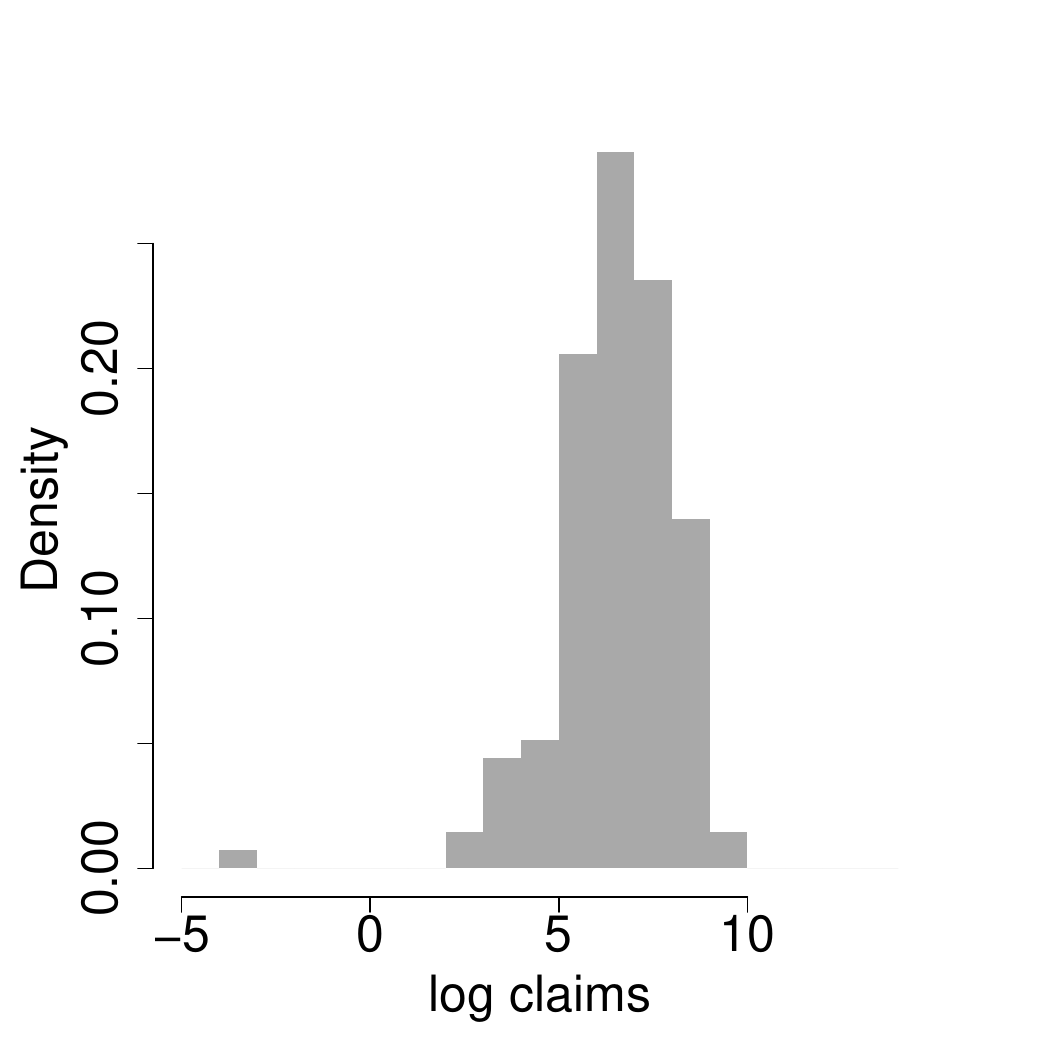} &   \includegraphics[width=4.5cm]{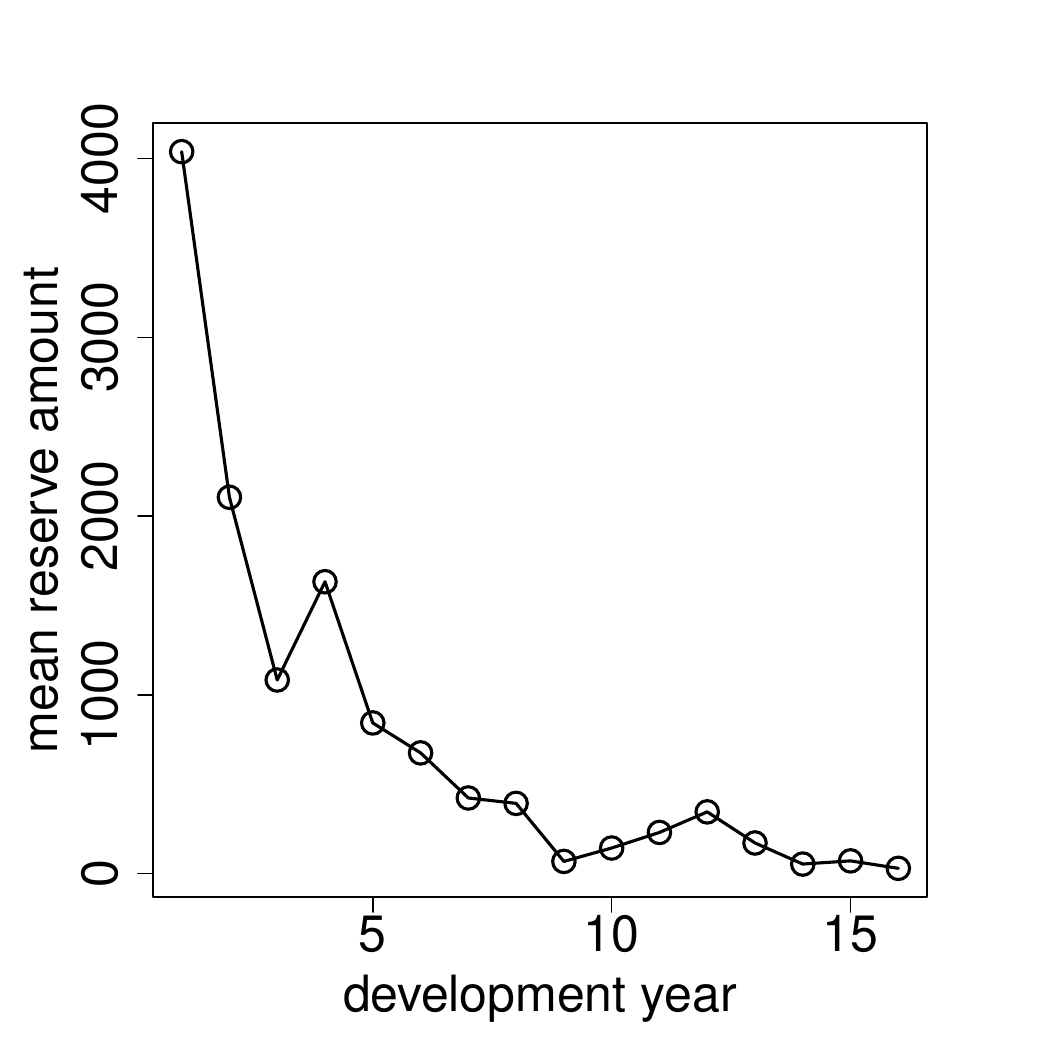}& \includegraphics[width=4.5cm]{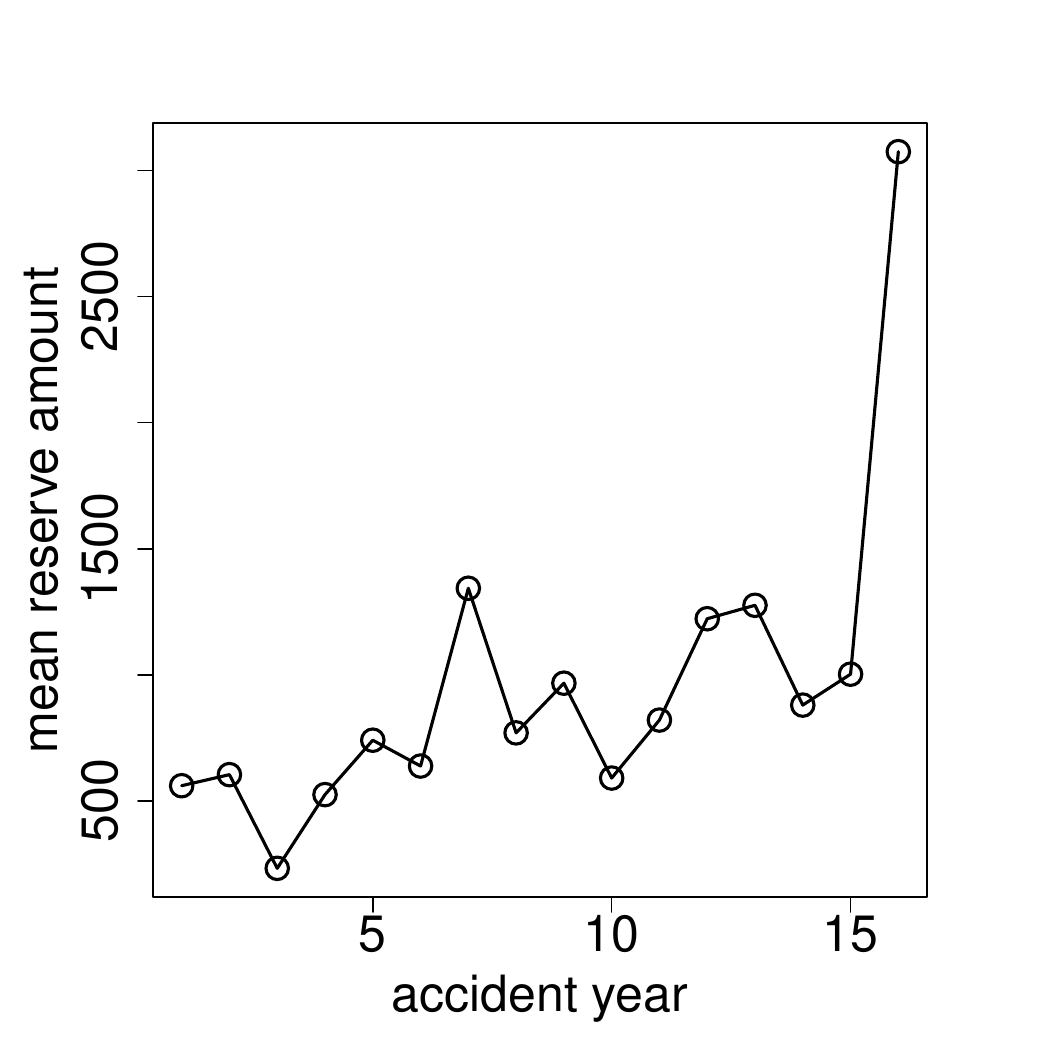} \\
    (a) log-claims distribution & (b)  development year & (c) accident year \\
    \end{tabular}
    \caption{Simulated dataset: (a) histogram of the log claims, (b) the empirical mean of the development lag year trend, and (c) the empirical mean of the accident year.}
    \label{apB:fig1}
\end{figure}

Figure \ref{apB:fig1} shows the behaviour of the simulated dataset in the log scale  (see panel (a)) indicating skewness and ticker tails. See that the behaviour of the claims is very irregular and contains some extreme outliers. Panels (b) and (c) of Figure \ref{apB:fig1} show the trend of effects development year and accident year in terms of the empirical mean, respectively. As we can see, in panel (b) the average of these trends decreases over development years. To the effect of the average accident year (panel (c)), we notice a volatility trend over the years.

\begin{figure}[H]
    \centering
    \begin{tabular}{cc}
    \includegraphics[width=5cm]{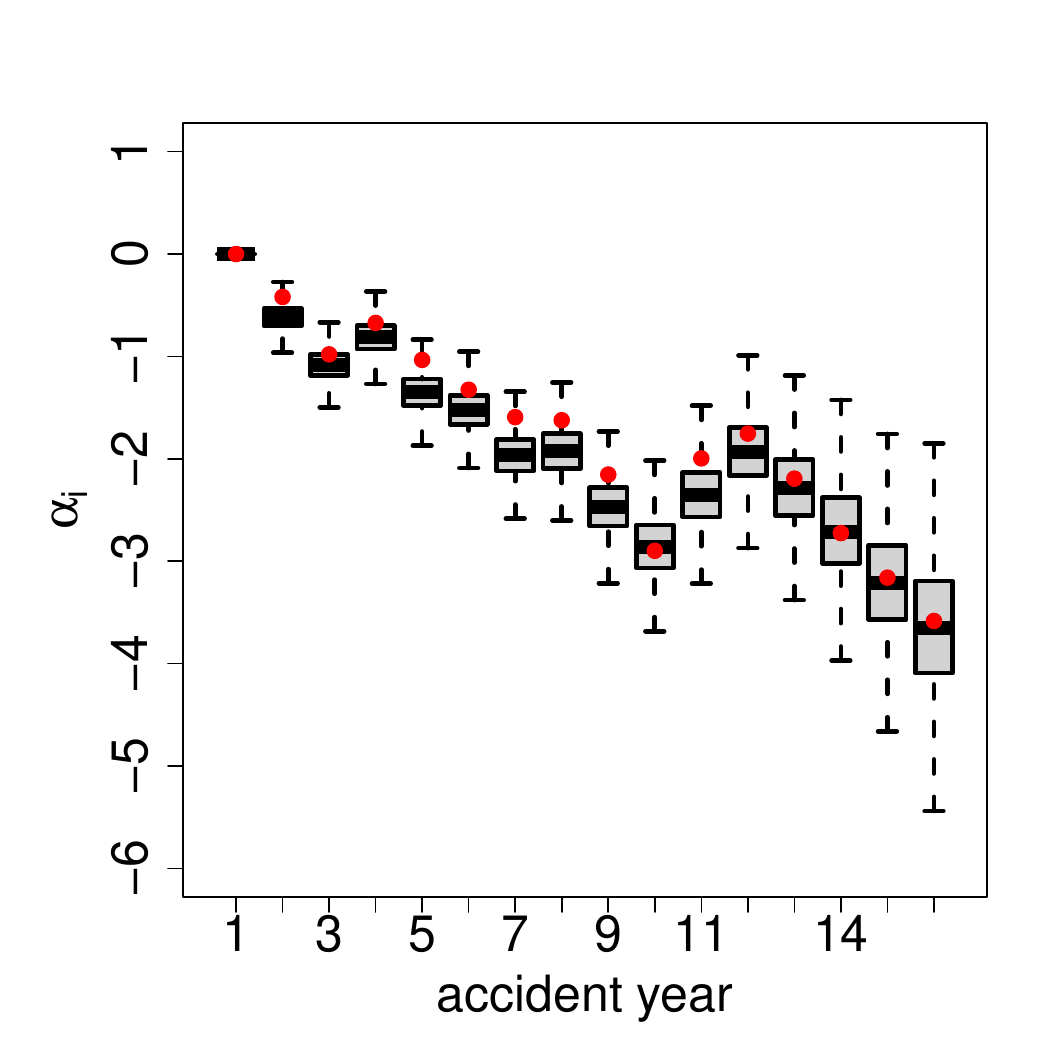}&
     \includegraphics[width=5cm]{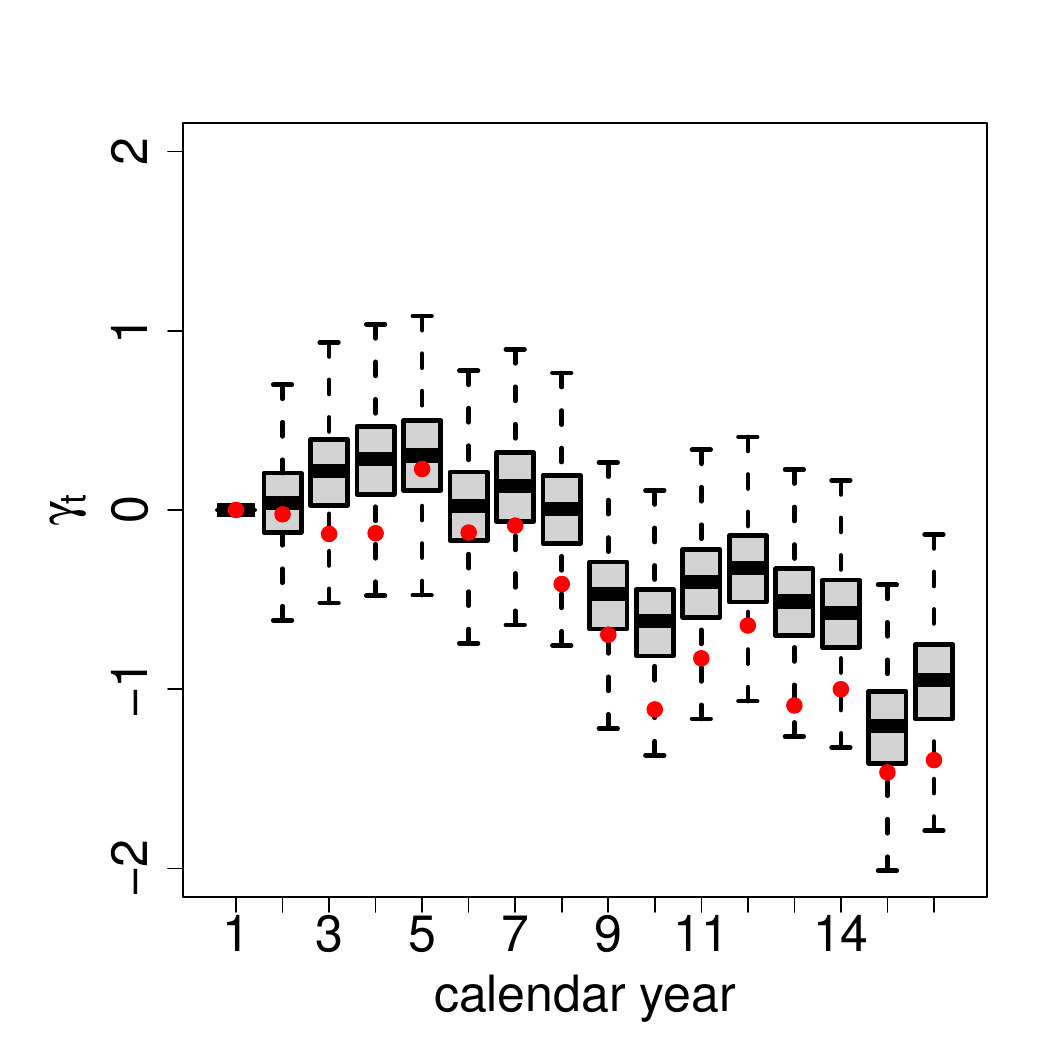} \\
     (a) & (b) \\
    \multicolumn{2}{c}{\includegraphics[width=10cm]{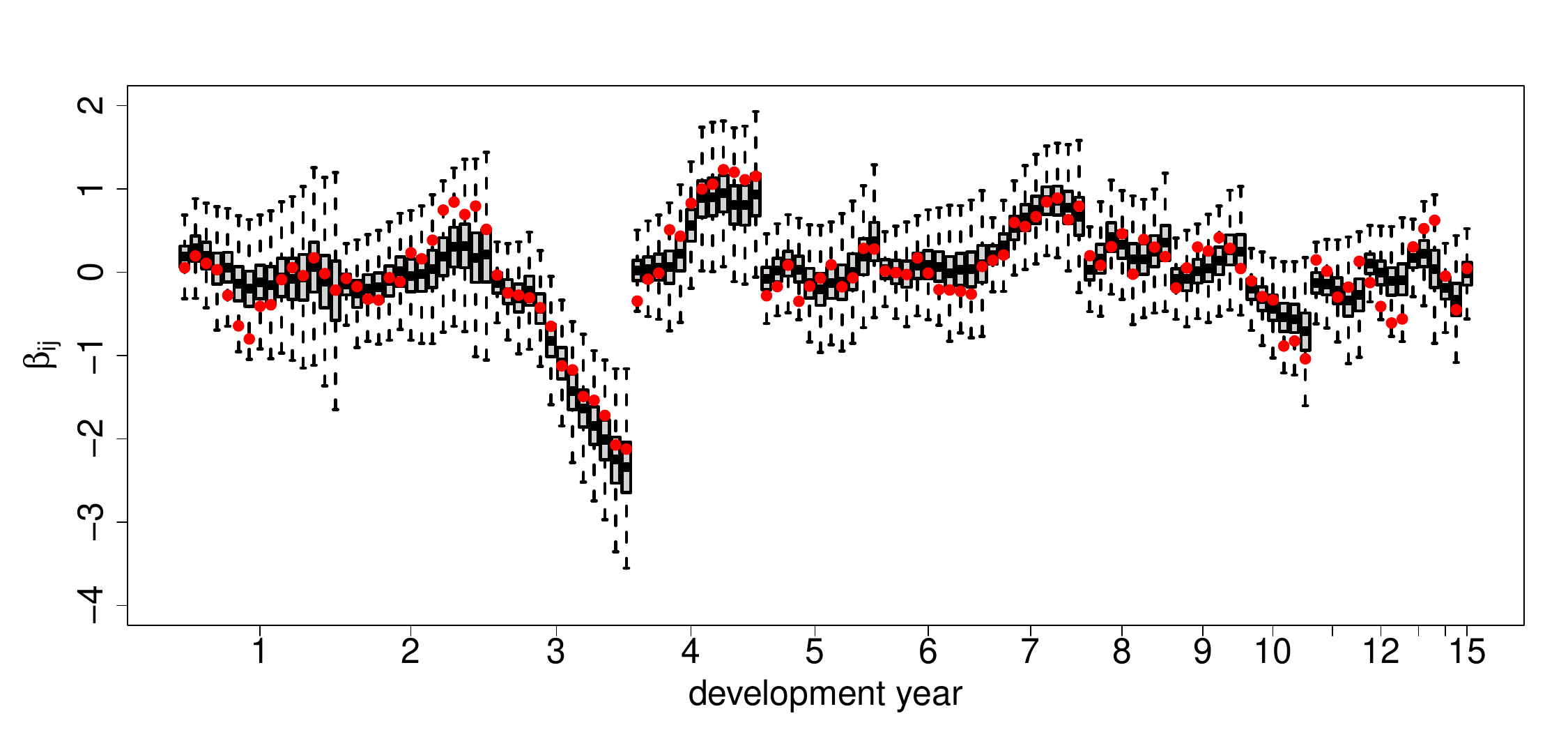}} \\
    \multicolumn{2}{c}{(c)}
    \end{tabular}
    \caption{Simulated dataset: posterior summaries for the dynamic mean effects $\alpha_{i}$, $\gamma_{t}$ and $\beta_{ij}$ under the Skew-t model. The box-plot represent the distribution of the parameters and the red dots represent the true values generated from the Skew-t process.}
    \label{apB:fig2}
\end{figure}

Posterior summaries are addressed for the competing models. The parameters to be estimated are the dynamic coefficients  ($\alpha_{i}$, $\beta_{ij}$, $\gamma_{t})$), the variance parameters ($\sigma^2_{\alpha}$, $\sigma^2_{\beta}$, $\sigma^2_{\gamma}$, $\sigma^2$), the function of the skew parameter $\rho$, the mixing parameter $\nu$ and the latent mixing variables ($\lambda_{ij}$, $T_{ij}$), $i=1, \dots, n$ and $j=1, \ldots, n$. Figure \ref{apB:fig2} presents the posterior distribution for the dynamic parameters $\alpha_i$, $\beta_{ij}$ and $\gamma_t$ compared with the true values (red dot), indicating that the estimates are well recovered by the Skew-t model. We have already expected this result since the data are generated from a Skew-t process. The first panel (a) shows a decreasing trend over accident years with low volatility, whereas, in panel (c), the state parameter $\beta_{ij}$ shows a trend of increase of the development year during the first two years and then in the third, the development year decrease abruptly over accident years. Notice that the third lag-year effect is low due to the fact the value of observations decreases over accident years. After, they turn to increase and maintain less volatility over the years. For the calendar year parameter, see an increasing trend over time and is relatively stable over in the first years (see panel (b)).

\begin{figure}[H]
    \centering
    \begin{tabular}{cc}
    \includegraphics[width=4cm]{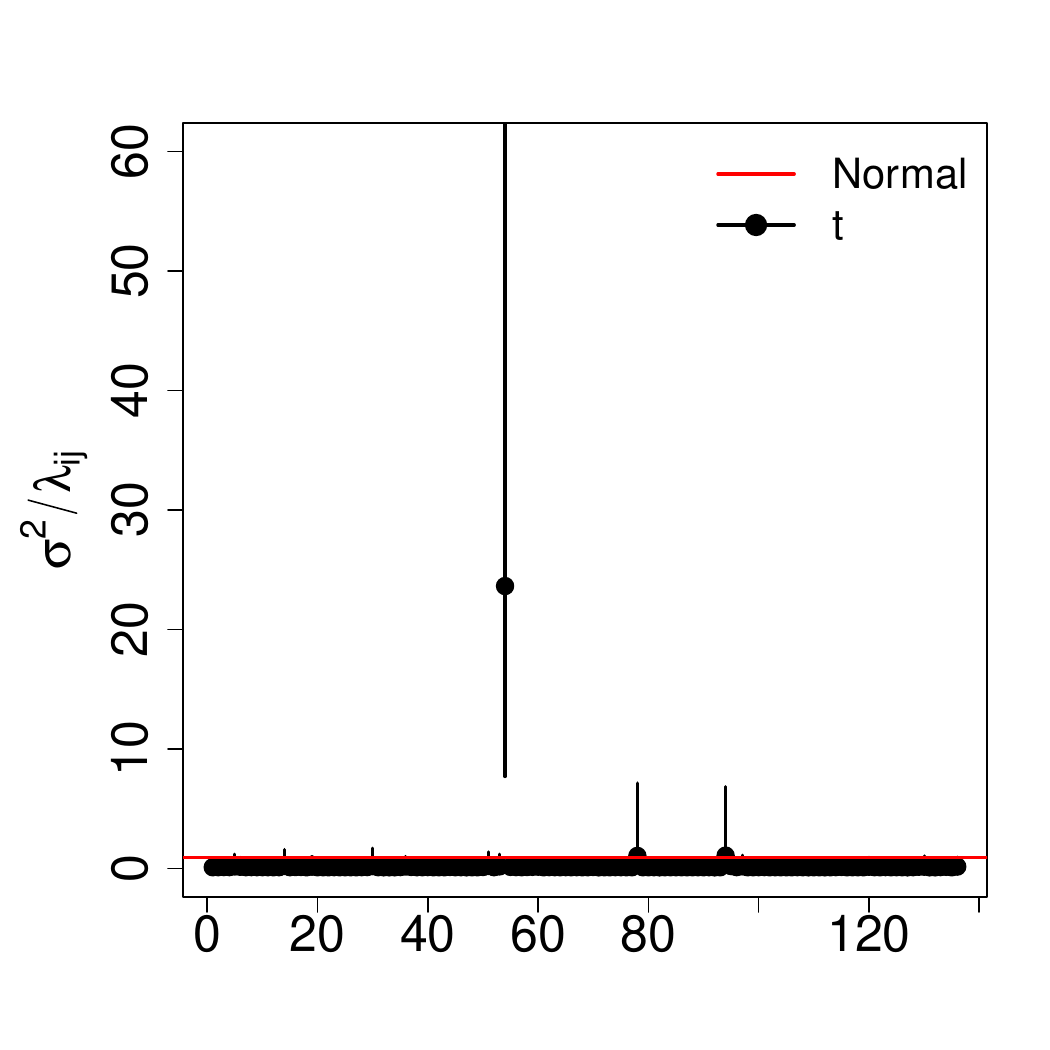}&  \includegraphics[width=4cm]{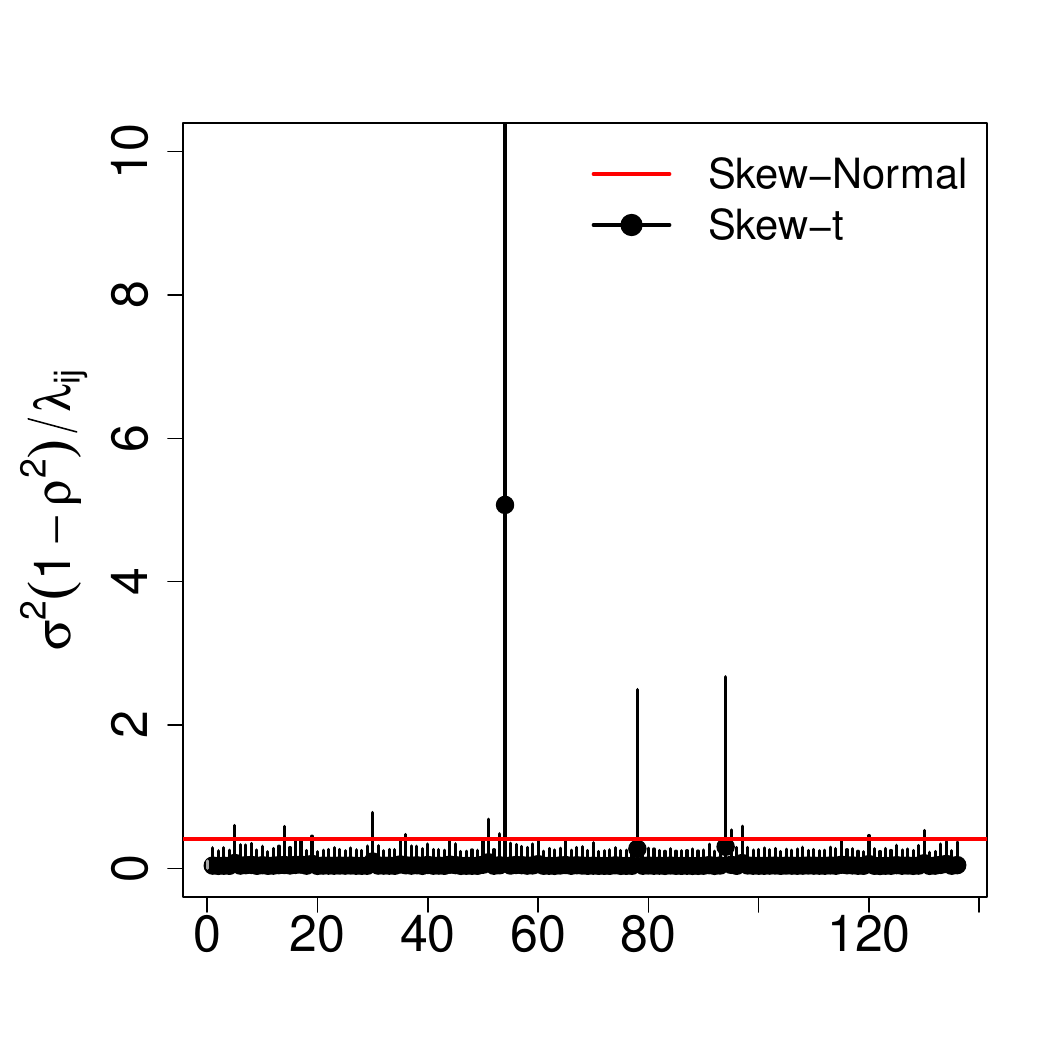}\\
    (a) $\sigma^2/\lambda_{ij}$ & (b) $\sigma^2(1-\rho^2)/\lambda_{ij}$\\
    \end{tabular}
    \caption{Simulated dataset: posterior median with 95\% posterior credible intervals for the variance estimated structure assuming (a) the non-skew models $\sigma^2/\lambda_{ij}$ and (b) the skew models $\sigma^2(1-\rho^2)/\lambda_{ij}$. The red line represents the variance of the Gaussian class.}
    \label{apB:fig3} 
\end{figure}
\begin{figure}[H]
    \centering
    \begin{tabular}{c}
    \includegraphics[width=10cm]{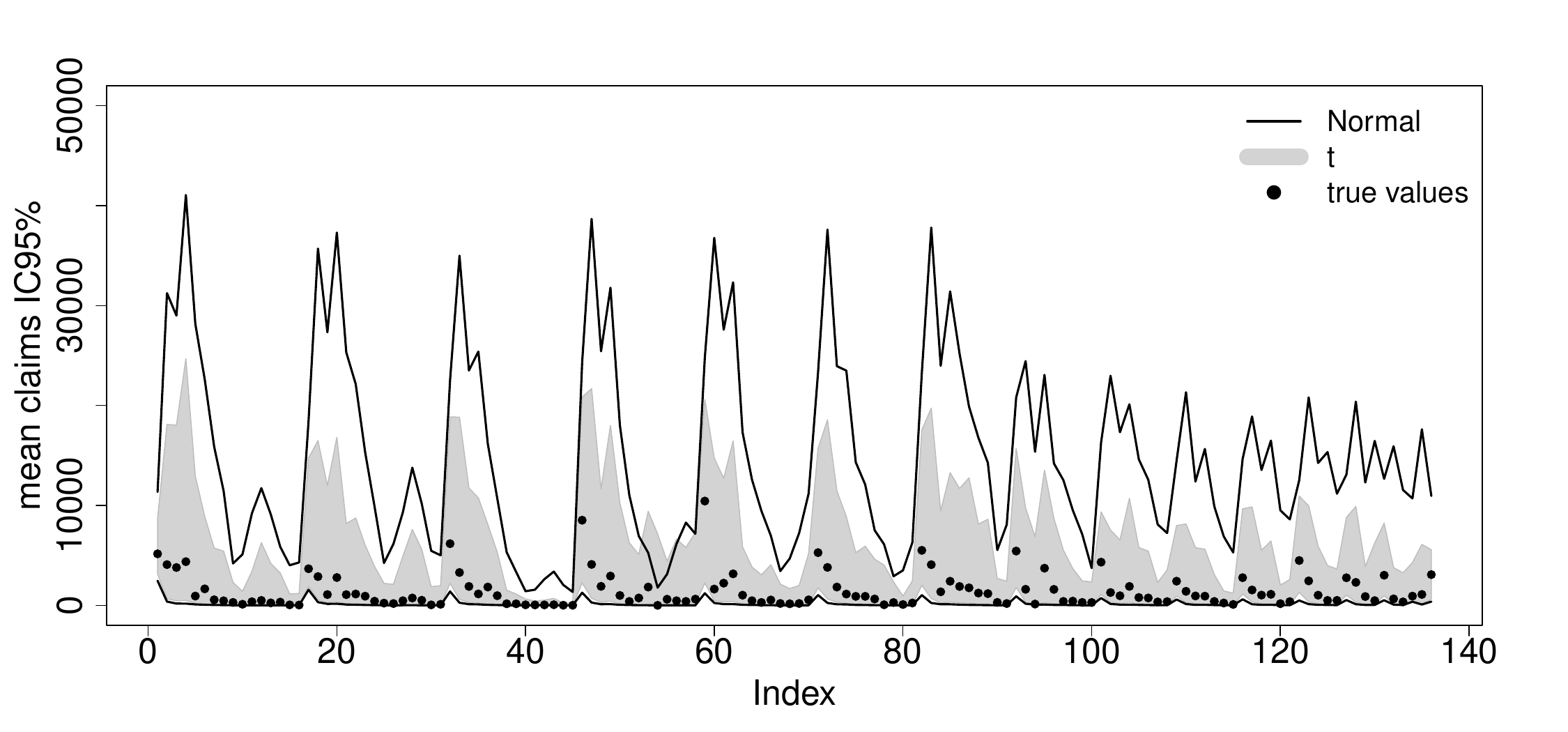} \\
     (a) non-Skew models \\   
     \includegraphics[width=10cm]{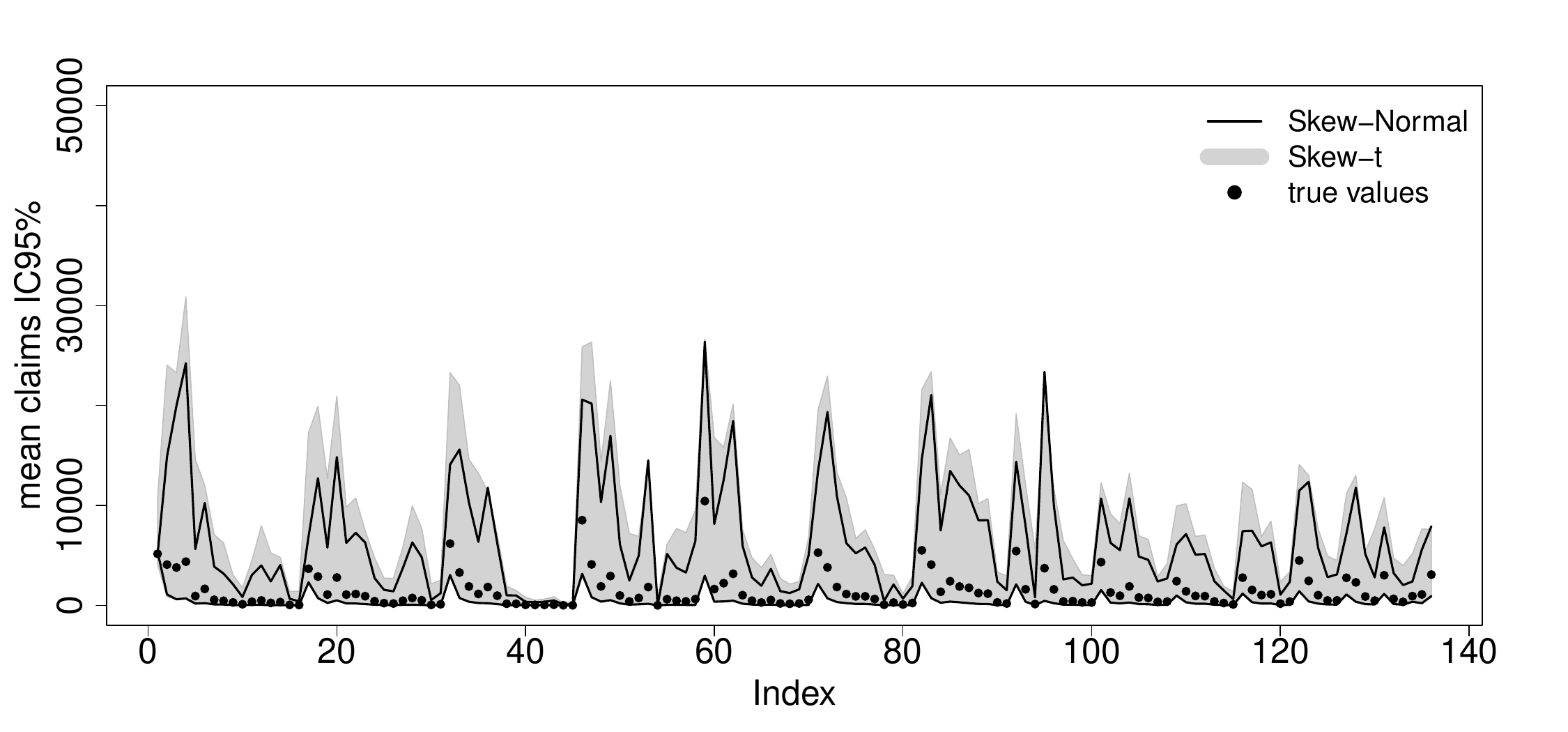} \\
     (b) Skew models
    \end{tabular}
    \caption{Simulated dataset: posterior claims with 95\% credible interval and true values (black dotted) for the dynamical mean $\mu_{ij}$, (a) the non-skew models and (b) the skew models. }
    \label{apB:fig4}
\end{figure}

In the most complete model given by equation (\ref{sec2:eq10}), the mean of $Z_{ij}$ conditional on $\lambda_{ij}$ and $T_{ij}$ is $\mu_{ij}= \mu+\alpha_i+\beta_{ij}+\gamma_{t}+\rho T_{ij}$ and the variance is $\sigma^2_{ij}= \sigma^2(1-\rho^2)/\lambda_{ij}$, from which we can deduce the variance structures over development and accident years. Figure \ref{apB:fig3} illustrates the variance $\sigma^2_{ij}$ estimated for the Gaussian, the Student-t, the Skew-Normal, and the Skew-Student-t models. Note that the variance is non-constant with some peaks over the development periods. As we can see, the class of the Student-t models can recover peaks of variability and potential outliers. Although the Skew-Normal model is able to capture changes in the skews of the distribution, the model does not take into account volatility over the development periods, which means $\lambda_{ij}=1$ for all $i=1, \ldots, n$ and $j=1, \ldots, n$ (see panel (b)). Notice that if we assume $\lambda_{ij}=1$ and $\rho=0$ the posterior variance for an usual dynamical Gaussian model is considered (see red line in panel (a)). The variance of the non-skew class is higher than the skew class (see axis $y$ in Figure \ref{apB:fig3}).   Table \ref{apB:tab2} shows the posterior summary for the Skew-t model. See that, the median posterior of $\rho$ is equal to -0.877 (with true value -0.890) indicating the Skew-t model captures the presence of skewness in the data. Panels of Figure \ref{apB:fig4} present the 95\% credible intervals posterior summaries of the mean $\mu_{ij}$ for the competing models compared with the true values (black dots) indicating that the proposed models recover the true mean over the period. Although the Gaussian model covers the true values, the model has a higher range of 95\% posterior credible intervals for $\mu_{ij}$ compared to the others.

\begin{figure}[H]
    \centering
    \begin{tabular}{cc}
    \includegraphics[width=4cm]{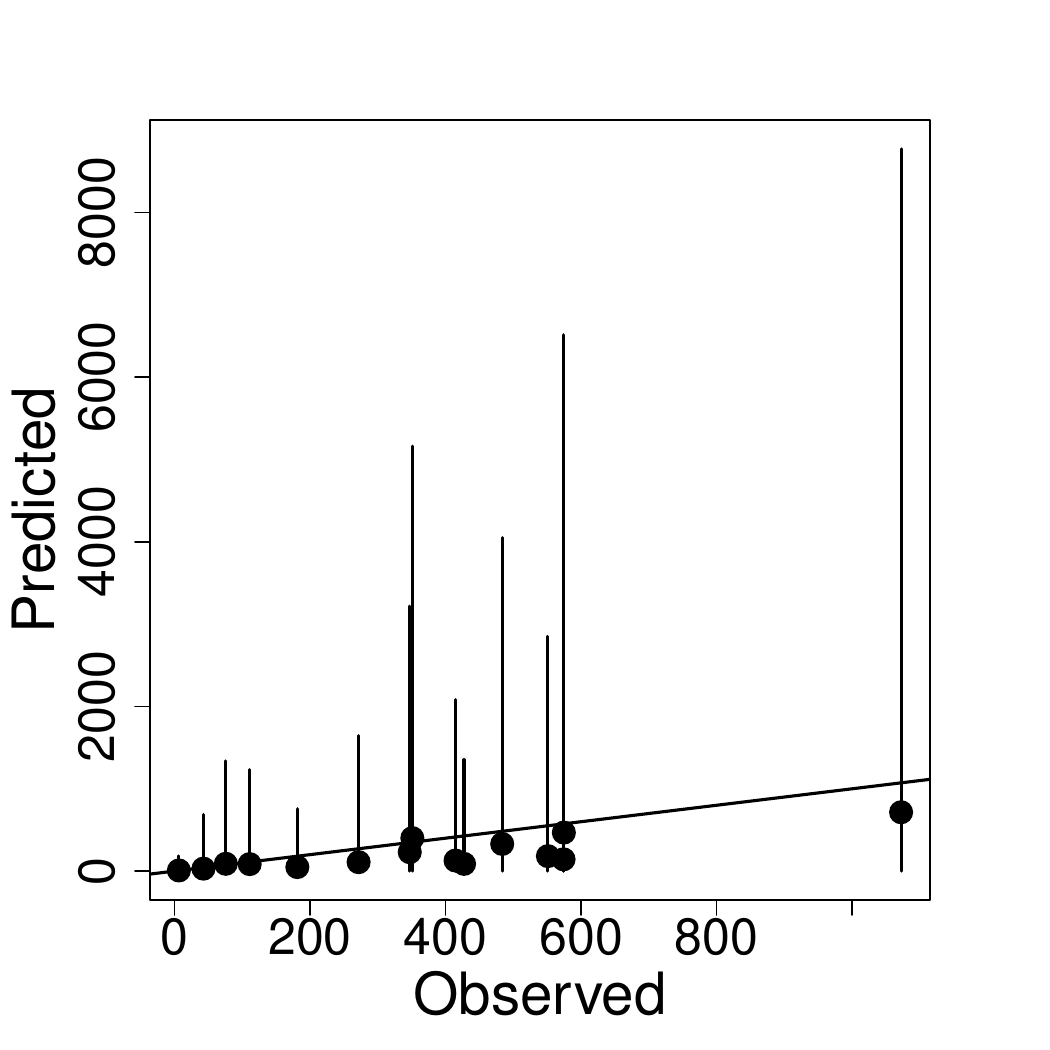}  &   \includegraphics[width=4cm]{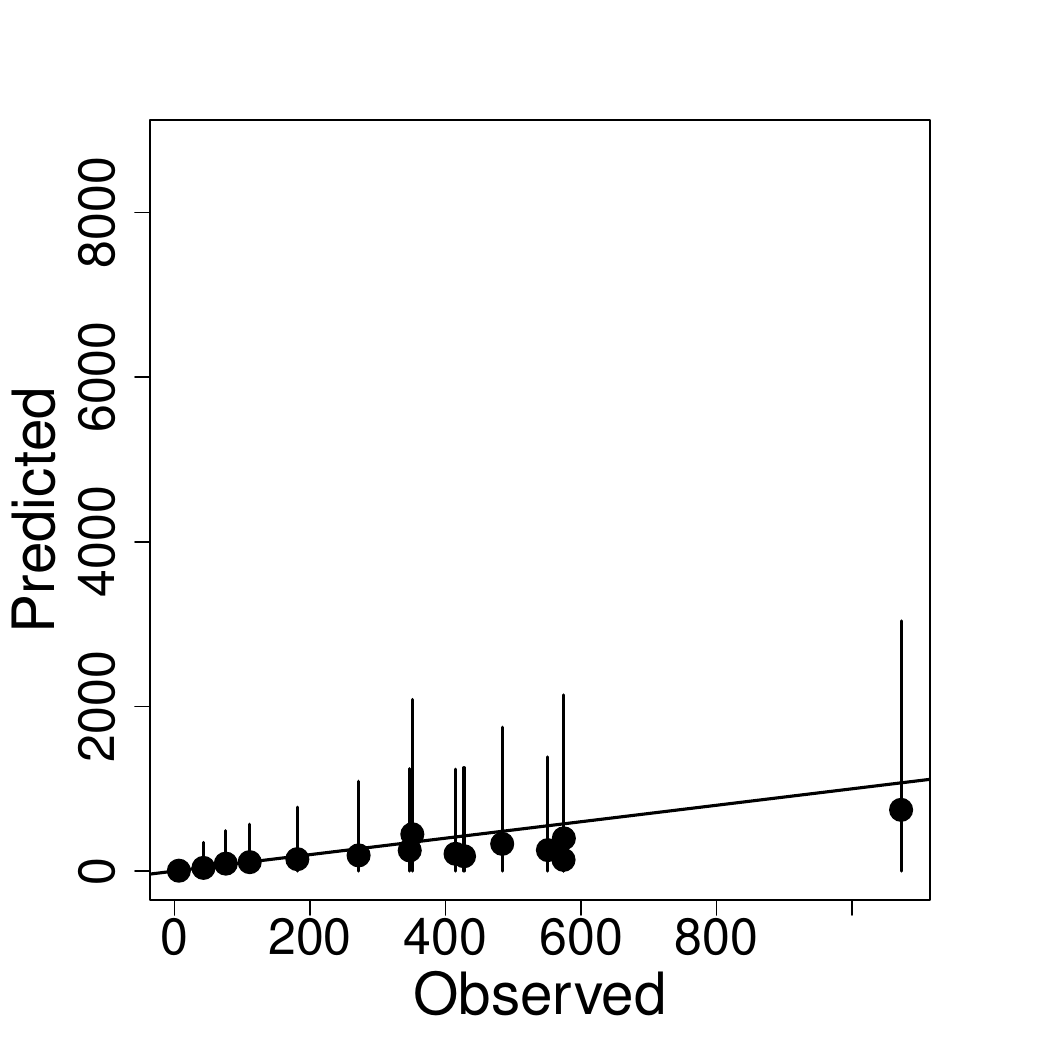} \\
     (a) Normal  & (b) Student-t\\   
     \includegraphics[width=4cm]{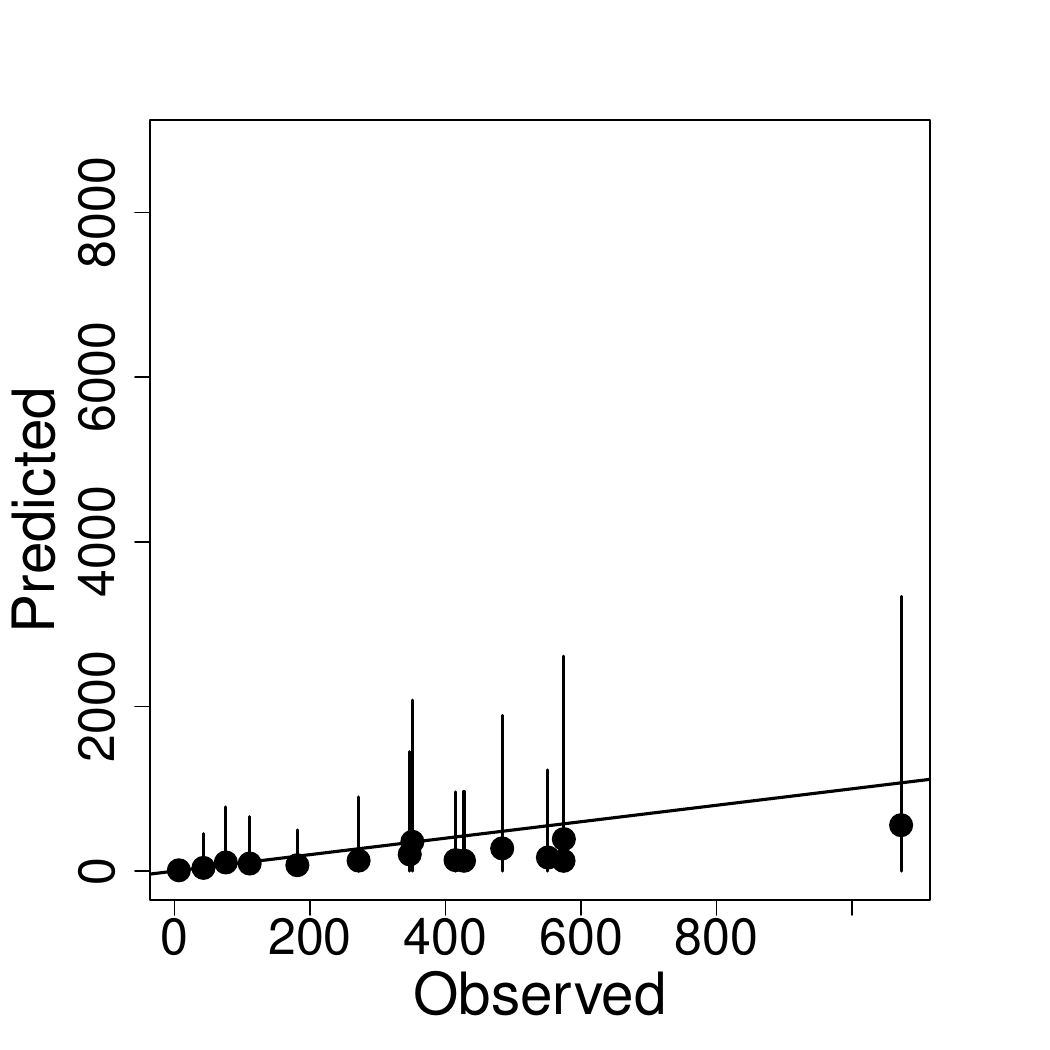} & \includegraphics[width=4cm]{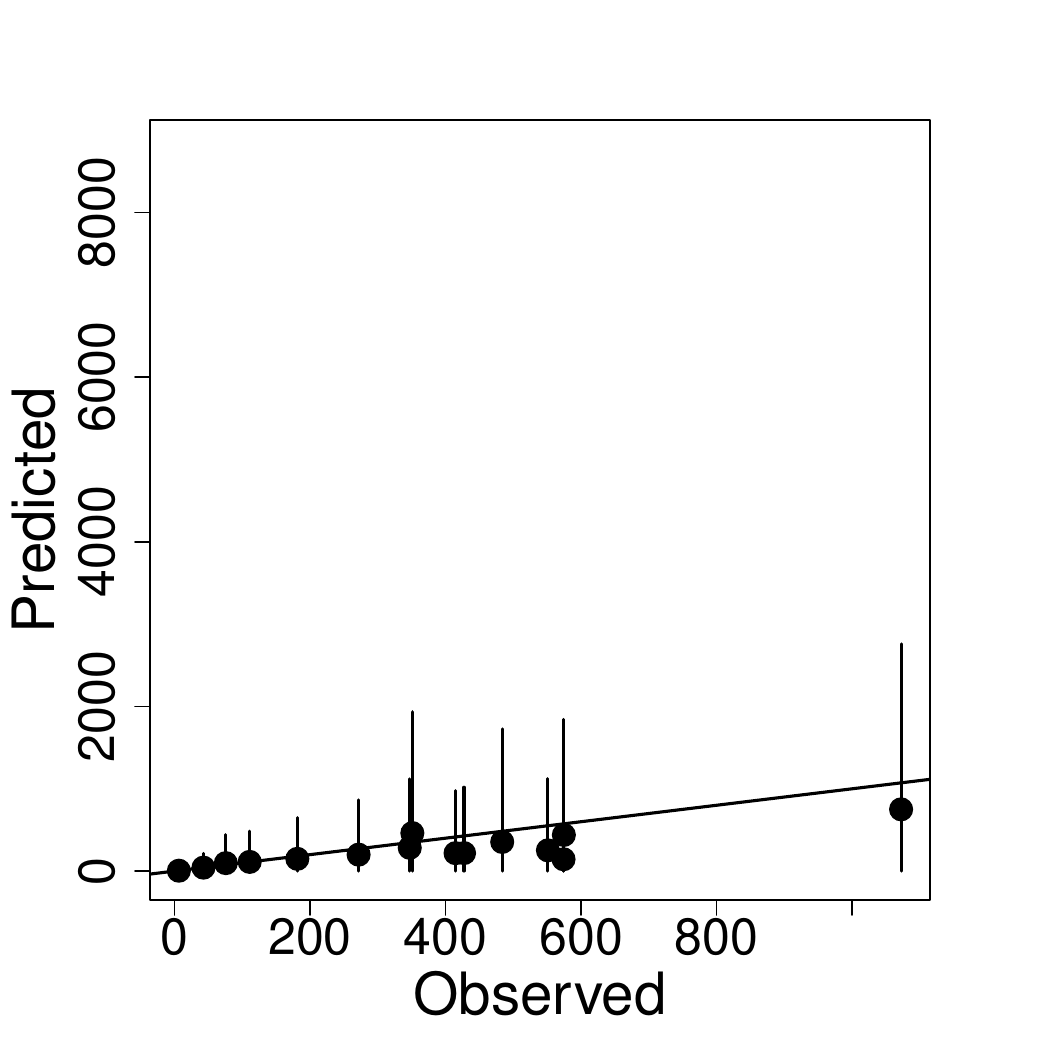}  \\
   (c) Skew-Normal & (d) Skew-t \\
    \end{tabular}
    \caption{Simulated dataset: Posterior predictive medians (dots) versus the true observed loss reserving for the first calendar year with vertical lines representing the 95\% credible predictive intervals. The diagonal line indicates $y = x$. (a) the Normal, (b) the Student-t, (c) the Skew-Normal, and (d) the Skew-t models. }
    \label{apB:fig5}
\end{figure}

\begin{figure}[H]
    \centering
    \begin{tabular}{c}
    \includegraphics[width=12cm]{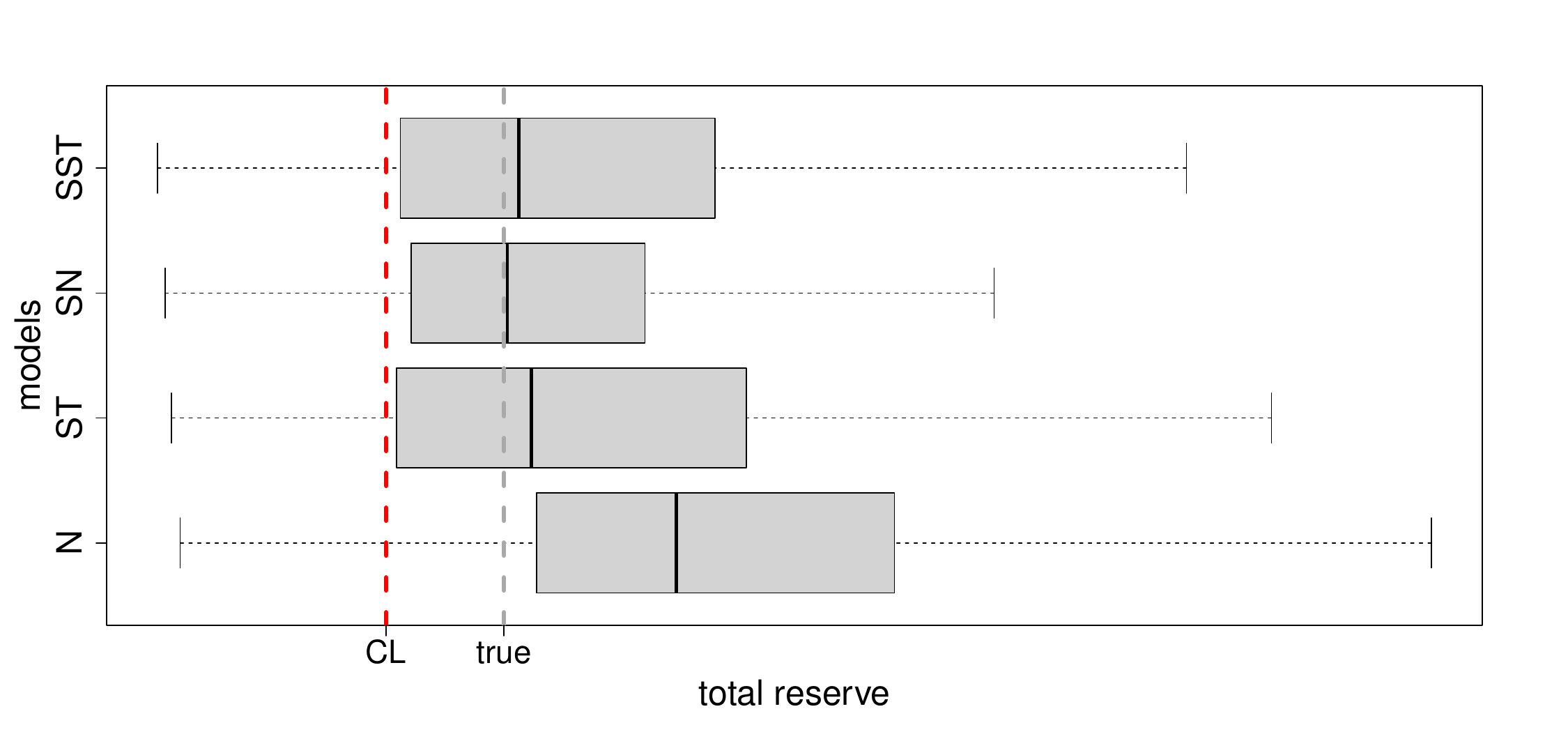} \\
      \end{tabular}
    \caption{Simulated dataset: predictive posterior of the total loss reserving for all competing models. The dashed grey line represents the true total loss reserve and the dashed red line represents the chain ladder reserve (CL).}
    \label{apB:fig6}
\end{figure}

For prediction purposes, we assess the amount of loss reserve for all competing models. Figure \ref{apB:fig5} plots the medians and 95\% credible intervals for the predictions of loss reserve for the first calendar year. The one-year ahead reserve could be important in short-term financial decisions for the insurance company, as they are the reserves for settling that are expected to occur soon.  The class of the Skew models provides interval widths that are smaller in the majority of first calendar year data when compared to the non-skewed models. Figure \ref{apB:fig6} shows the predictive posterior total loss reserving distribution for all competing models compared with the true total reserve and chain ladder reserve. The uncertainty is worse for the non-skew class, mainly for the Gaussian model. Besides that, the chain ladder method underestimates the loss reserving producing an amount smaller than would be expected of 22,835.28 (dashed red line). Remember that the true total reserve is 33,047.39. In fact, the chain ladder technique does not include any calendar year effects and is unsuitable for volatile data. In the actuarial context, this would be an aggravating factor due to the fact that the insurance company does not save enough to constitute an efficient reserve for soon payments. For the dynamical linear models, the total reserve posterior median and 95\% credible interval are $Y_{\nabla \mathcal{N}}= 47,993.08 (18,354.42; 155,179.7)$, $Y_{\nabla \mathcal{T}}=35,407.88 (11,387.48; 181,571.9)$, $Y_{\nabla \mathcal{SN}}=33,339.87 (13,486.54; 103,837.2) $, $Y_{\nabla \mathcal{ST}}= 34,339.96 (12,106.10; 146,763.2)$.  Table \ref{apB:tab1} shows model comparison based on RMSPE and scoring rules criteria indicating that the Skew-t model provides better performance when compared to others. This suggests that taking into account heterogeneous variance and skewness is essential in order to produce good predictions in loss reserving context, especially for extreme values.

\begin{table}[H]
    \centering
    \caption{Simulated dataset: Posterior summary for the Skew-t model compared with the true values.}
    \label{apB:tab2}
    \begin{tabular}{|l|cccc|}
    \hline 
    Parameters  & True value & Median             & S.E. & IC95\%\\
    \hline
   $\mu$               & 9.00 & 8.815 &0.284  &  (8.328 , 9.478)\\
   $\rho$              &-0.89 &-0.877 &0.211  & (-0.996 , -0.187)\\
   $\nu$               & 3.00 & 4.579 &1.458  &  (2.606 , 8.234)\\
   $\sigma^2$          & 0.14 & 0.162 &0.080  &  (0.061 , 0.366)\\
   $\sigma_{\alpha}^2$ & 0.13 & 0.249 &0.139  &  (0.105 , 0.638)\\
   $\sigma_{\beta}^2$  & 0.05 & 0.060 &0.021  &  (0.030 , 0.112)\\
   $\sigma_{\gamma}^2$ & 0.13 & 0.094 &0.061  &  (0.033 , 0.258)\\
   \hline
    \end{tabular}
\end{table}

\begin{table}[H]
    \centering
    \caption{Simulated dataset: Model comparison based on the Root Mean Square Prediction Error (RMSPE), the Interval Score (IS), the Width of Credibility Interval (WCI), and the Continuous Ranking Predictive Score criteria for the predicted total loss reserve (the lower triangle) under all fitted models.}
    \label{apB:tab1}
    \begin{tabular}{|l|cccc|}
    \hline 
    \textbf{Competing}  & \multicolumn{4}{c|}{\textbf{Measures}}\\
    \textbf{Models}     & average IS & average WCI & RMSPE & average CRPS  \\
    \hline
   % $\mathcal{N}$  & 5.686 & 5.686 & 1.607 & 0.547\\
   % $\mathcal{T}$  & 5.172 & 5.172 & 1.447 & 0.466\\
   % $\mathcal{SN}$ & 5.136 & 5.136 & 1.474 & 0.495\\
   % $\mathcal{ST}$ & \textbf{5.118} & \textbf{5.118} & \textbf{1.431} & \textbf{0.460}\\
   
    $\mathcal{N}$  & 5.307 & 5.307 & 16.957 &  0.545\\
    $\mathcal{ST}$  & 4.452 & 4.452 & 14.497 &  0.437\\
    $\mathcal{SN}$ & 4.732 & 4.732 & 15.408 & 0.489\\
    $\mathcal{SST}$ & \textbf{4.421} & \textbf{4.421} & \textbf{14.354} & \textbf{0.434}\\
   
   \hline
    \end{tabular}
\end{table}

\newpage
\subsection{Case study: a real dataset from Chan et al. (2008)}\label{sec3.2}
We investigate the same dataset considered by \cite{Choy2008} and \cite{Jennifer16}. The data refers to the claim amounts paid to the insureds of an insurance company during the period from 1978 to 1995 (see Appendix \ref{apD}, Table \ref{tabapC:1}). 
There are $n=171$ observations in the upper triangle. Given a policy year, the amount of claims paid follows an increasing trend in the first 4 to 6 lag years and then a decreasing trend thereafter (\cite{Choy2008}). For prediction evaluation, we consider a cross-validation proposal where $13 \times 13$ observations were used for the inference procedure, being that the last five diagonals (calendar year) we left out for validation (for more details, see Appendix \ref{apD} - highlighted in bold). Now, our data is composed of $n= 91$ observations (upper triangle, $Y_{\Delta}$) and the remaining representing the lower triangle (total reserve, $Y_{\nabla}$ highlighted in dark and light grey).

\begin{figure}[H]
    \centering
    \begin{tabular}{ccc}
    \includegraphics[width=4.5cm]{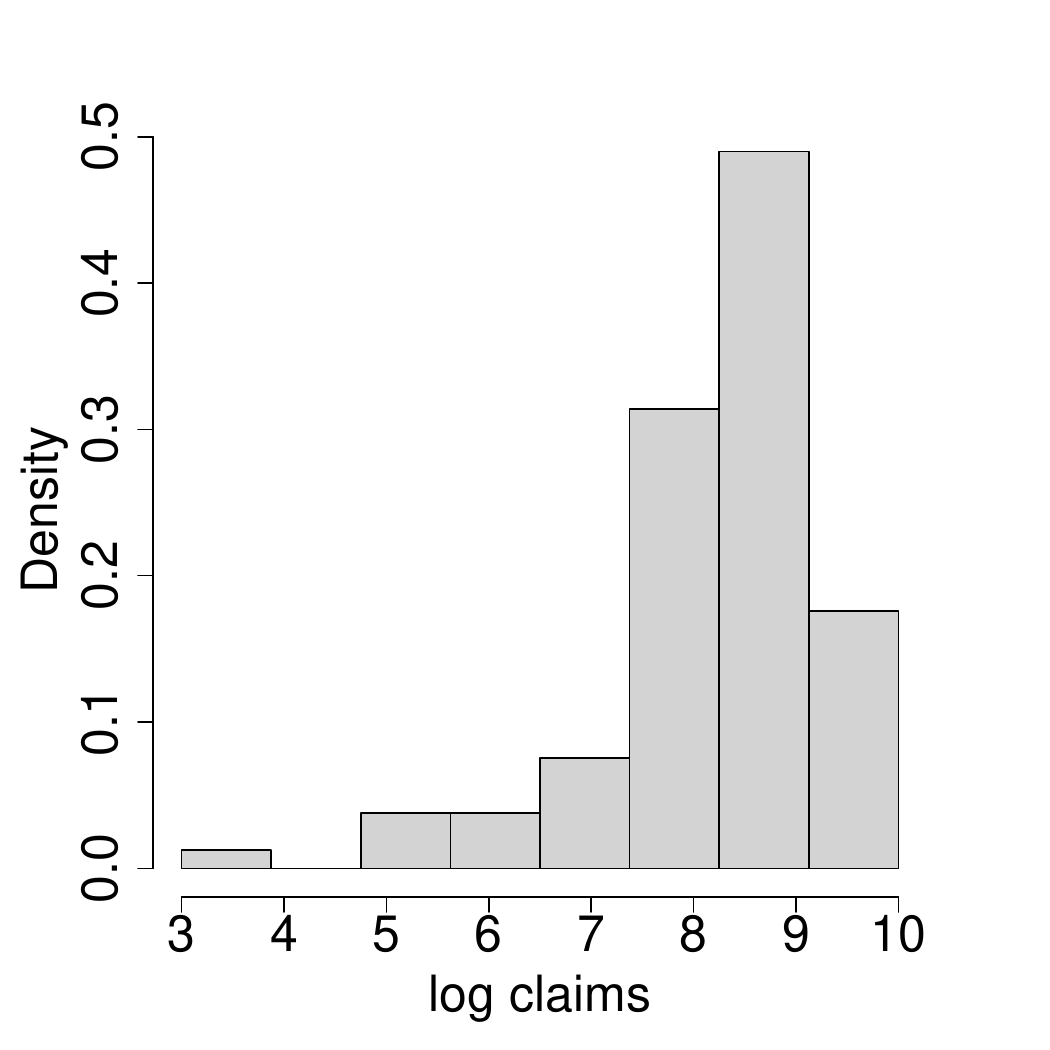} & 
    \includegraphics[width=4.5cm]{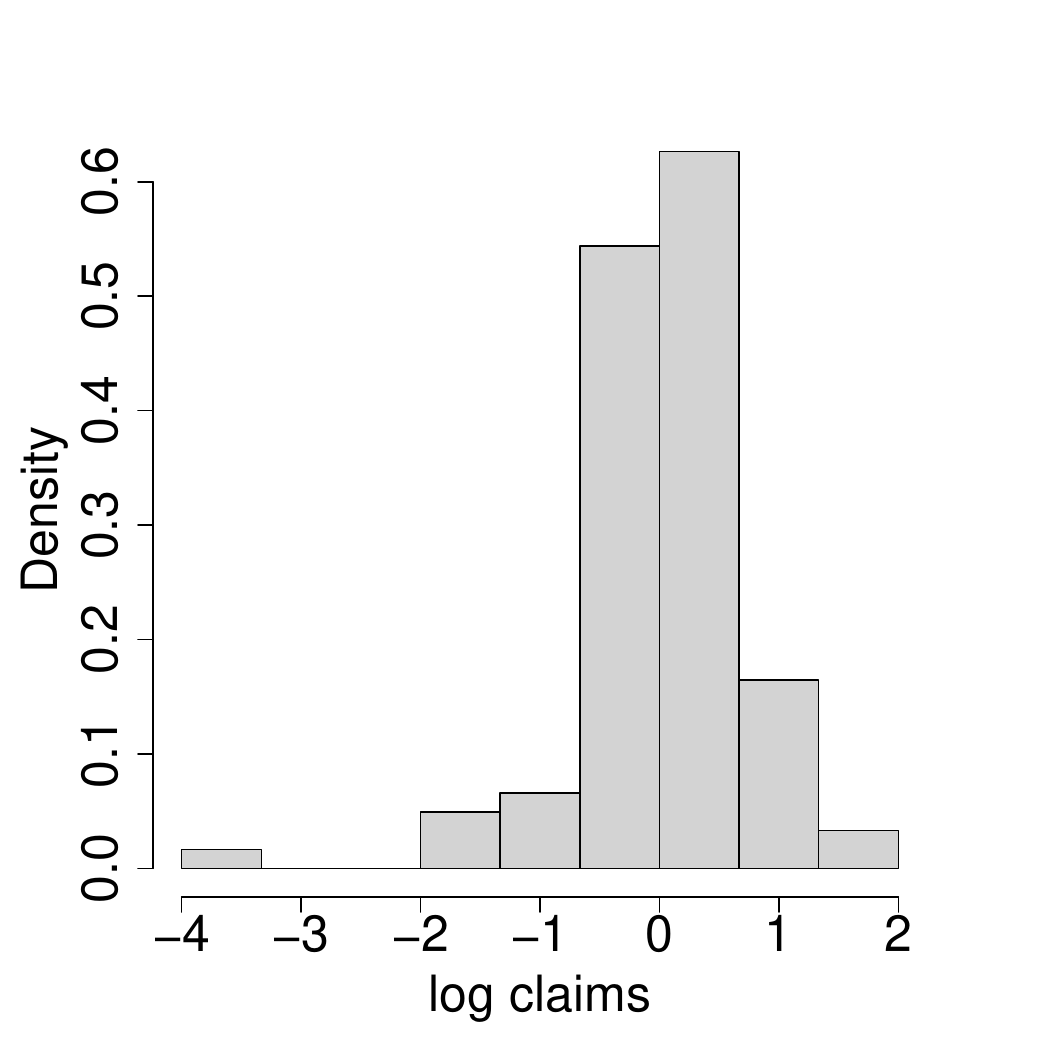}& 
     \includegraphics[width=4.5cm]{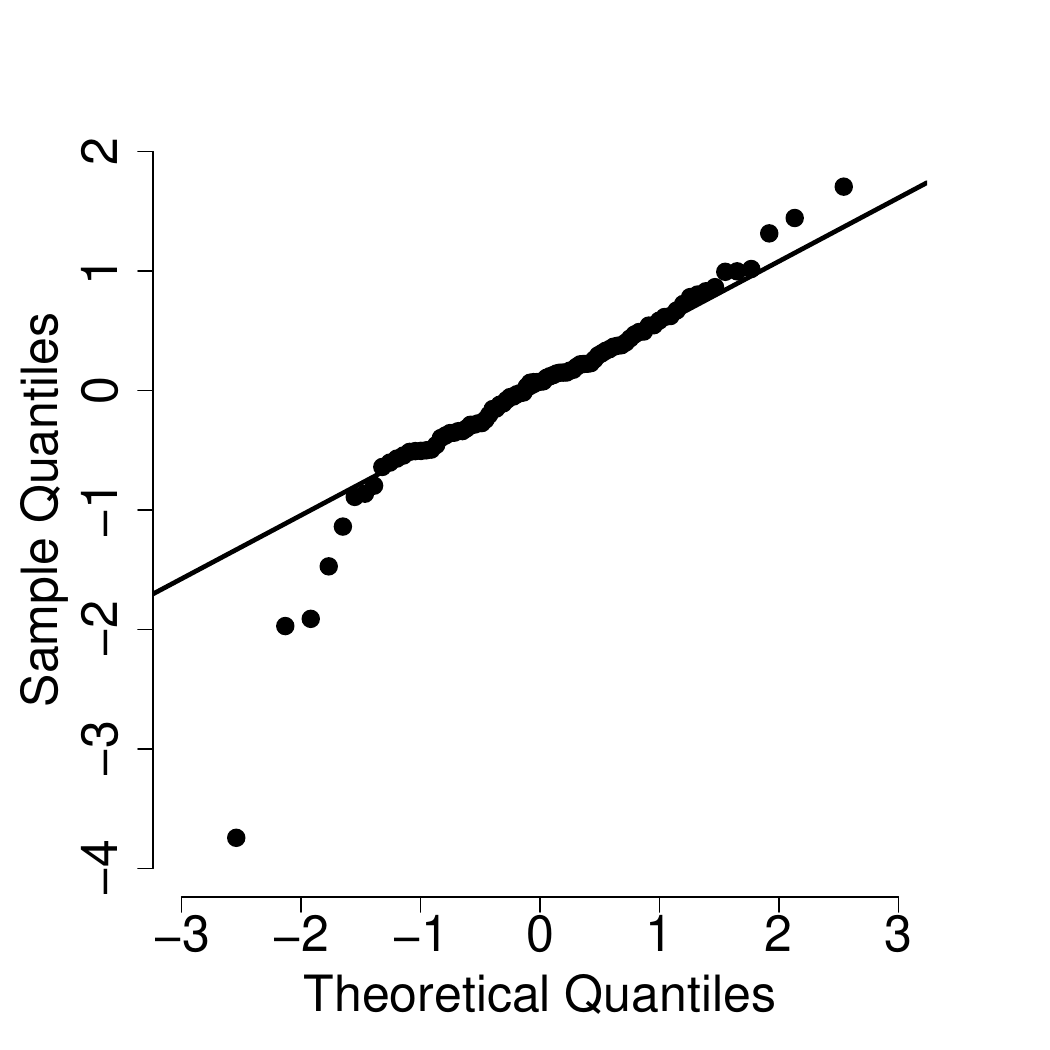} \\
    (a) log-claims data & (b) residual empirical median & (c)  q-q plot of residual  \\
    \end{tabular}
    \caption{Case study: (a) the histogram of log-claims distribution (b) the residual median of the Gaussian model and (c) the Normal q-q plot of residuals for the dataset.}
    \label{sec32:fig1}
\end{figure}

The mean, median, standard deviation, skewness, and excess kurtosis of the log-transformed data log($Y_{ij}$) are 8.214  8.450  1.079 -1.759 and 3.767, respectively.  The behaviour of the claims is irregular and contains potential outliers. %For instance, {\color{red}the amount of 15,546 dollars in the 4-th development year of the policy year 1992.  chechar se eh referente 13 por 13}
Panel (a) of Figure \ref{sec32:fig1} presents the distribution of the log claims suggesting that skewness and heavy tails may be accommodated in the modelling process.  

Initially, we fit a Gaussian model, which means $\lambda_{ij}=1$, for all $i, j=1, \ldots, n$ and $\kappa=0$. Panels (b) and (c) of Figure \ref{sec32:fig1} show the residual empirical median and the theoretical quantiles for the Gaussian distribution versus the empirical quantiles based on the residuals from the Gaussian model suggesting that the model with constant variance and without asymmetry might be unsuitable to predict the loss reserving. Note that the empirical excess kurtosis and skewness of the residuals are 3.767 and -1.759, respectively. These features corroborate that we need to consider a model that is able to capture this behaviour. 
 Given the results from Figure \ref{sec32:fig1}, we fit all competing models described in Table \ref{tabmodels}. The parameters to be estimated are related to dynamical mean $\mu_{ij}$ that is $(\mu, \alpha_i, \beta_{ij},\gamma_t)$, the variance parameters $(\sigma^2, \sigma^2_{\alpha}, \sigma^2_{\beta}, \sigma^2_{\gamma})$, the mixing parameters $(\nu, \rho)$, and the latent variables $(\lambda_{ij}, T_{ij})$. 

Sensitivity to the prior parameter specification is an important step in the inference procedure. An unsuitable specification of the prior parameter can lead a misleading results. For this, we investigate the performance of the parameters based on a sensitivity analysis by considering different specifications of the prior distributions. The proposed priors are presented in Table \ref{tab:priors} and the choice of the prior distributions are detailed in \cite{gelman06}, \cite{Goudarzi2018}, and \cite{rodr18} for the Skew-t model. The prior sensitivity analysis for the Skew-t model is shown in Table \ref{tab:priorsens} through the posterior median, 95\% credible interval and the predicted posterior total reserve. See that there are no significant differences for the specifications of $\mu$, $\rho$, $\sigma^2$, $\sigma_{\alpha}^2$, $\sigma_{\beta}^2$ and $\sigma_{\gamma}^2$ distributions. Although there are significant differences in the choice of the $\nu$ parameter ($\mathcal{G}(12,0.8)$, $Exp(0.3)$, $\mathcal{U}(2,40)$ and Jeffreys prior distribution proposed by \cite{Fonseca08}), there is no impact on the total reserve for several scenarios. {The following results shown in this study are based on scenario 2. }

\begin{table}[ht]
    \caption{Case study: Prior distributions scenario for the Skew-t model.}
    \label{tab:priors}
    \centering
    \begin{tabular}{|lllll|}\hline
              & \multicolumn{4}{c|}{Prior distributions}\\\cline{2-5}
        scenario & $\mu$ & $\rho$ & $\nu$ & $\sigma^2,\sigma_{\alpha}^2,\sigma_{\beta}^2$ and $\sigma_{\gamma}^2$\\\cline{2-5}
        1&$\mathcal{N}(0,100)$ & $\mathcal{B}e(1,1)$ & $\mathcal{G}(12,0.8)$ & $\mathcal{G}(0.01,0.01)$\\
        2&$\mathcal{N}(0,100)$ & $\mathcal{B}e(1,1)$ & $\mathcal{G}(12,0.8)$ & $\mathcal{G}(0.001,0.001)$\\
        3&$\mathcal{N}(0,100)$ & $\mathcal{B}(0.5,0.5)$ & $\mathcal{G}(12,0.8)$ & $\mathcal{G}(0.001,0.001)$\\
        4&$\mathcal{N}(0,100)$ & $\mathcal{B}e(1,1)$ & $\mathcal{U}(2,40)$ & $\mathcal{G}(0.001,0.001)$\\
        5&$\mathcal{N}(0,100)$ & $\mathcal{B}e(1,1)$ & $Exp(0.3)$ & $\mathcal{G}(0.001,0.001)$\\
        6&$\mathcal{N}(0,100)$ & $\mathcal{B}e(1,1)$ & Jeffreys & $\mathcal{G}(0.001,0.001)$\\\hline
    \end{tabular}
\end{table}

\begin{table}[ht]
    \caption{Case study: Prior sensitivity analysis for the Skew-t model.}
    \label{tab:priorsens}
    \centering
    \adjustbox{max width=\textwidth}{
    \begin{tabular}{|lllllllll|}\hline
               & \multicolumn{8}{c|}{Parameters}\\\cline{2-9}
         scenario & Reserve & $\mu$ & $\rho$ & $\sigma^2$ & $\nu$ & $\sigma_{\alpha}^{2}$ & $\sigma_{\beta}^{2}$ & $\sigma_{\gamma}^{2}$\\\cline{2-9}
         \multirow{2}{*}{1}     &\multirow{2}{*}{630165.9}& 8.891 & -0.960 & 0.278 & 11.187 & 0.113 & 0.061 & 0.131\\
         &&(8.229,9.624)&(-0.998,-0.337)&(0.111,0.545)&(5.175,21.119)&(0.039,0.365)&(0.028,0.126)&(0.053,0.389)\\
         \multirow{2}{*}{2}     &\multirow{2}{*}{621282.4}& 8.921 & -0.959 & 0.282 & 11.150 & 0.112 & 0.059 & 0.129\\
         &&(8.206,9.679)&(-0.999,-0.455)&(0.112,0.556)&(4.958,21.064)&(0.038,0.342)&(0.027,0.122)&(0.050,0.391)\\
         \multirow{2}{*}{3}     &\multirow{2}{*}{613902.6}& 8.955 &-0.989  &0.323  &11.462  &0.103  &0.058  &0.127\\
         &&(8.246,9.743)&(-1.000,-0.794)&(0.140,0.607)&(5.207,21.401)&(0.035,0.324)&(0.027,0.120)&(0.049,0.388)\\
         \multirow{2}{*}{4}     &\multirow{2}{*}{591693.9}& 8.773 &-0.896  &0.106  &2.947   &0.091  &0.058  &0.118\\
         &&(8.132,9.566)&(-0.996,-0.216)&(0.034,0.353)&(2.035,11.827)&(0.030,0.289)&(0.027,0.117)&(0.046,0.374)\\
         \multirow{2}{*}{5}     &\multirow{2}{*}{586052.7}& 8.734 &-0.875  &0.091  &2.655   &0.087  &0.058  &0.114\\
         &&(8.122,9.557)&(-0.994,-0.209)&(0.031,0.264)&(2.025,5.890)&(0.029,0.279)&(0.028,0.116)&(0.044,0.348)\\
         \multirow{2}{*}{6}     &\multirow{2}{*}{593969.0}& 8.743 &-0.876  &0.088  &2.558   &0.086  &0.058  &0.115\\
         &&(8.119,9.590)&(-0.995,-0.181)&(0.031,0.273)&(2.021,5.636)&(0.027,0.275)&(0.027,0.121)&(0.042,0.346)\\\hline
    \end{tabular}
    }
\end{table}

\begin{figure}[H]
    \centering
    \begin{tabular}{cccc}
    \includegraphics[width=3.6cm]{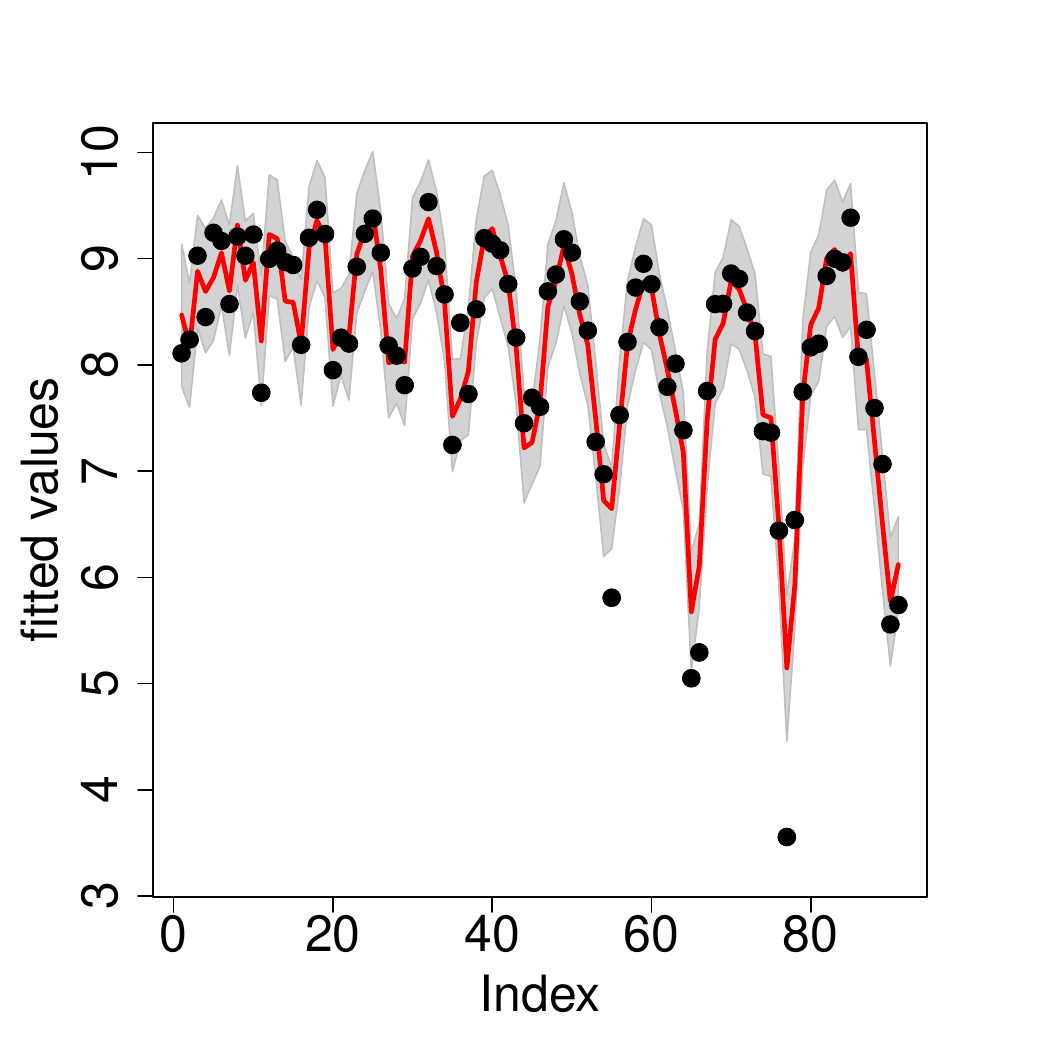}  &   \includegraphics[width=3.6cm]{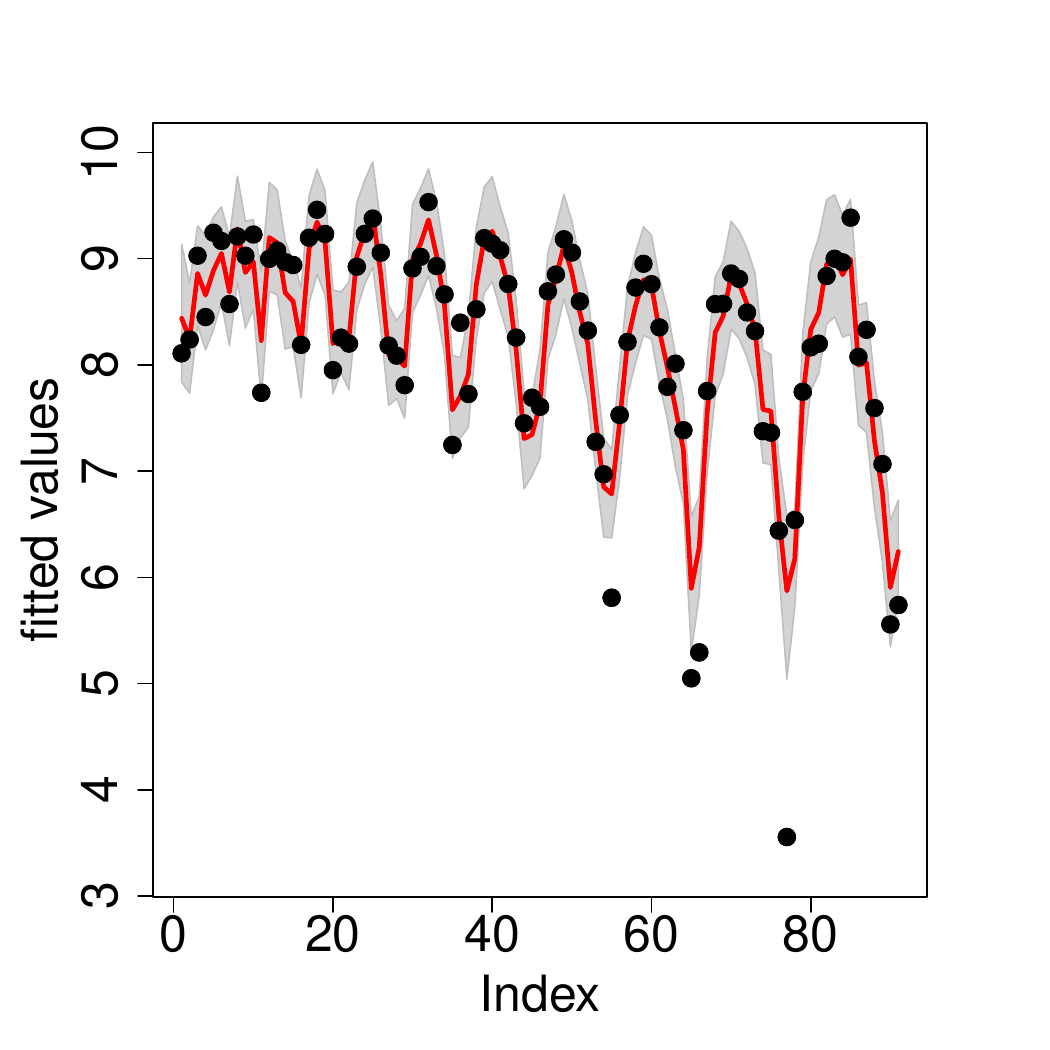} & \includegraphics[width=3.6cm]{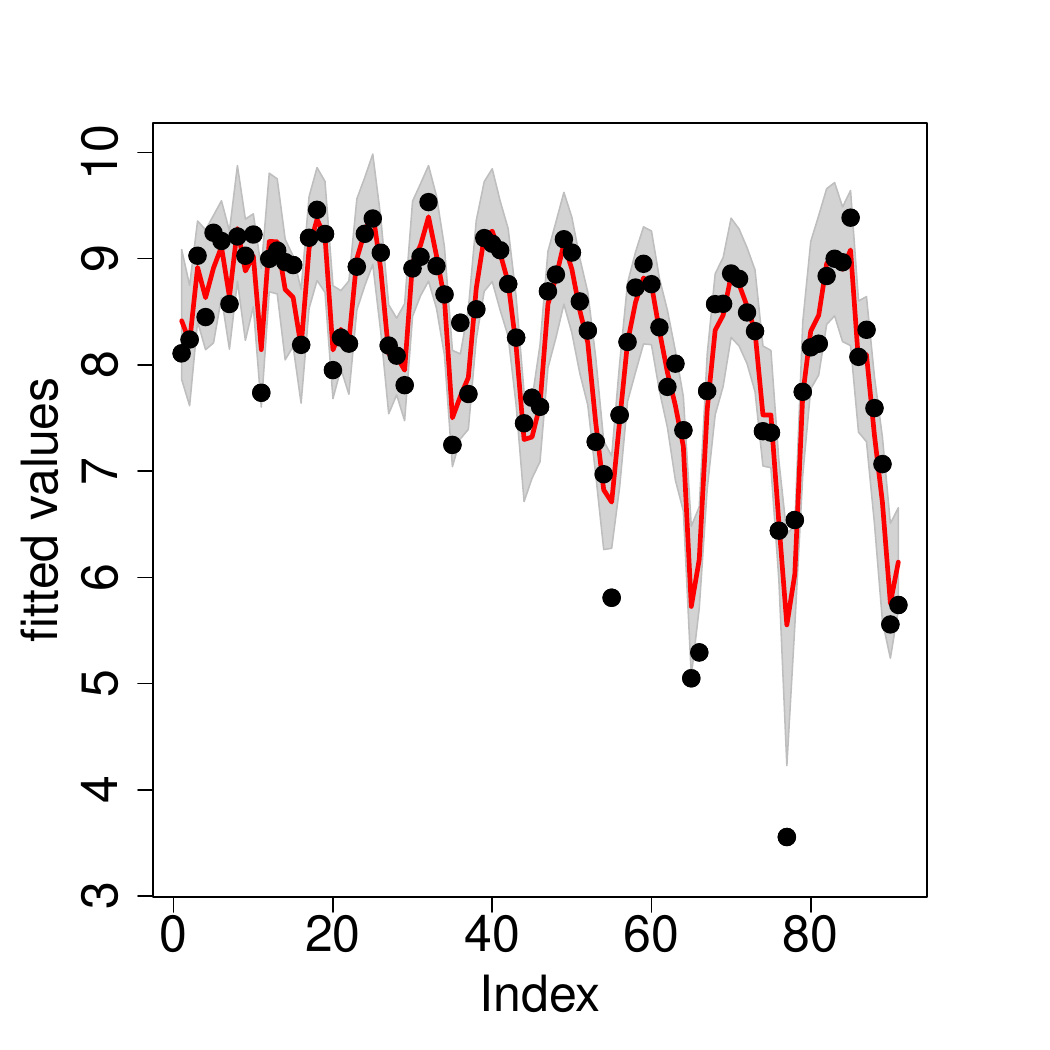} & \includegraphics[width=3.6cm]{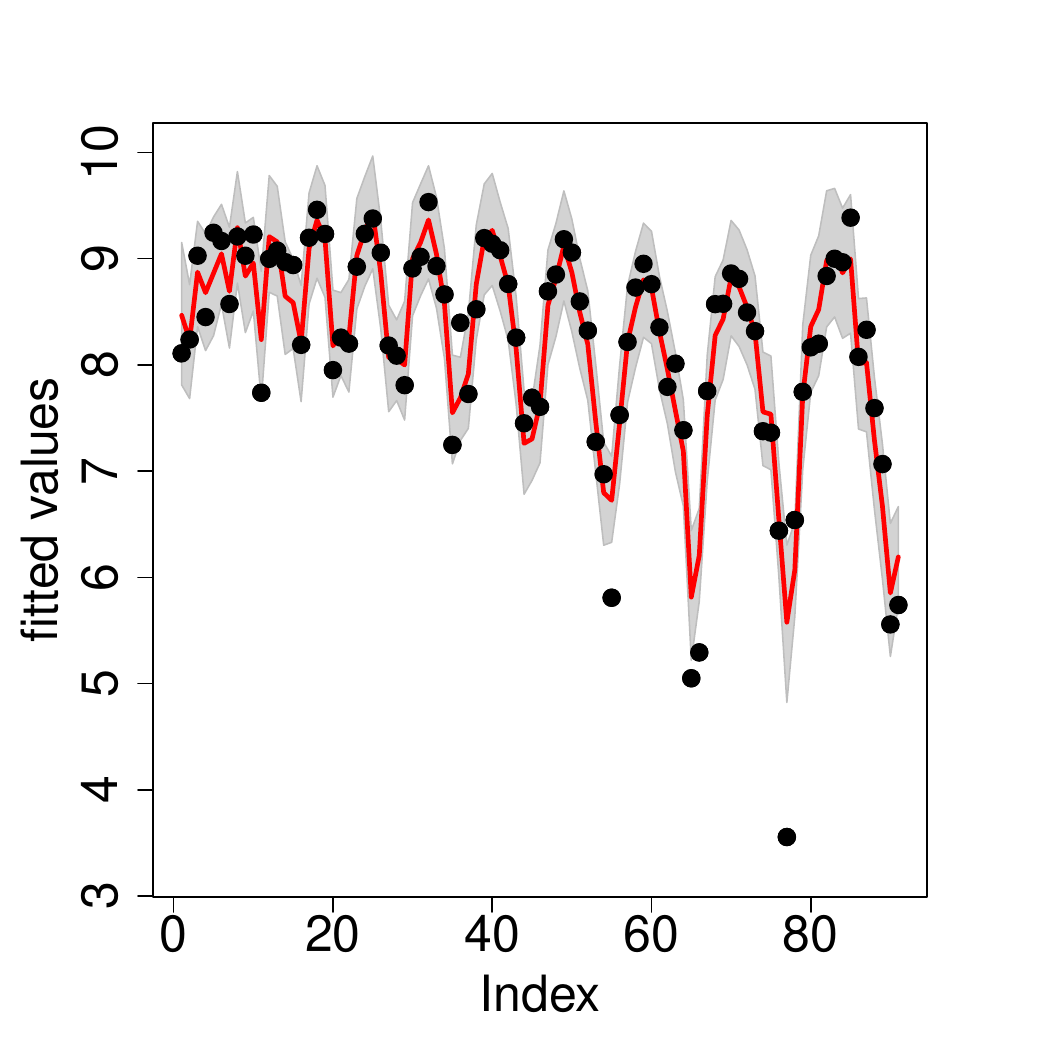}\\
     (a) Normal  & (b) Student-t & (c) Slash & (d) Variance Gamma\\
    \includegraphics[width=3.6cm]{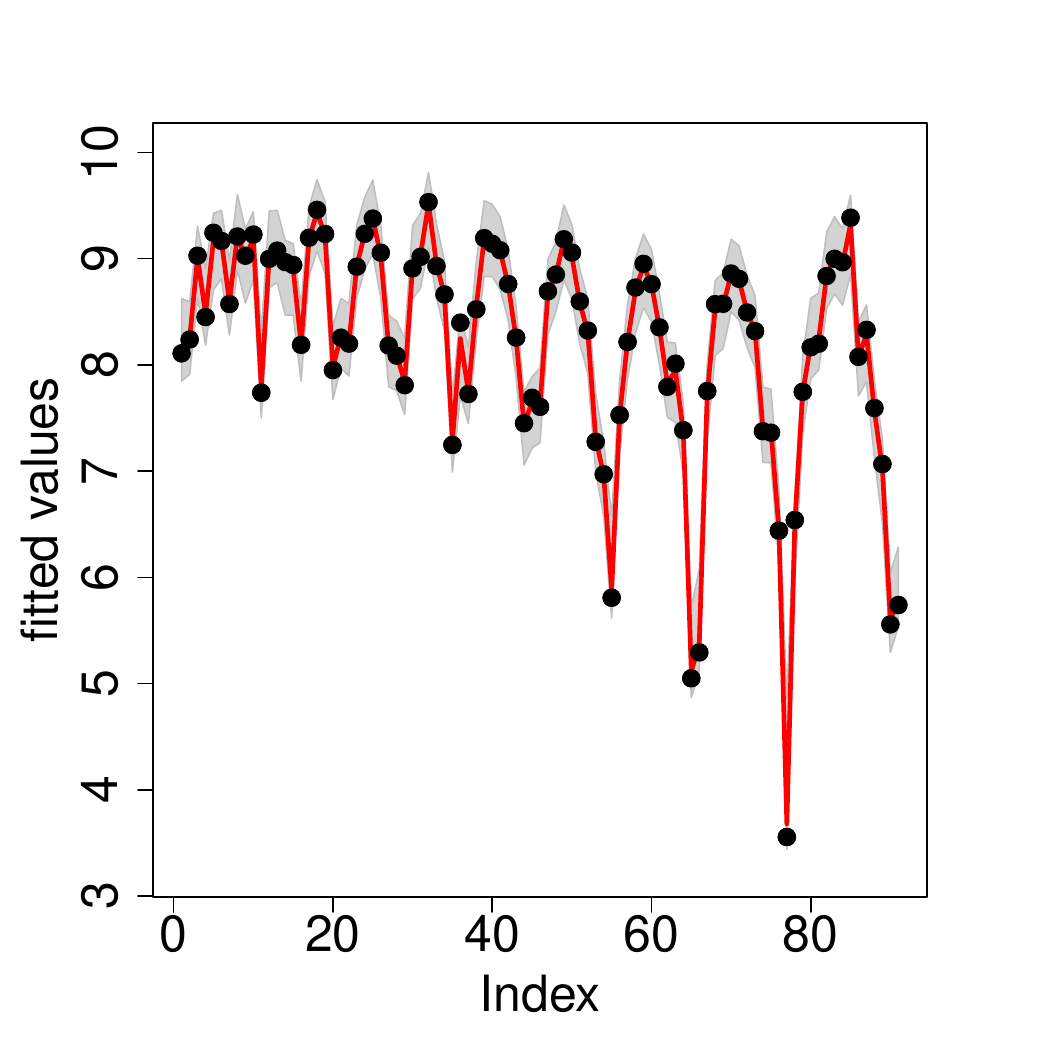}  &   \includegraphics[width=3.6cm]{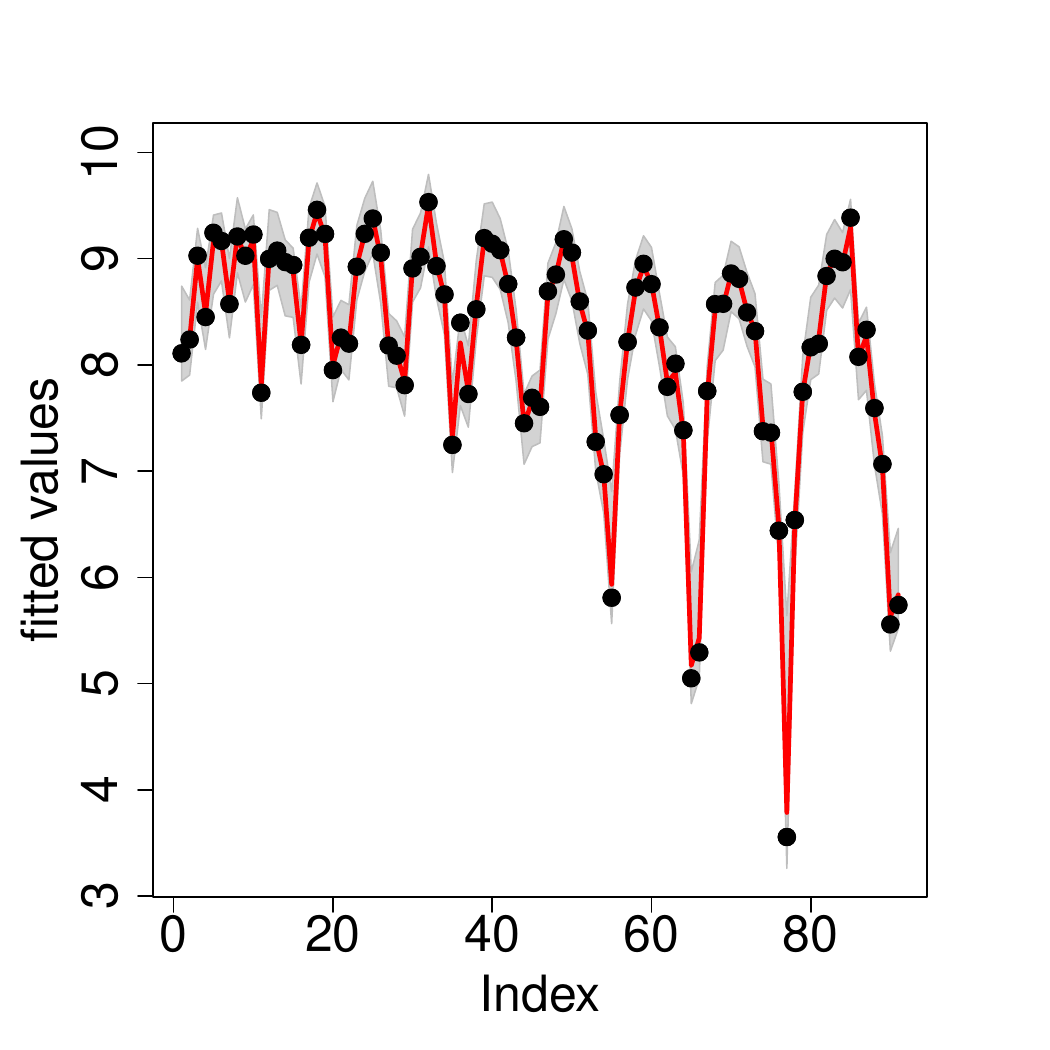} & \includegraphics[width=3.6cm]{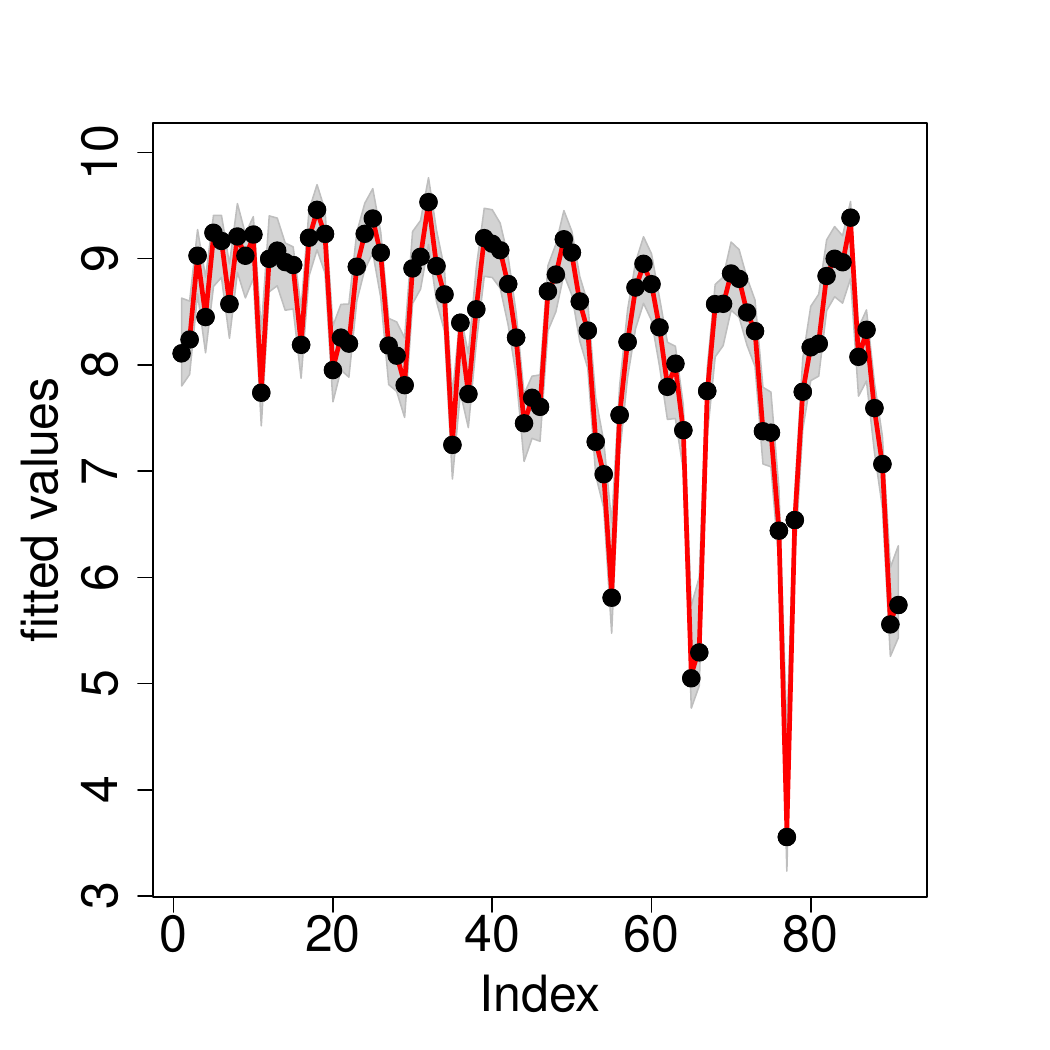} & \includegraphics[width=3.6cm]{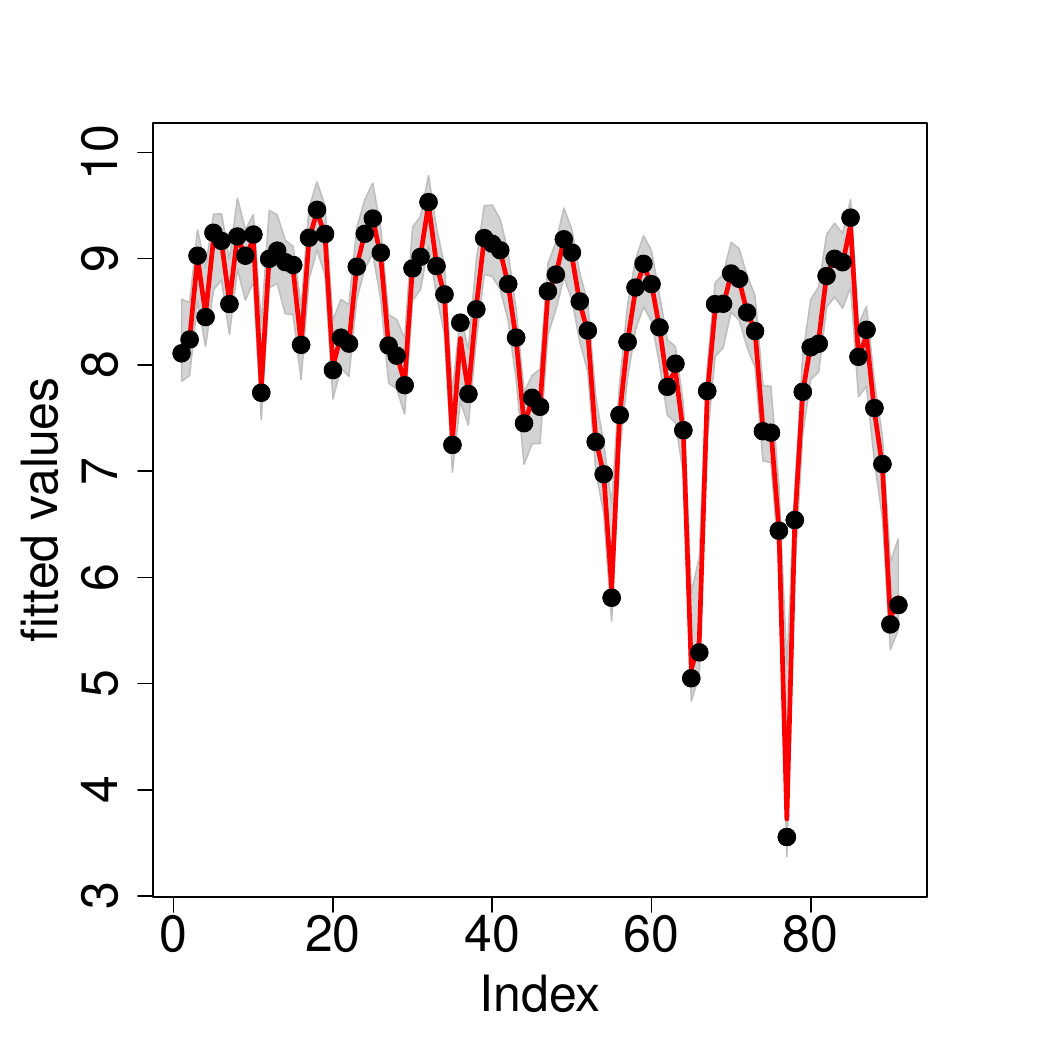}\\
     (e) Skew Normal  & (f) Skew-t & (g) Skew Slash & (h) Skew Variance Gamma\\
    \end{tabular}
    \caption{Case study: Posterior fit value medians (red line) with 95\% credible interval (grey) versus the true observed (black dots) $Y_{\Delta}$. (a) the Normal model, (b) the Student-t model, (c) the Slash model, (d) the Variance-Gamma model, (e) the Skew-Normal model, (f) the Skew-t model, (g) the Skew Slash model and (h) the Skew Variance-Gamma model.}
    \label{sec32:fig2}
\end{figure}

\begin{figure}[H]
   \centering
   \includegraphics[width=12cm]{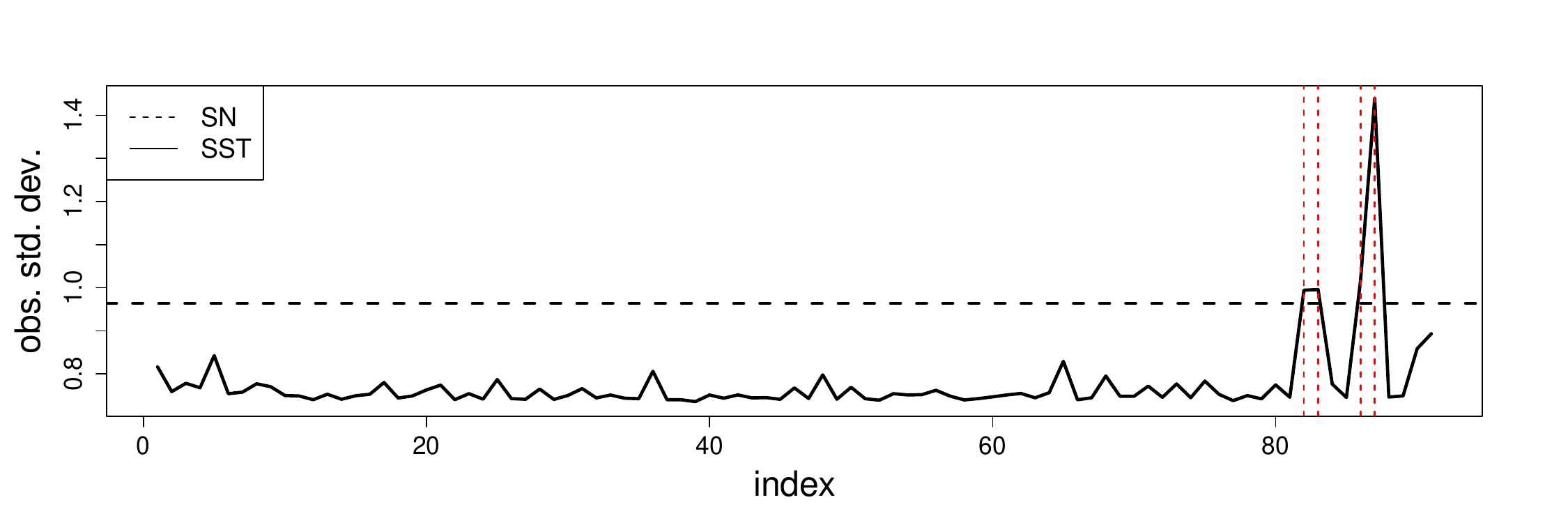}\\
   \includegraphics[width=12cm]{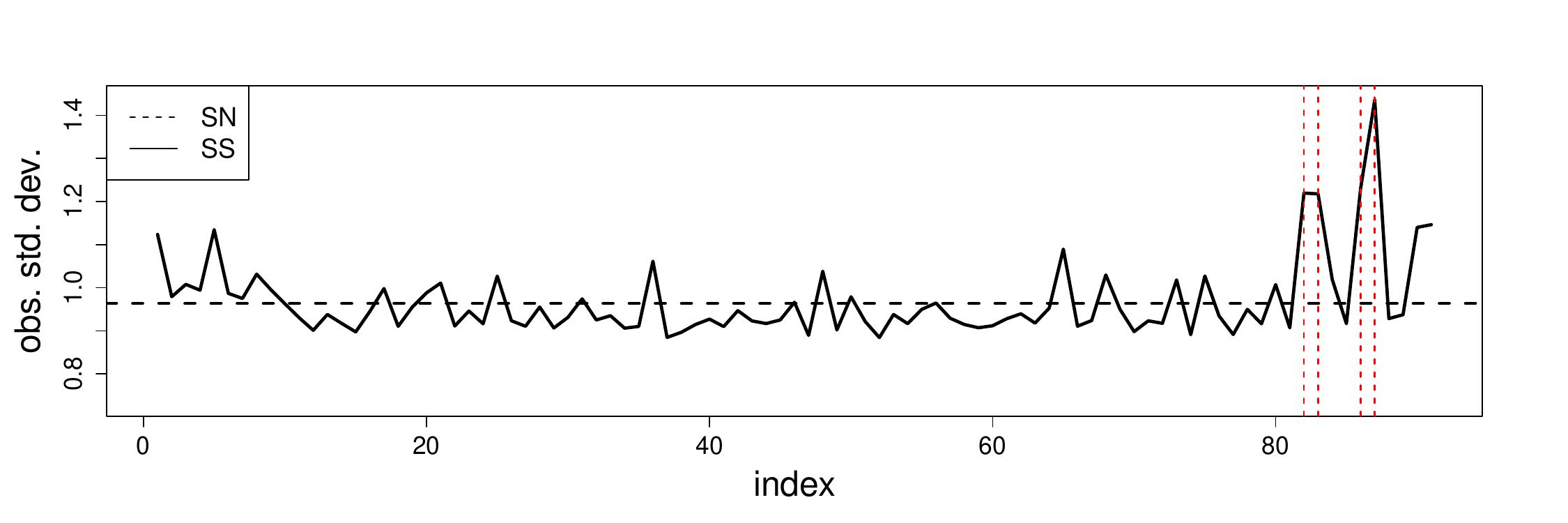}\\
   \includegraphics[width=12cm]{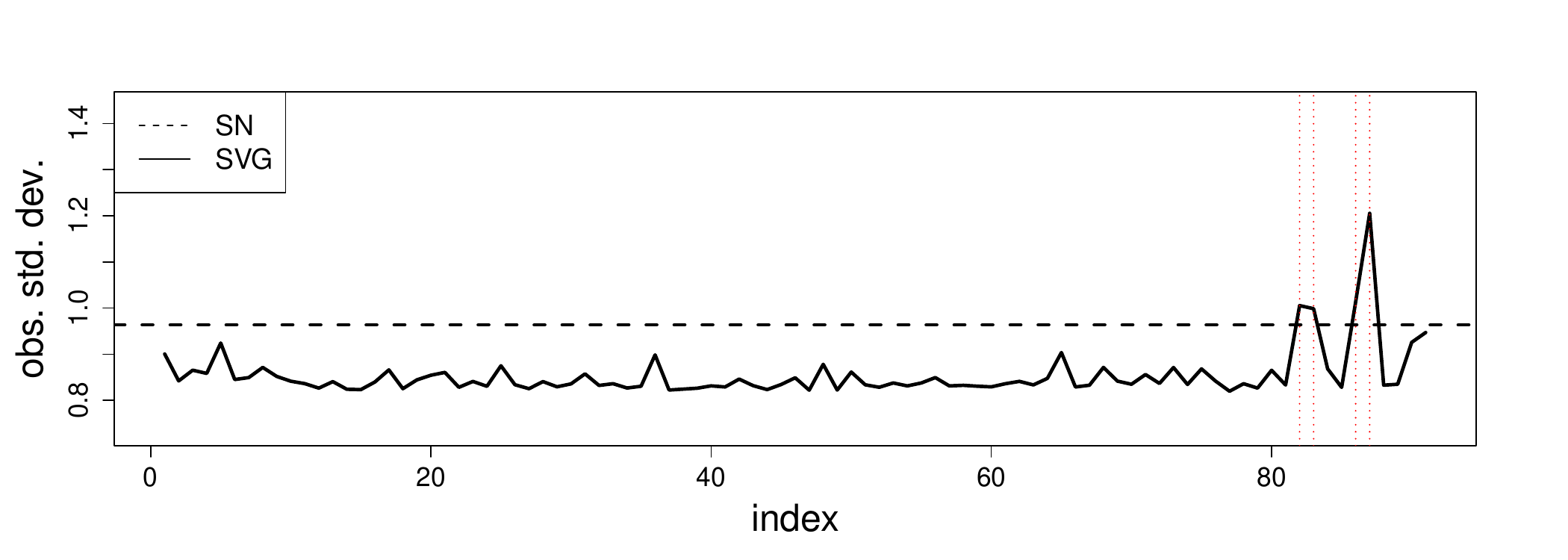}
   \caption{Case study: Posterior smoothed mean of observational standard deviation $\lambda_{ij}$ with Skew Student t distribution $(\mathcal{SVG})$ (top panel), Skew Slash distribution $(\mathcal{SS})$ (center panel) and Skew Variance Gamma distribution $(\mathcal{SVG})$ (bottom panel). The vertical red dotted lines indicates the points $Y_{1,10}=333$, $Y_{2,10}=156$, $Y_{1,11}=199$ and $Y_{2,11}=35$. The horizontal dotted line indicates the observational standard deviation estimate of the Skew Normal model.}
    \label{sec32:fig3}
\end{figure}

\begin{figure}[H]
   \centering
   \includegraphics[width=5.0cm]{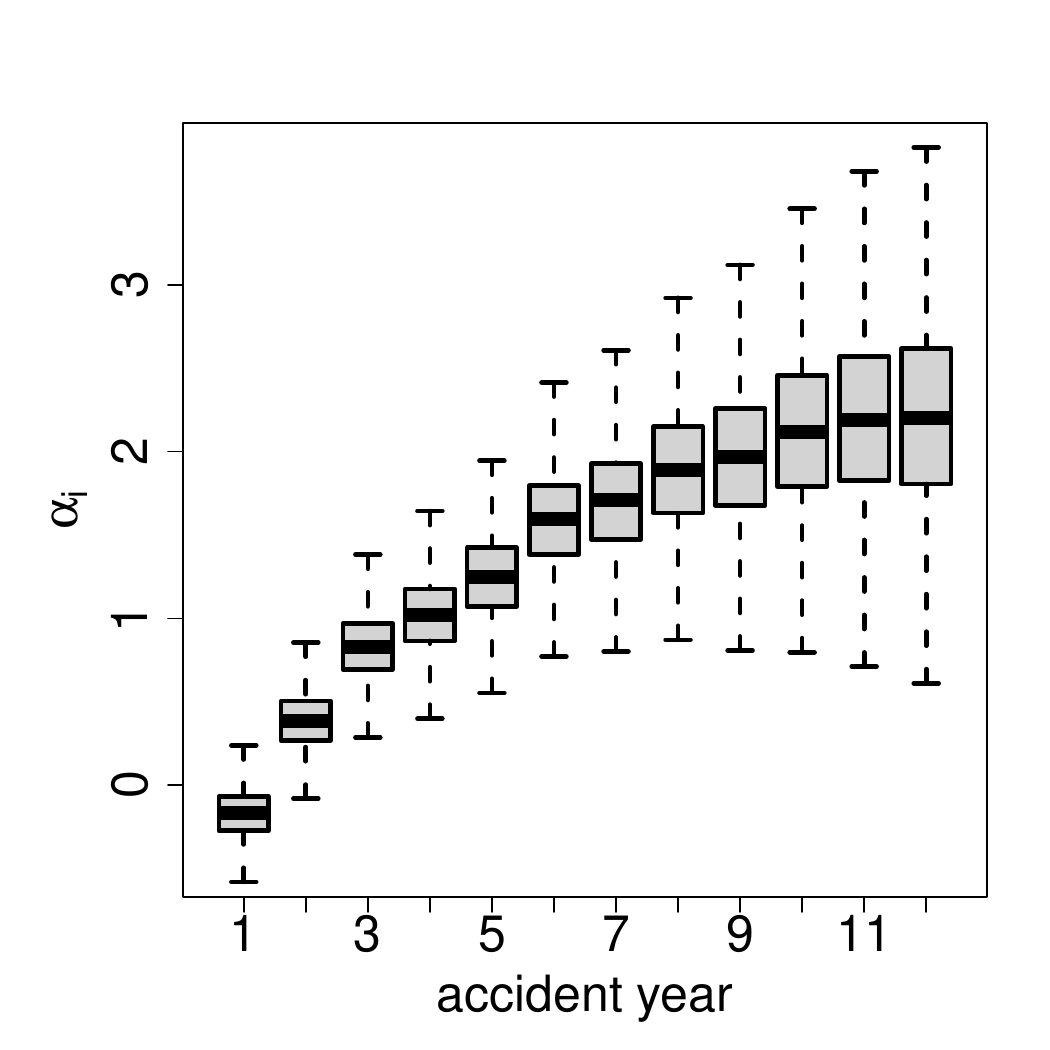}
   \includegraphics[width=5.0cm]{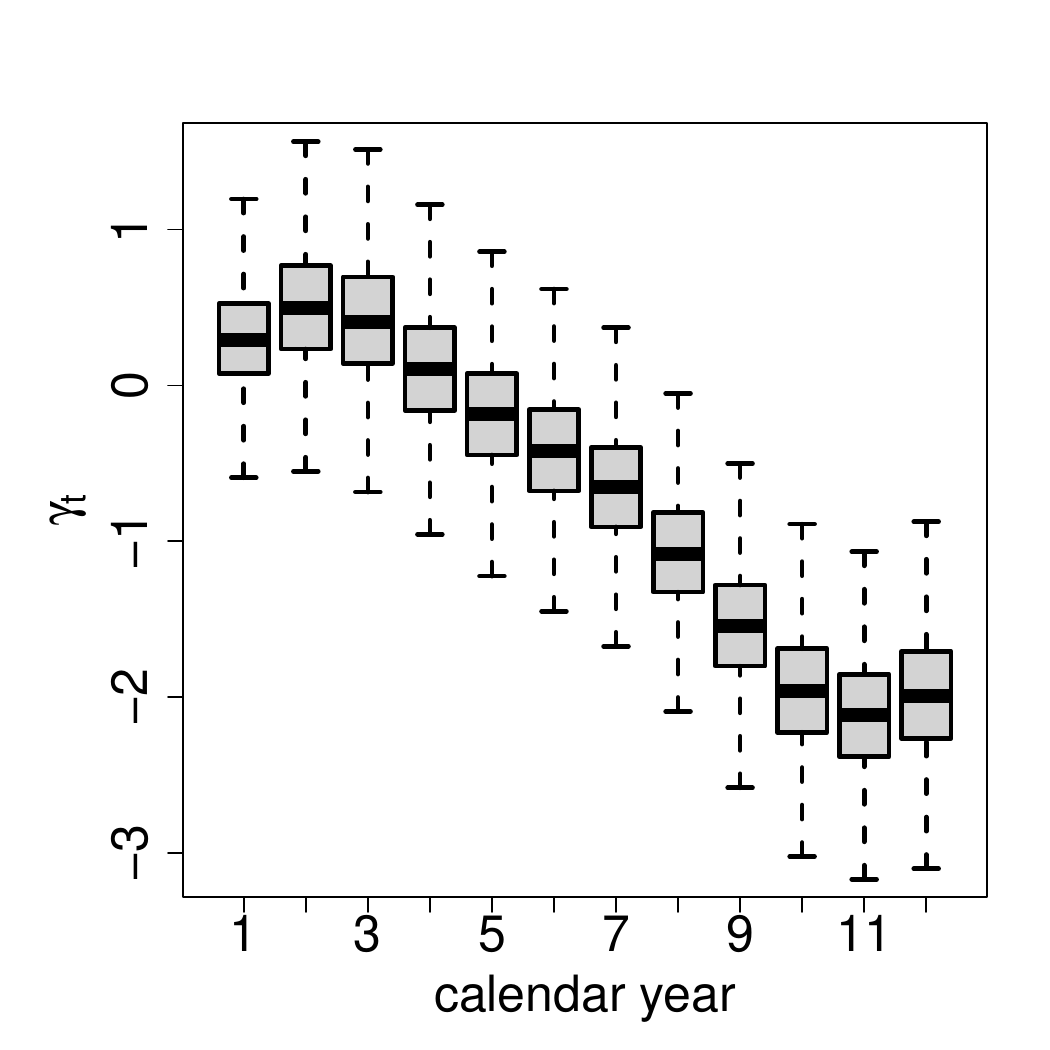}\\
   \includegraphics[width=10cm]{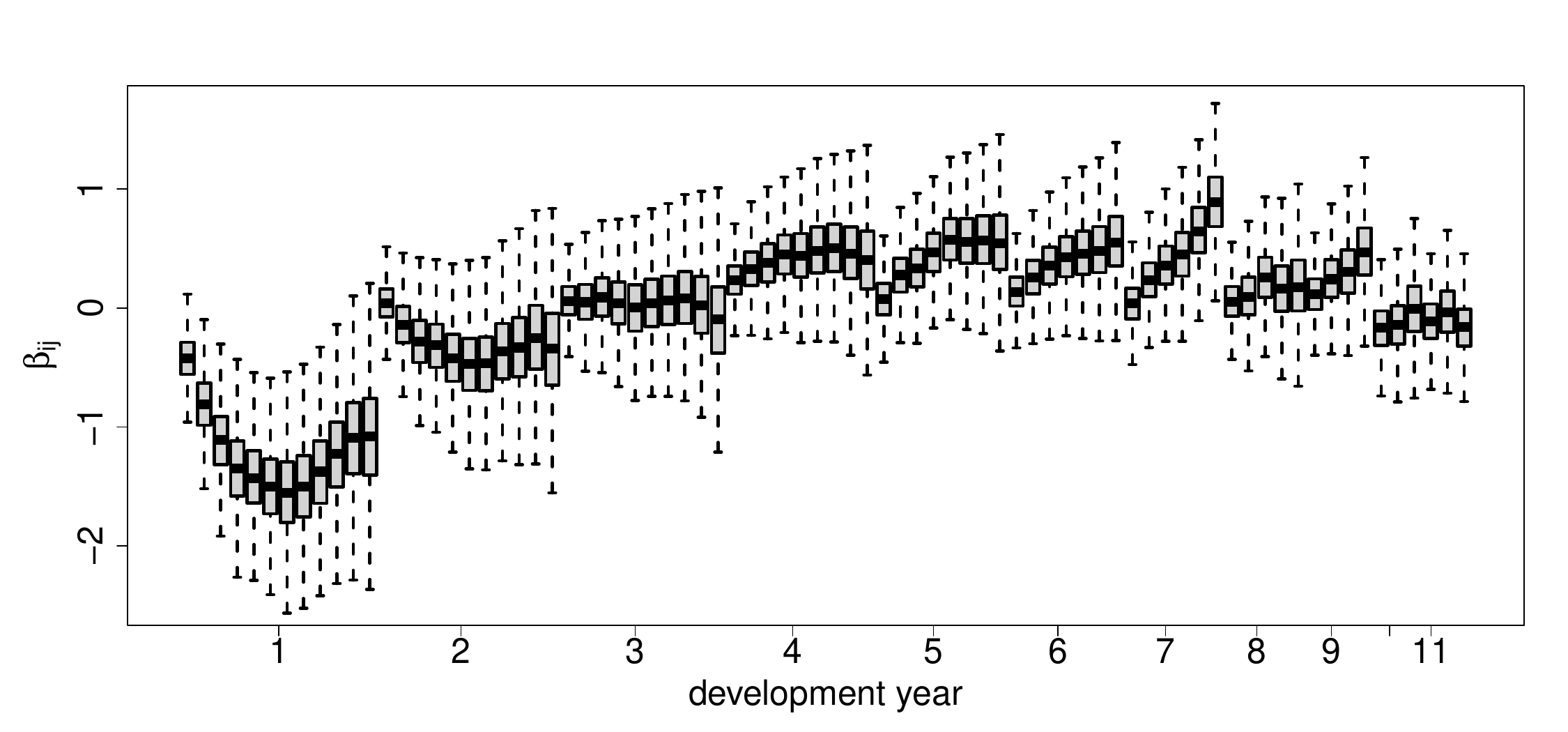}
   \caption{Case study: box-plot for the posterior distribution of the accident year ($\alpha_{i}$), calendar year effect ($\gamma_{t}$) and $\beta_{ij}$ (development year) for the Skew-t model.}
   \label{sec32:fig4}
\end{figure}

Posterior summaries for all competing models are provided by Figures \ref{sec32:fig2}, \ref{sec32:fig3}, and \ref{sec32:fig4}. Panels in Figure \ref{sec32:fig2} show the posterior median and the limits of the 95\% credible interval (in grey) compared to the observed values. The skew class exhibits better performance, effectively capturing the impact of atypical observations in the log claims when compared to the non-skew class.

Figure \ref{sec32:fig3} presents the behaviour of the estimated standard deviation for the proposed skew class. It is possible to identify four points (highlighted in dashed red line), $Y_{1,10}=333$, $Y_{2,10}=156$, $Y_{1,11}=199$, and $Y_{2,11}=35$, common to all panels, where models with heavy tails recognize these observations as potential outliers. The Skew-Normal model inflates the variability for all observations to adjust the standard deviation, while models that take into account heavy tails are able to identify atypical observations and precisely increase the necessary variability of a given log-claim. Figure \ref{sec32:fig4} presents the box-plots of the effects of accident year $\alpha_i$, calendar year $\gamma_t$ and the interaction between accident and development years $\beta_{ij}$. The dynamics of the three effects show significant behaviour, it can be seen that the values of $\alpha_i$'s show an increasing trend, the values of $\gamma_t$'s show a decreasing trend and the values of $\beta_{ij}$'s show increasing, decreasing and even almost constant variations for a given year of development.

\begin{figure}[H]
    \centering
    \begin{tabular}{cccc}
    \includegraphics[width=3.6cm]{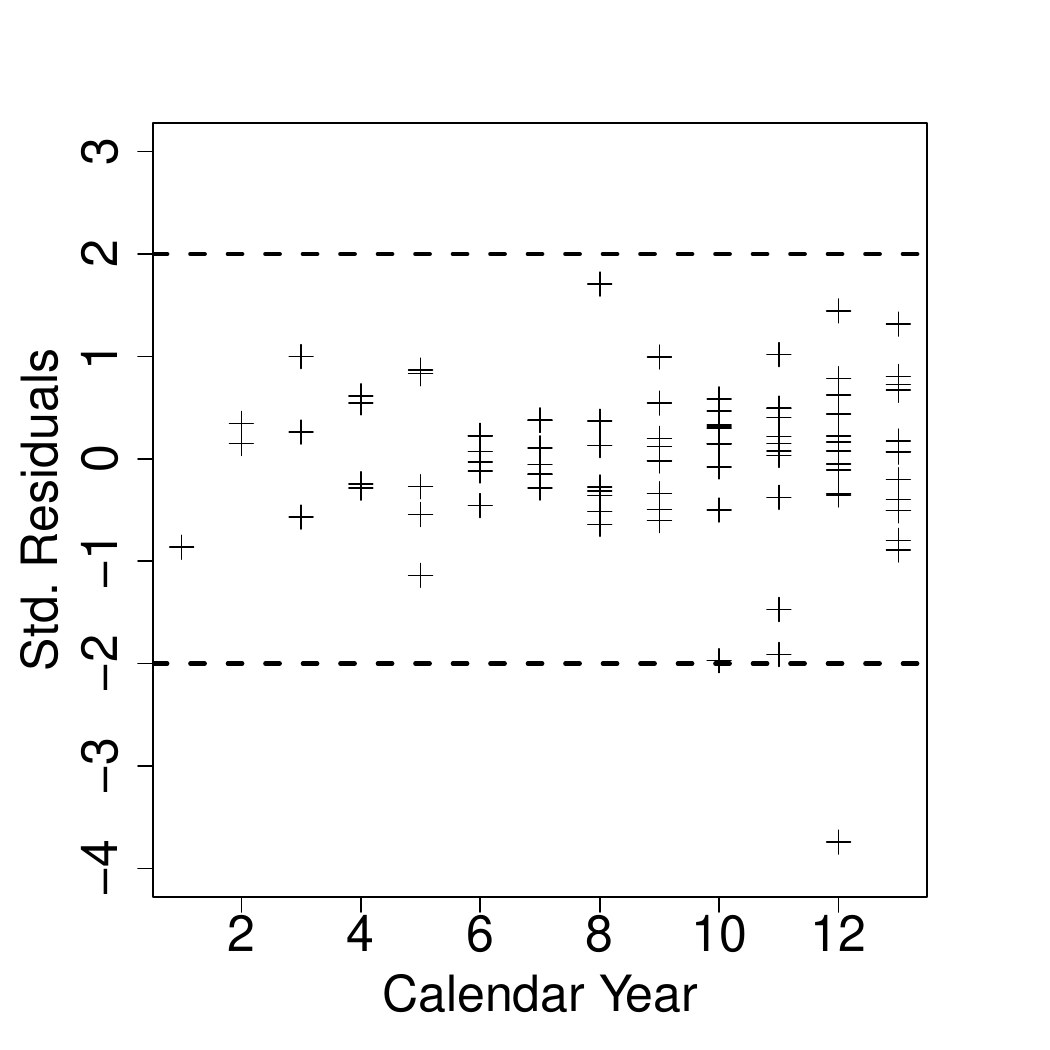}  &   \includegraphics[width=3.6cm]{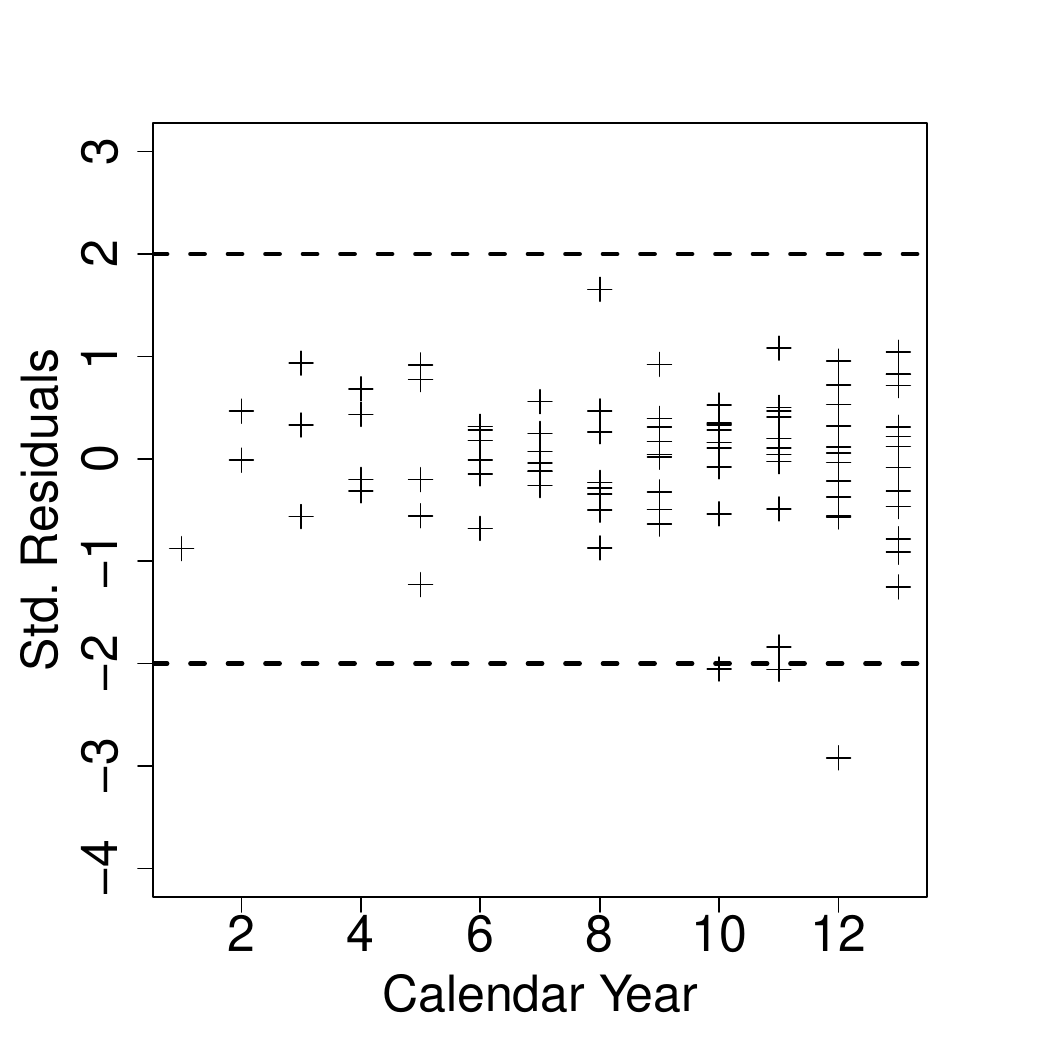} & \includegraphics[width=3.6cm]{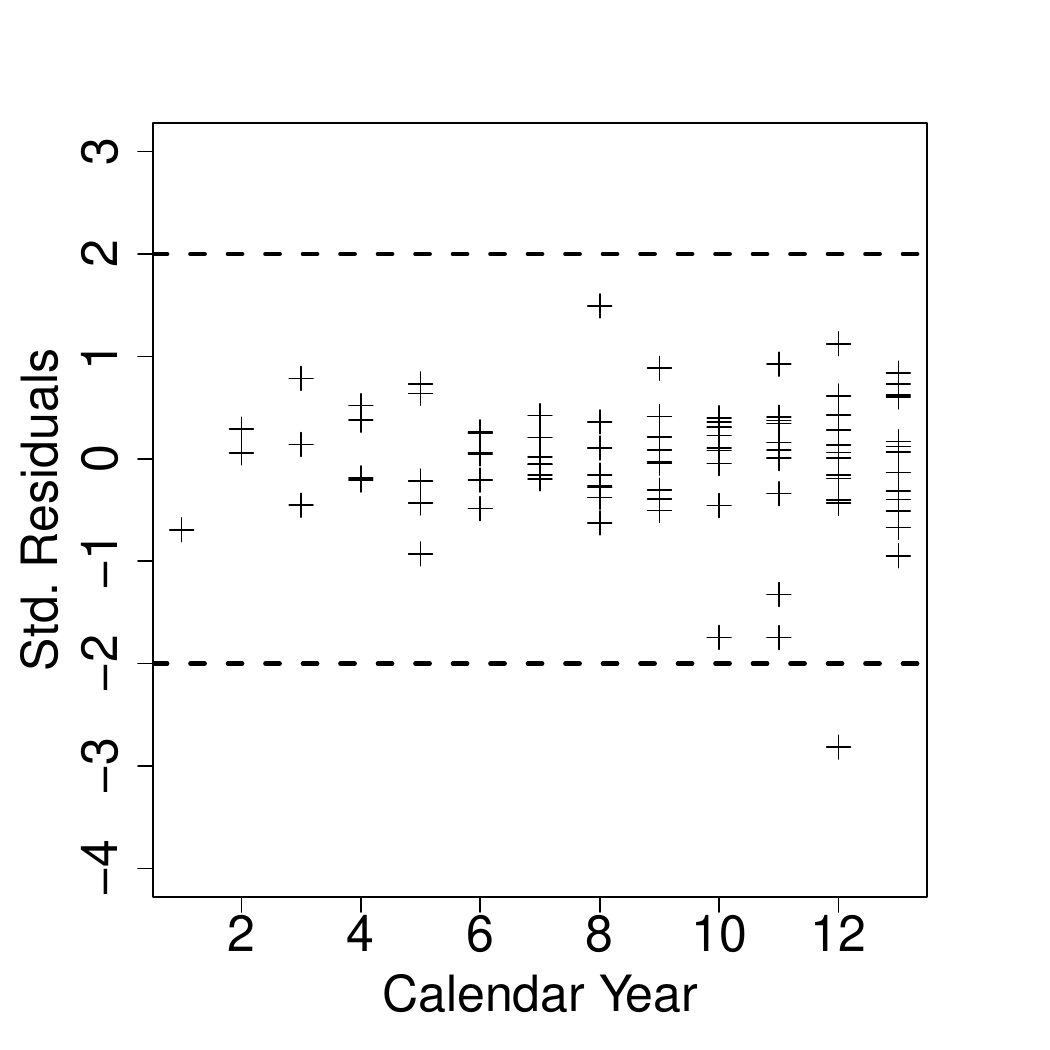} & \includegraphics[width=3.6cm]{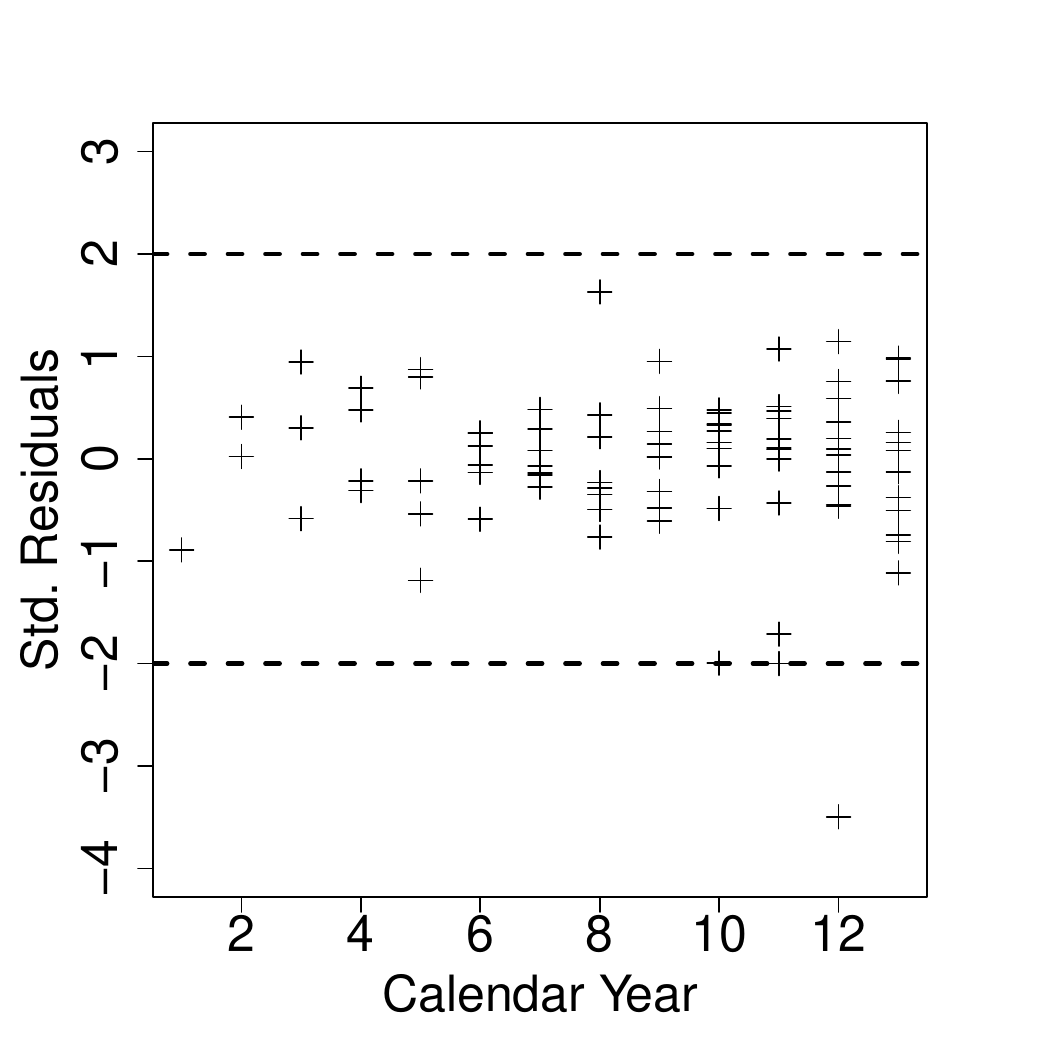}\\
     (a) Normal  & (b) Student-t & (c) Slash & (d) Variance Gamma\\
    \includegraphics[width=3.6cm]{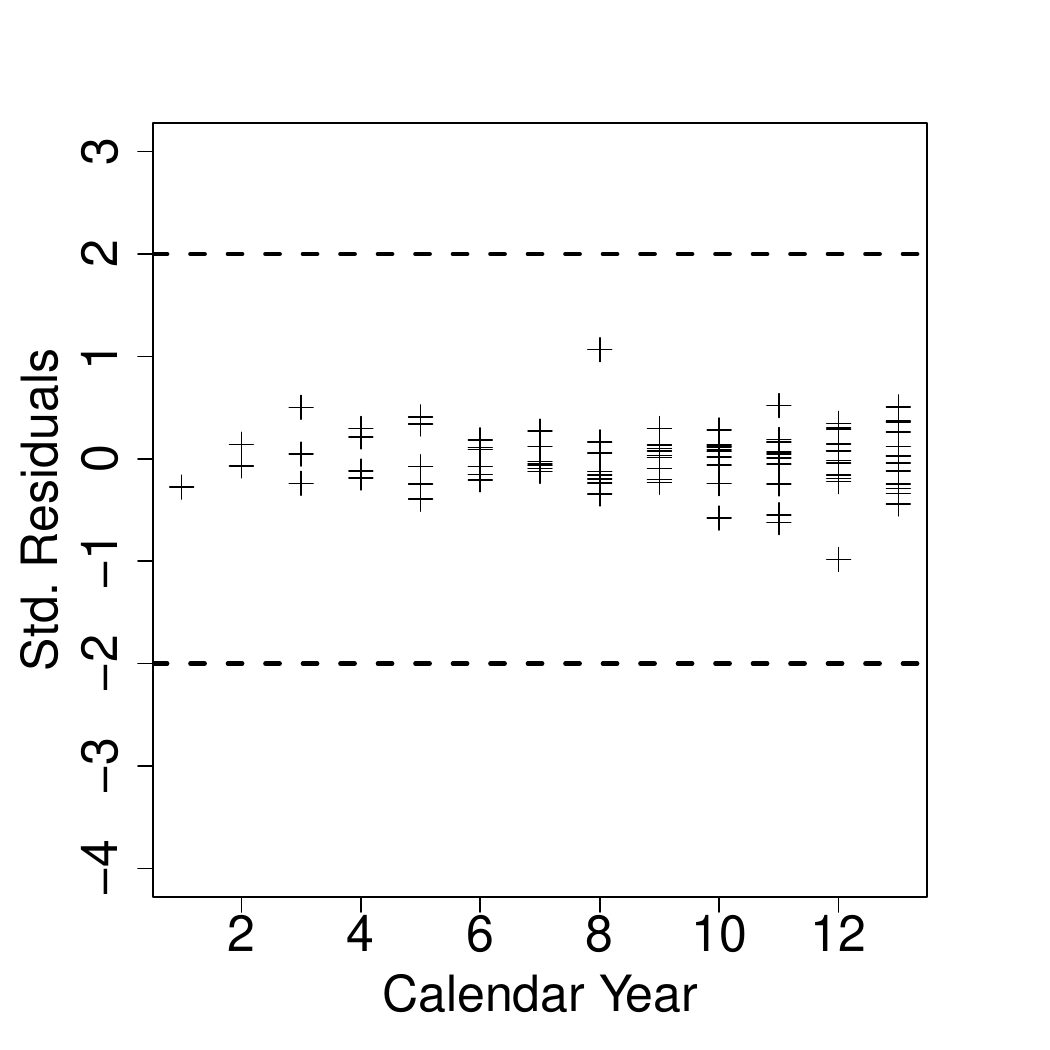}  &   \includegraphics[width=3.6cm]{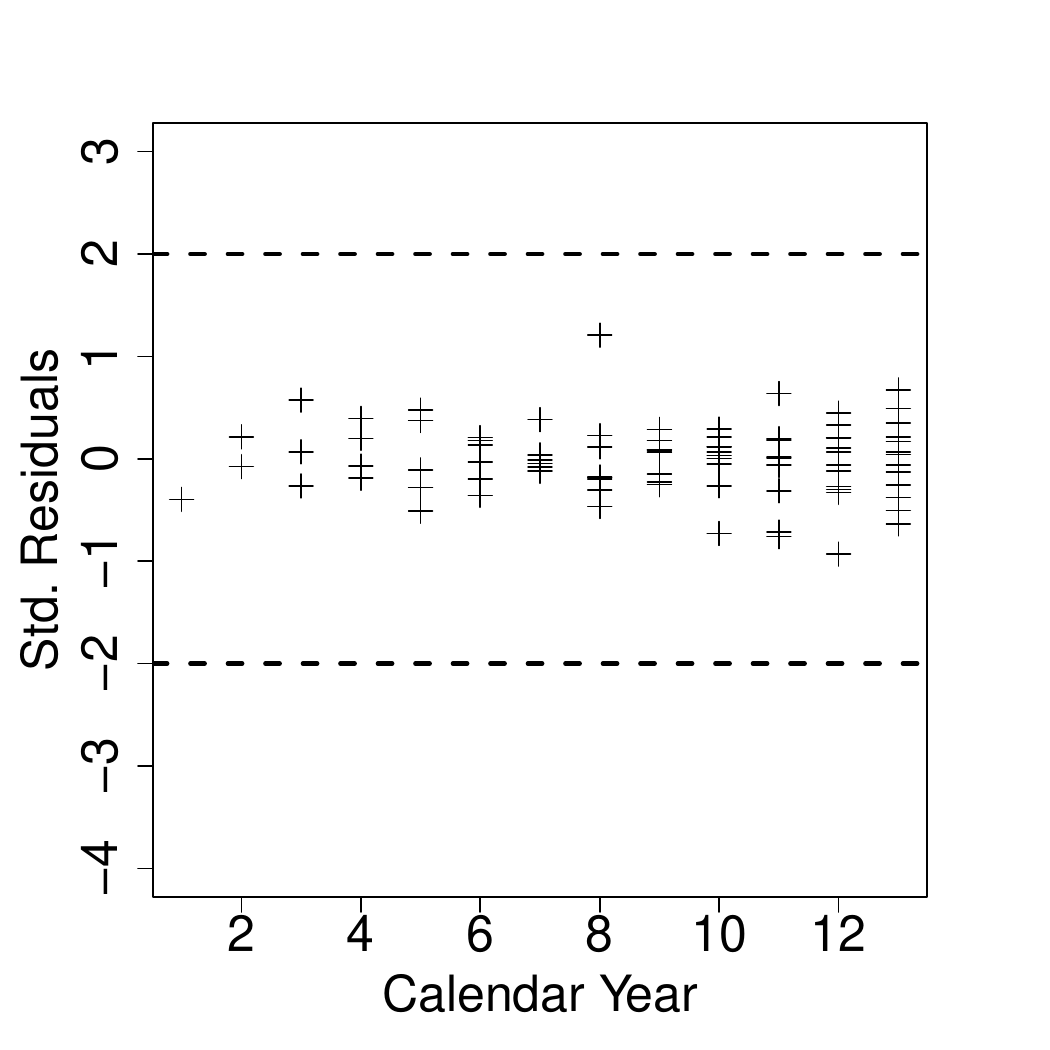} & \includegraphics[width=3.6cm]{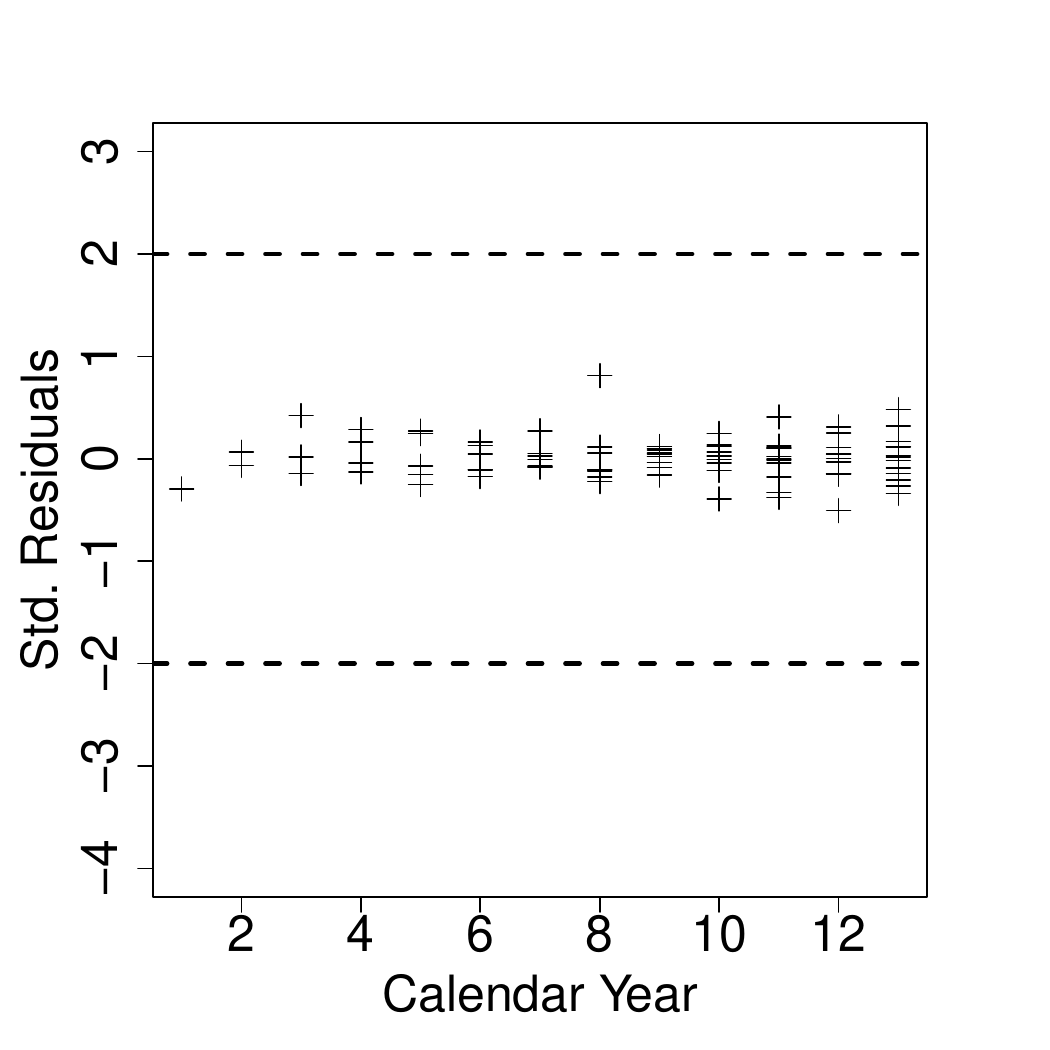} & \includegraphics[width=3.6cm]{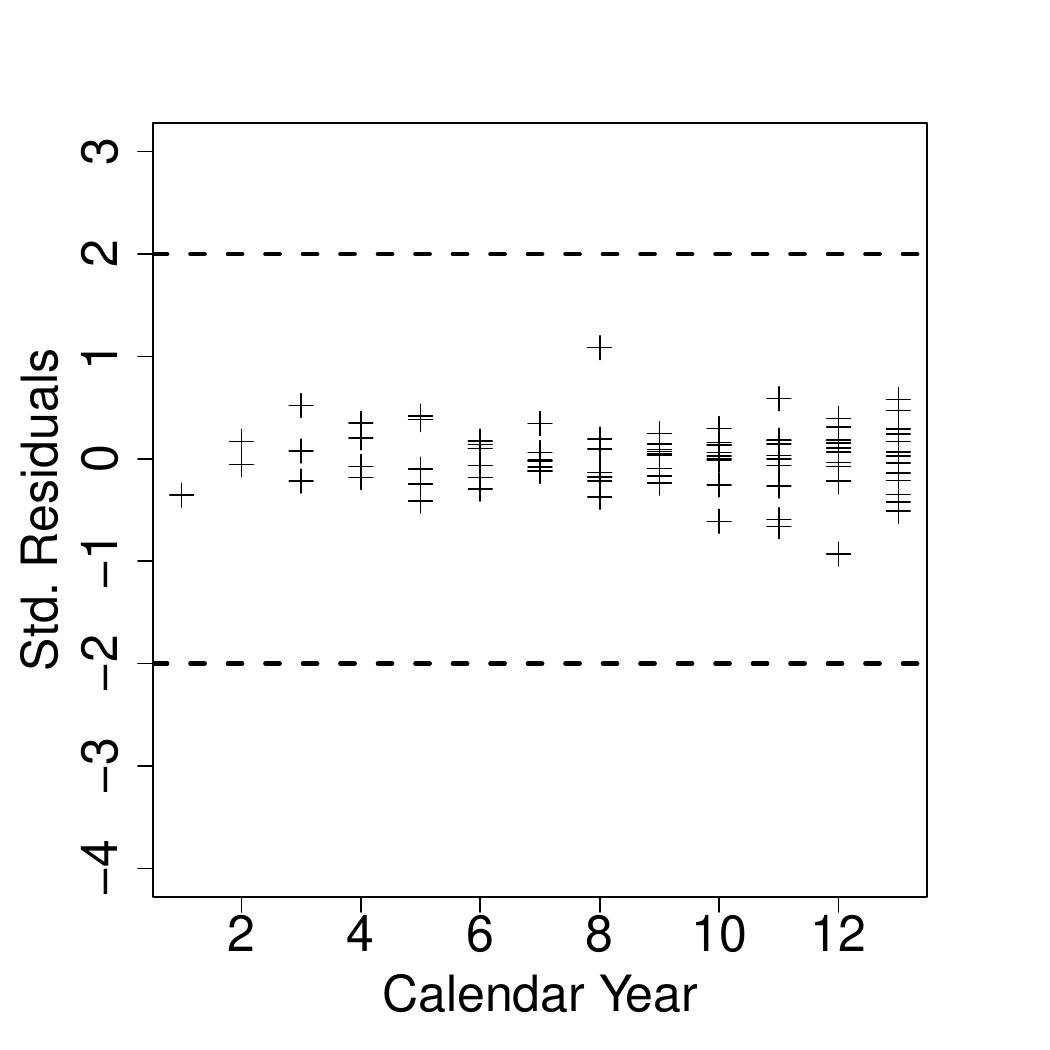}\\
     (e) Skew Normal  & (f) Skew-t & (g) Skew Slash & (h) Skew Variance Gamma\\
    \end{tabular}
    \caption{Case study: the standardized residuals for all competing models.}
    \label{sec32:fig8}
\end{figure}

In order to check the presence of outliers in the dataset, we consider computing the Bayesian standardized residuals (see Section \ref{sec2.5}) for all competing models. The panels in Figure \ref{sec32:fig8} display the residuals, indicating that there are no outliers for the skew class, whereas outliers are evident in the non-skew models, being unsuitable to accommodate this behaviour of the data.

\begin{figure}[H]
    \centering
    \begin{tabular}{cccc}
    \includegraphics[width=3.6cm]{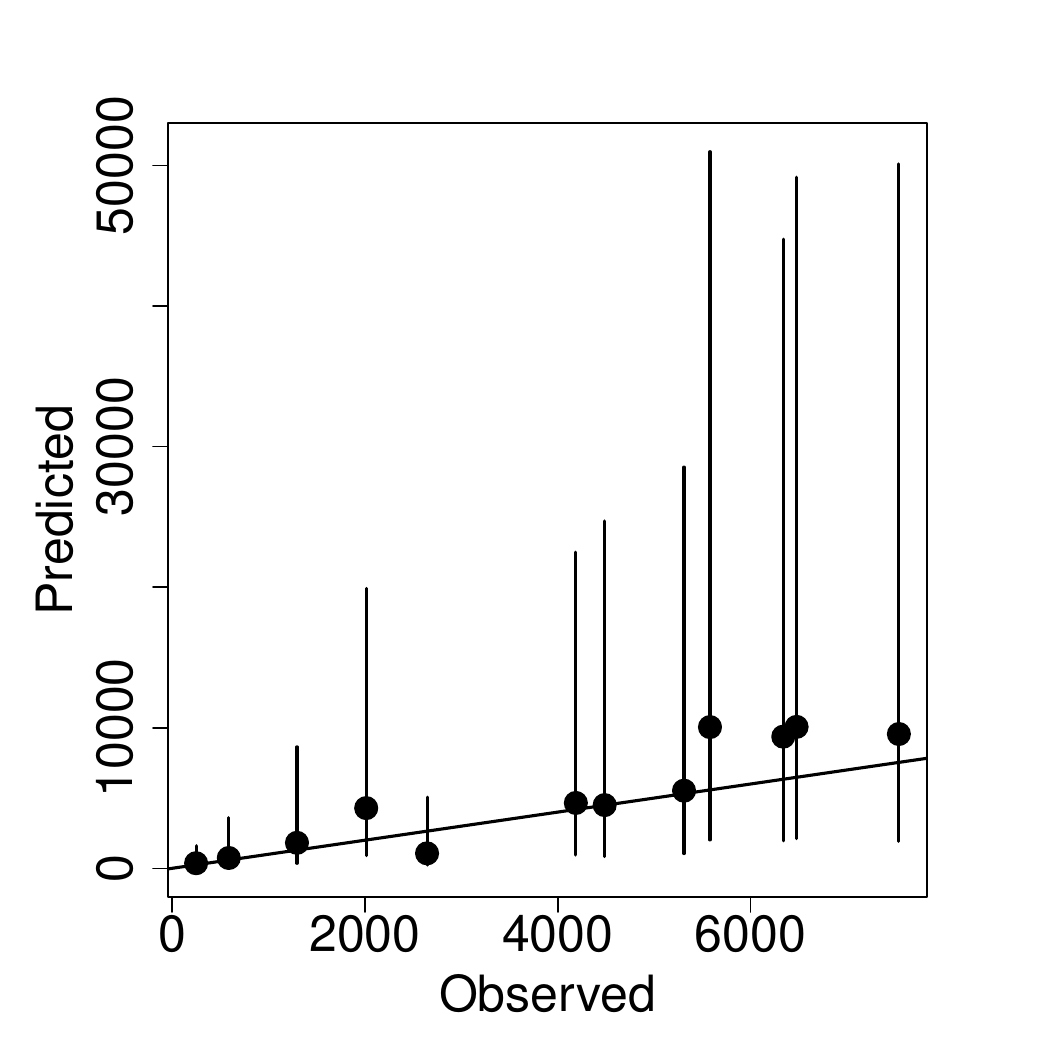}  &   \includegraphics[width=3.6cm]{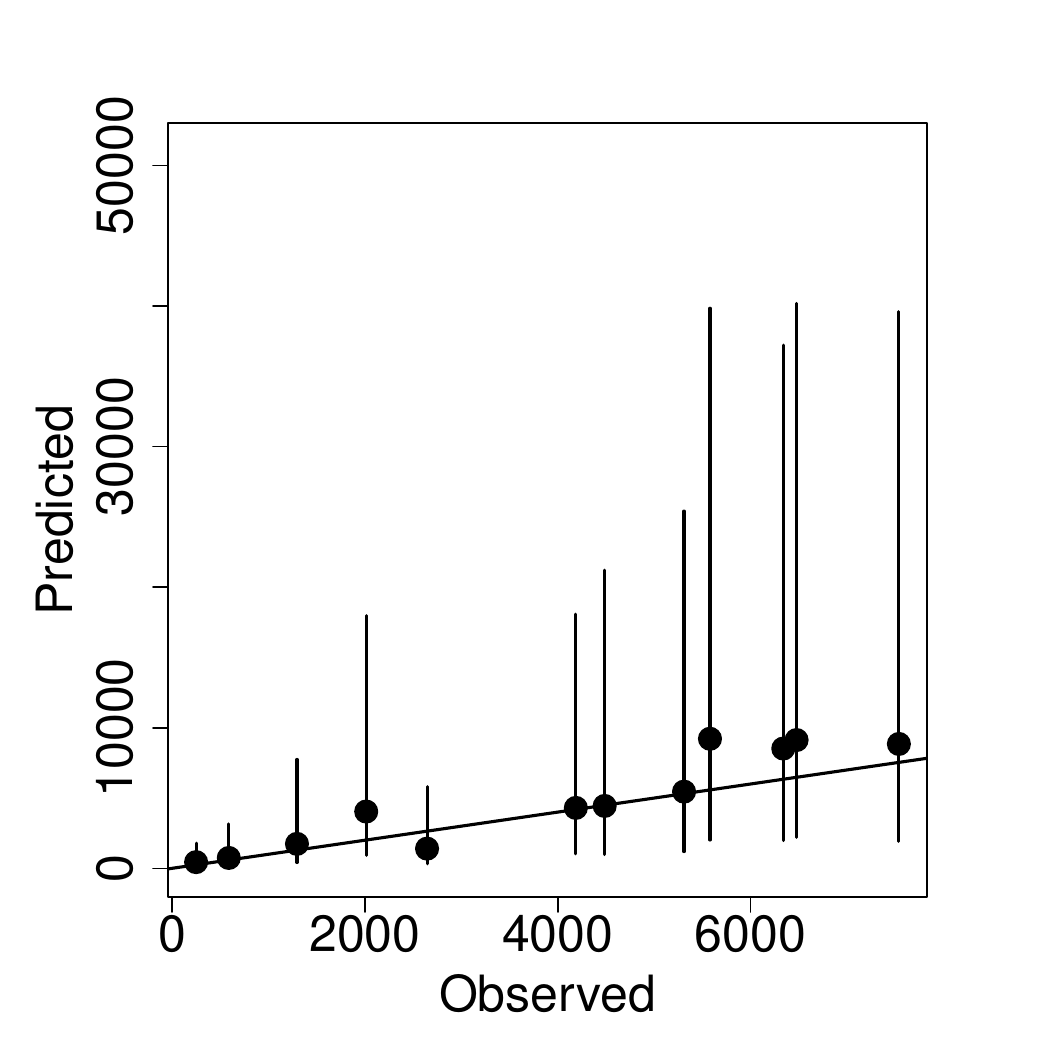} & \includegraphics[width=3.6cm]{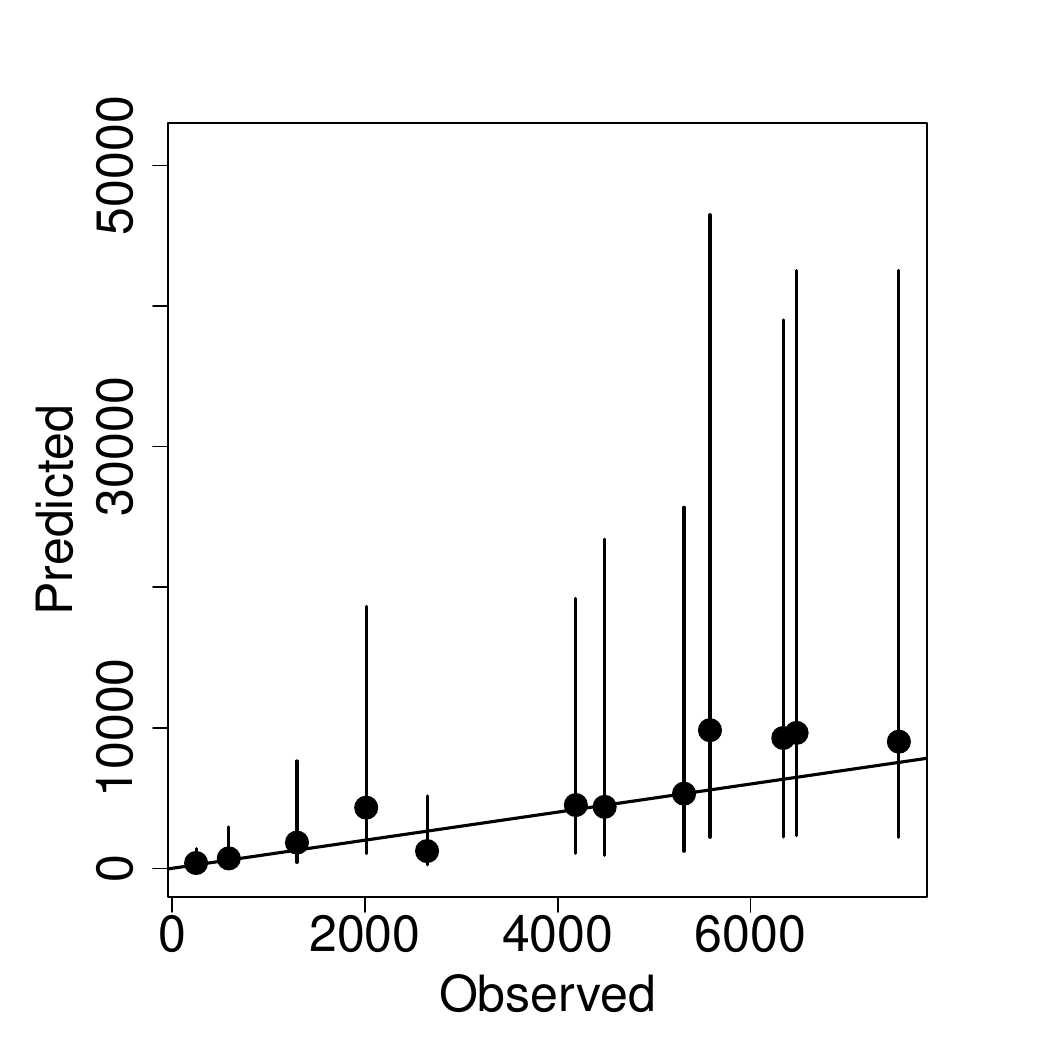} & \includegraphics[width=3.6cm]{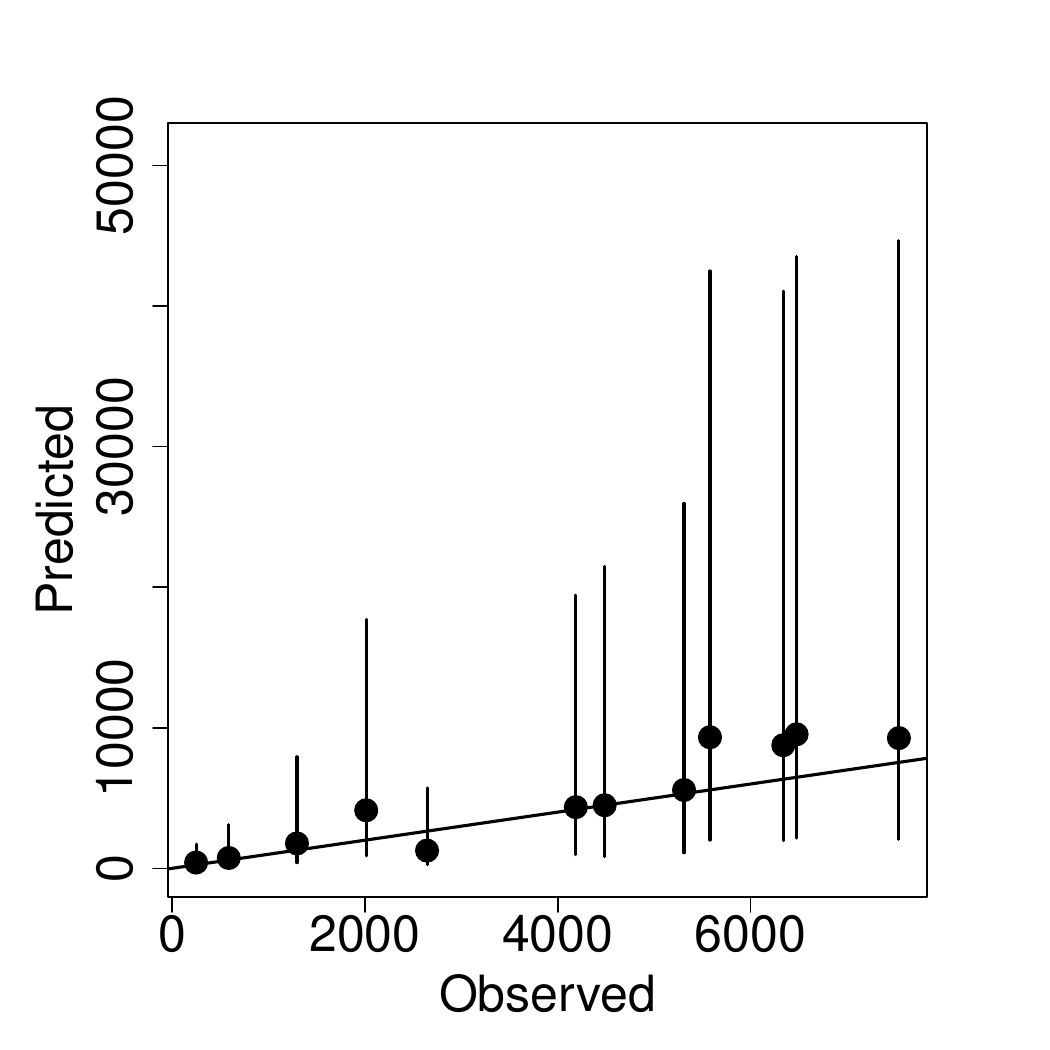}\\
     (a) Normal  & (b) Student-t & (c) Slash & (d) Variance Gamma\\
    \includegraphics[width=3.6cm]{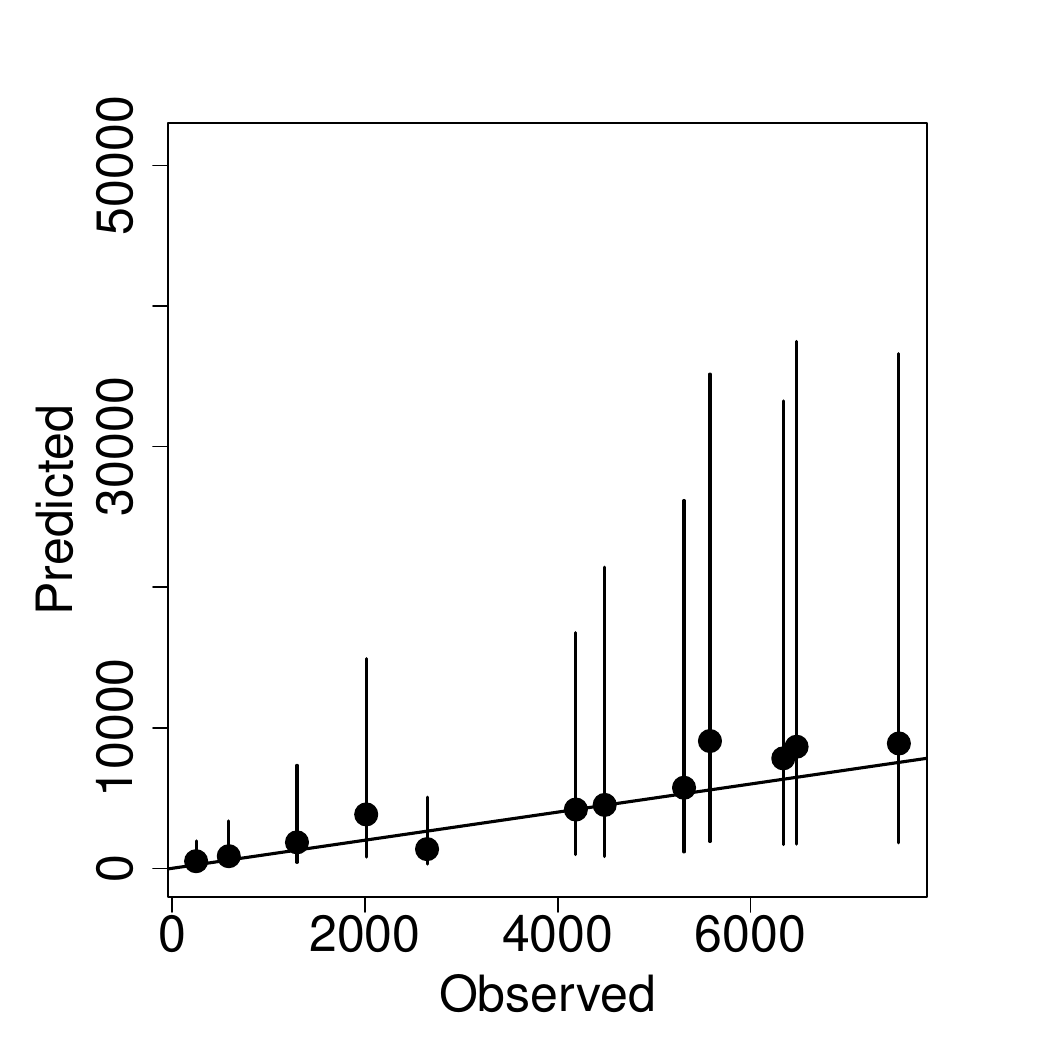}  &   \includegraphics[width=3.6cm]{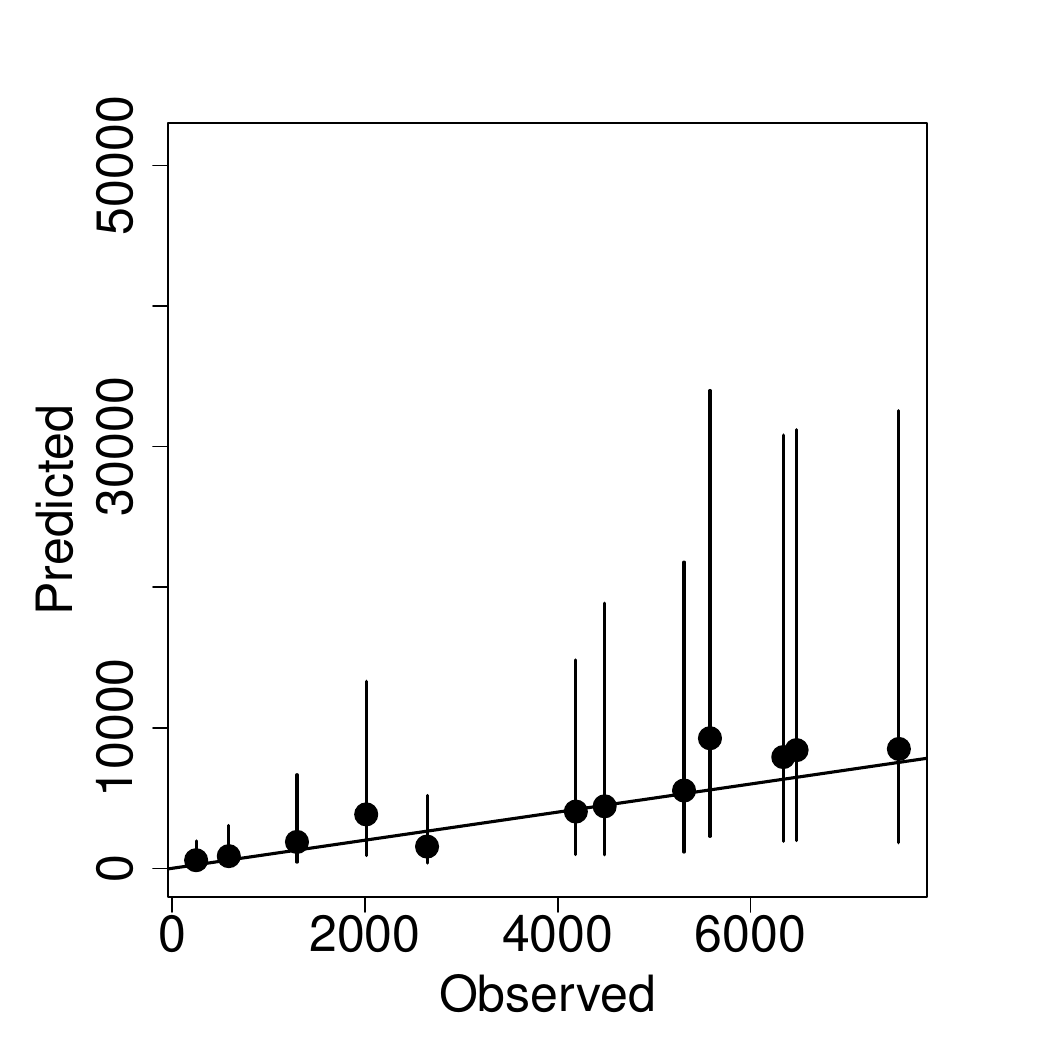} & \includegraphics[width=3.6cm]{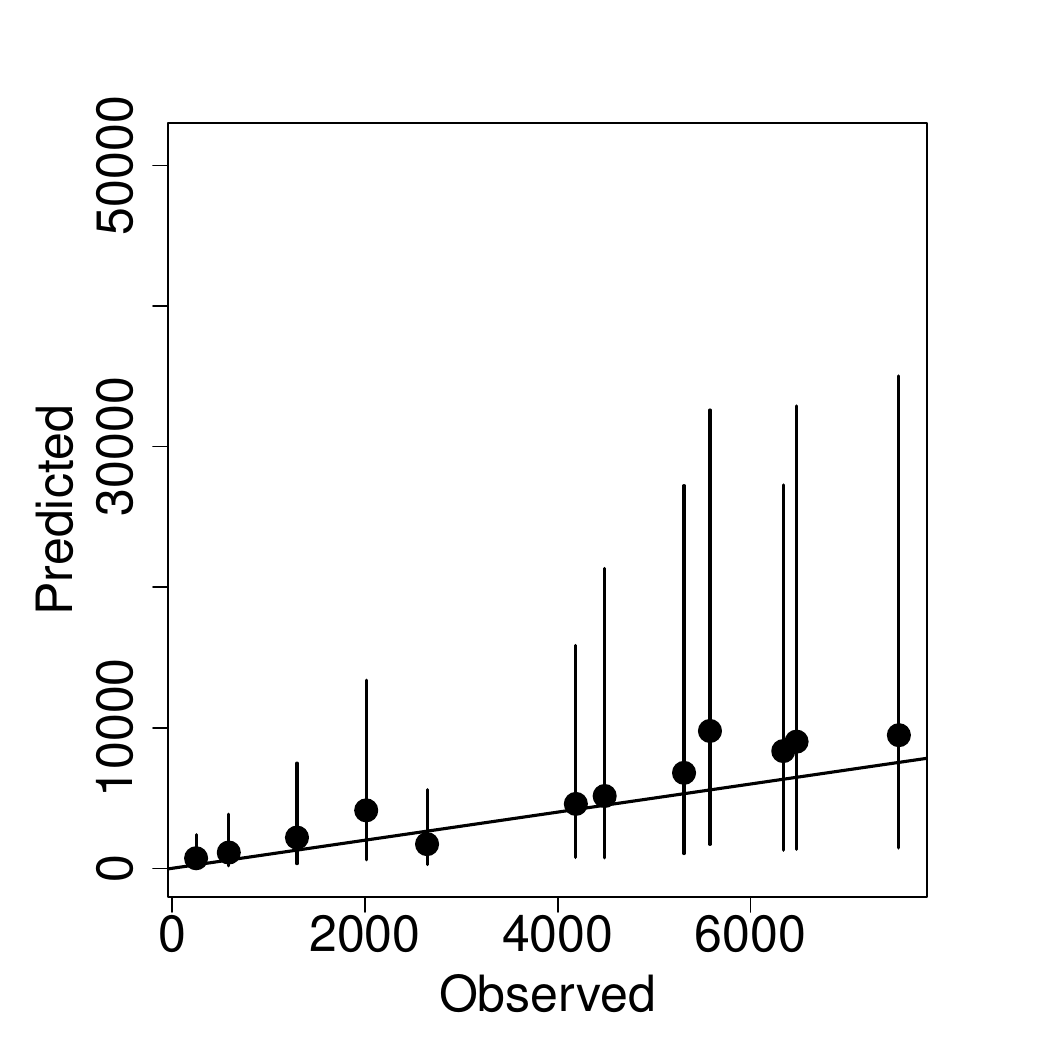} & \includegraphics[width=3.6cm]{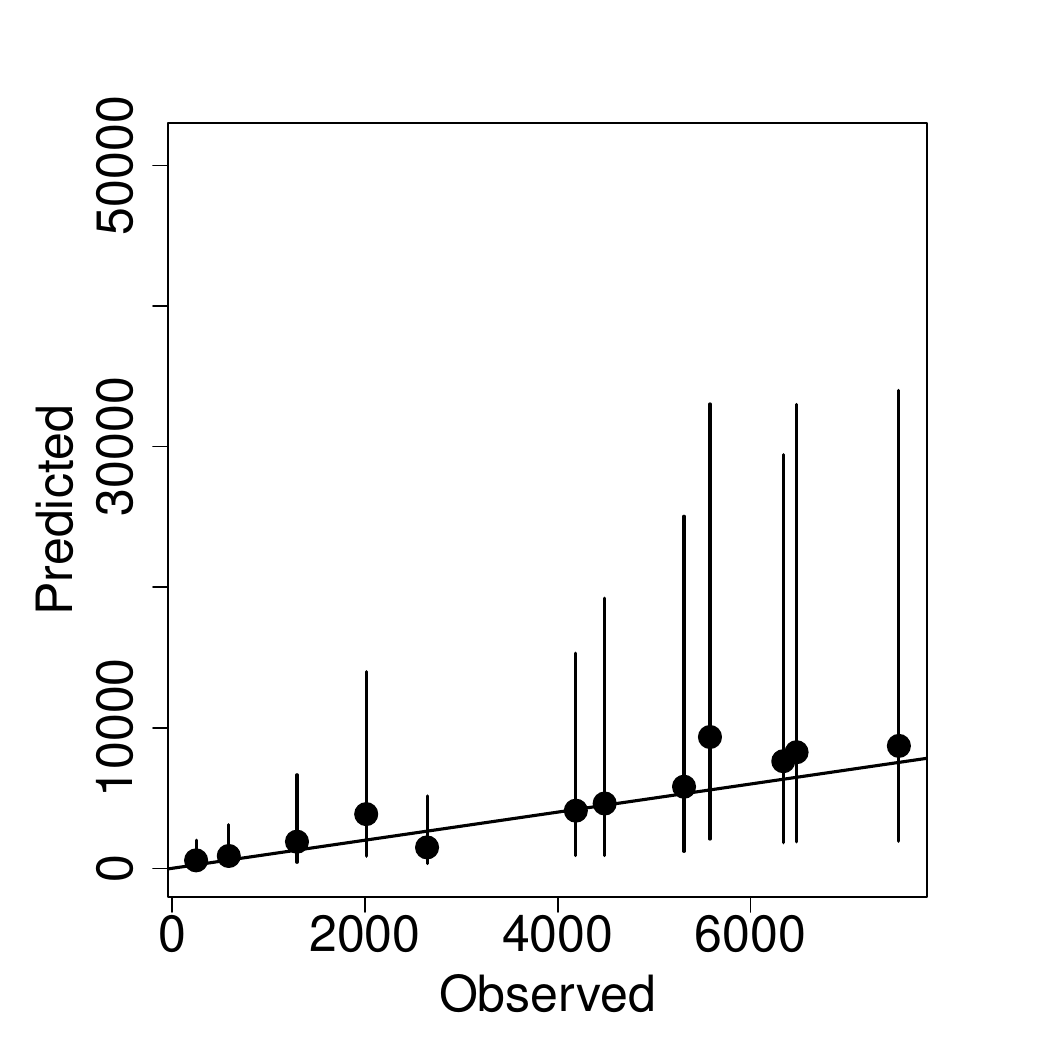}\\
     (e) Skew Normal  & (f) Skew-t & (g) Skew Slash & (h) Skew Variance Gamma\\
    \end{tabular}
    \caption{Case study: Posterior predictive medians (dots) versus the true observed loss reserving for the calendar year 1991 with vertical lines representing the 95\% credible predictive intervals. The diagonal line indicates $y = x$. (a) the Normal model, (b) the Student-t model, (c) the Slash model, (d) the Variance-Gamma model, (e) the Skew-Normal model, (f) the Skew-t model, (g) the Skew Slash model and (h) the Skew Variance-Gamma model. }
    \label{sec32:fig5}
\end{figure}

\begin{figure}[H]
    \centering
    \begin{tabular}{cccc}
    \includegraphics[width=3.6cm]{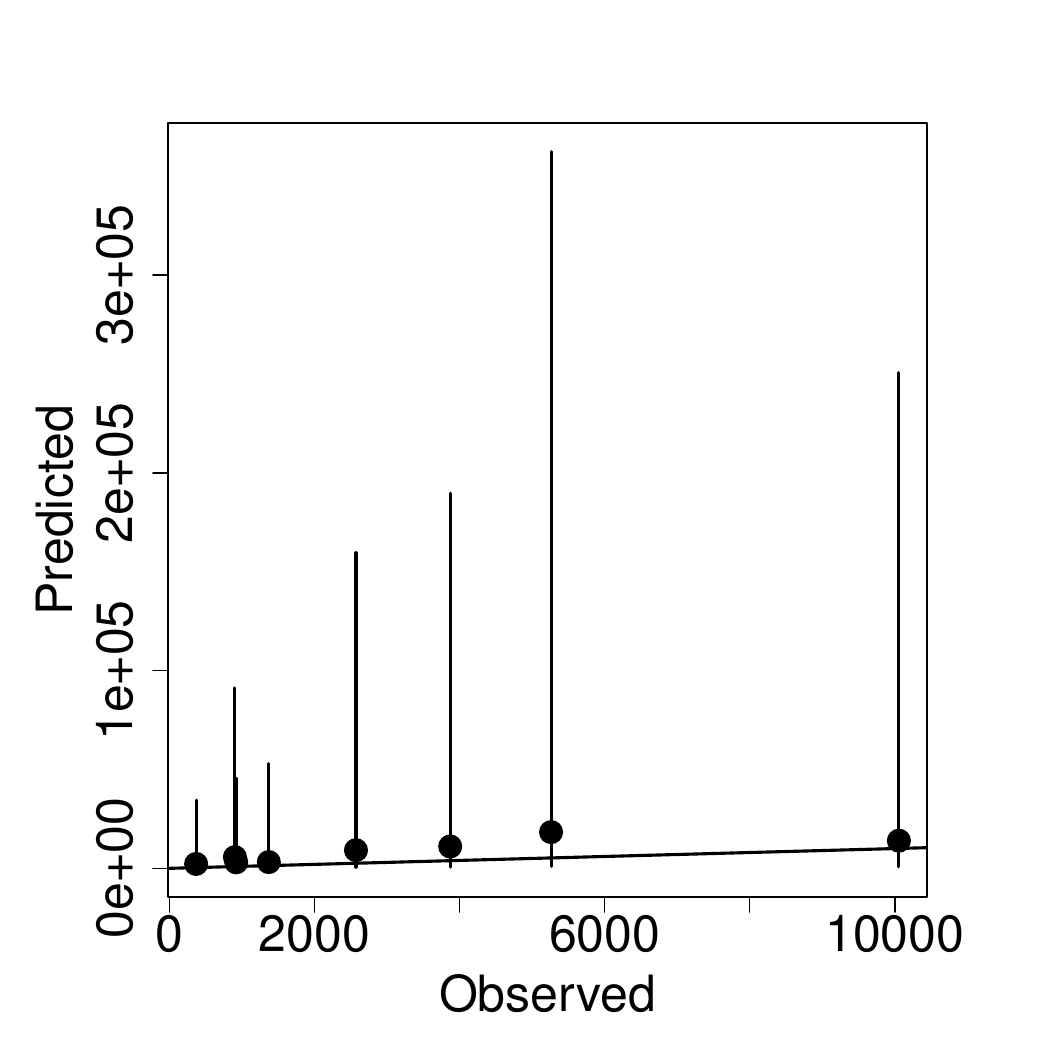}  &   \includegraphics[width=3.6cm]{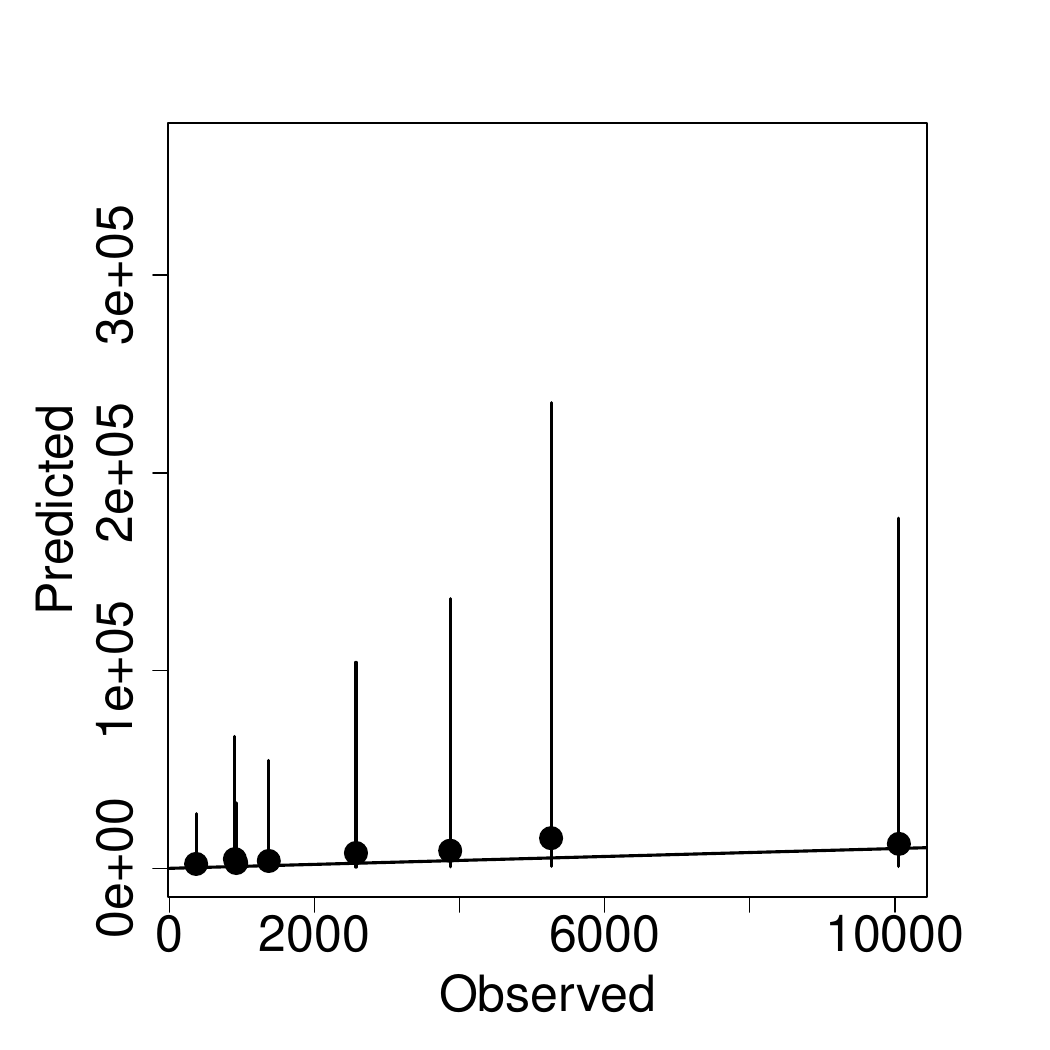} & \includegraphics[width=3.6cm]{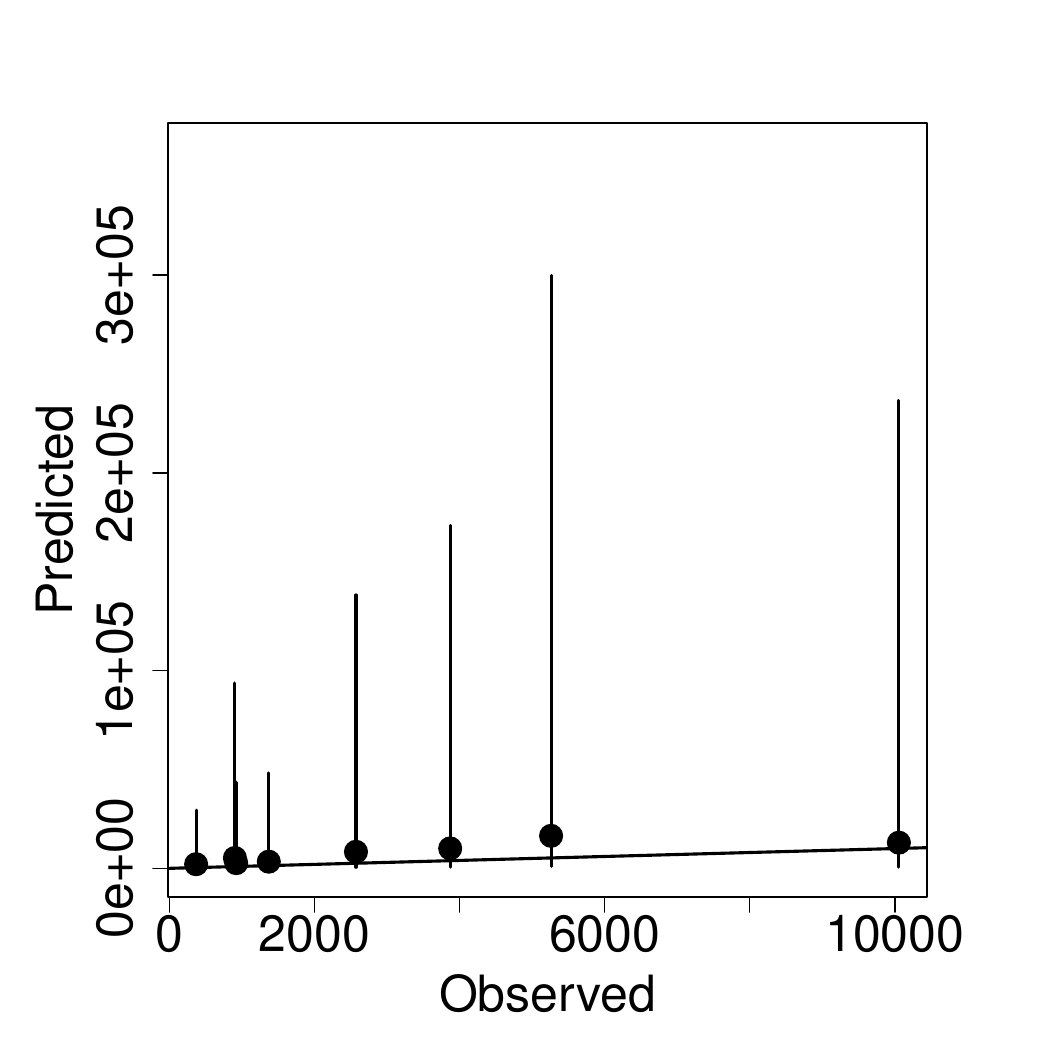} & \includegraphics[width=3.6cm]{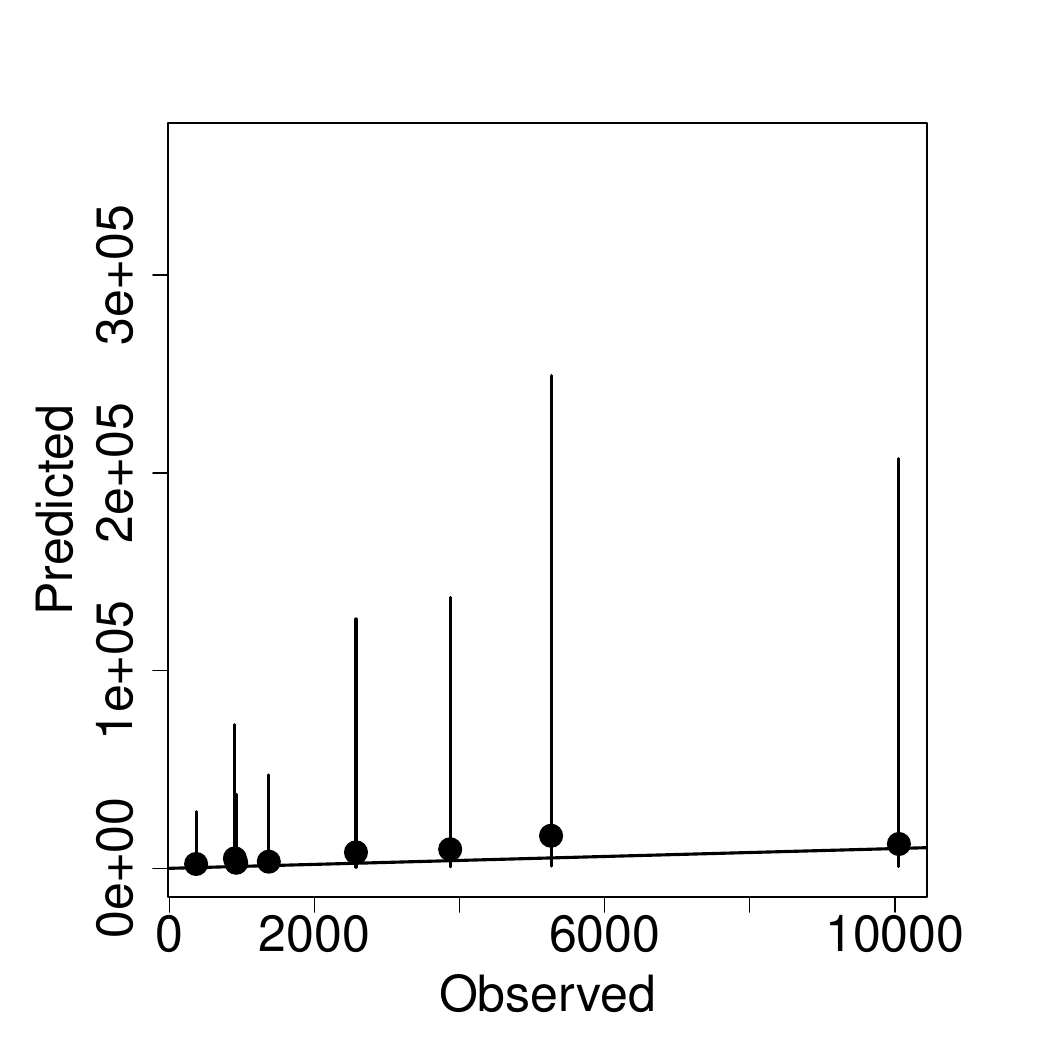}\\
     (a) Normal  & (b) Student-t & (c) Slash & (d) Variance Gamma\\
    \includegraphics[width=3.6cm]{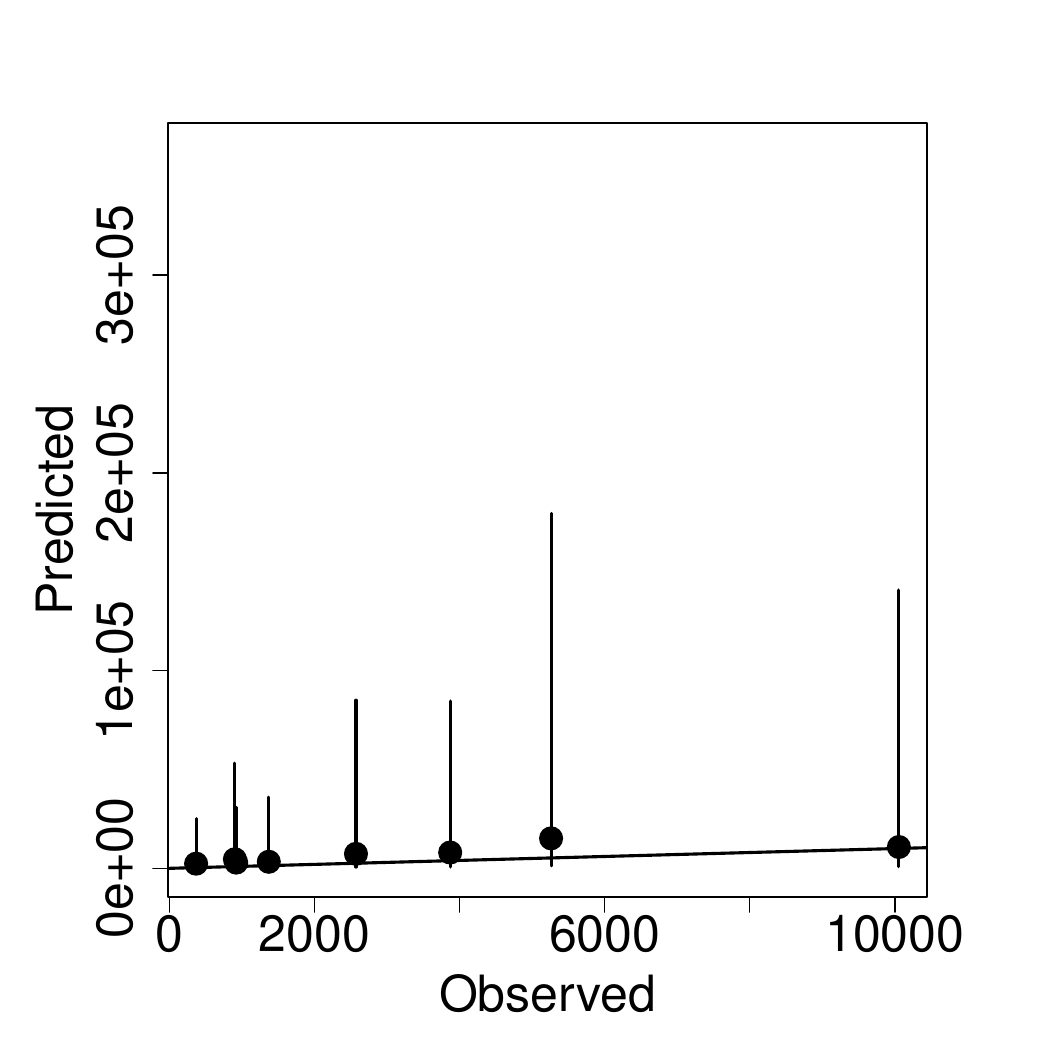}  &   \includegraphics[width=3.6cm]{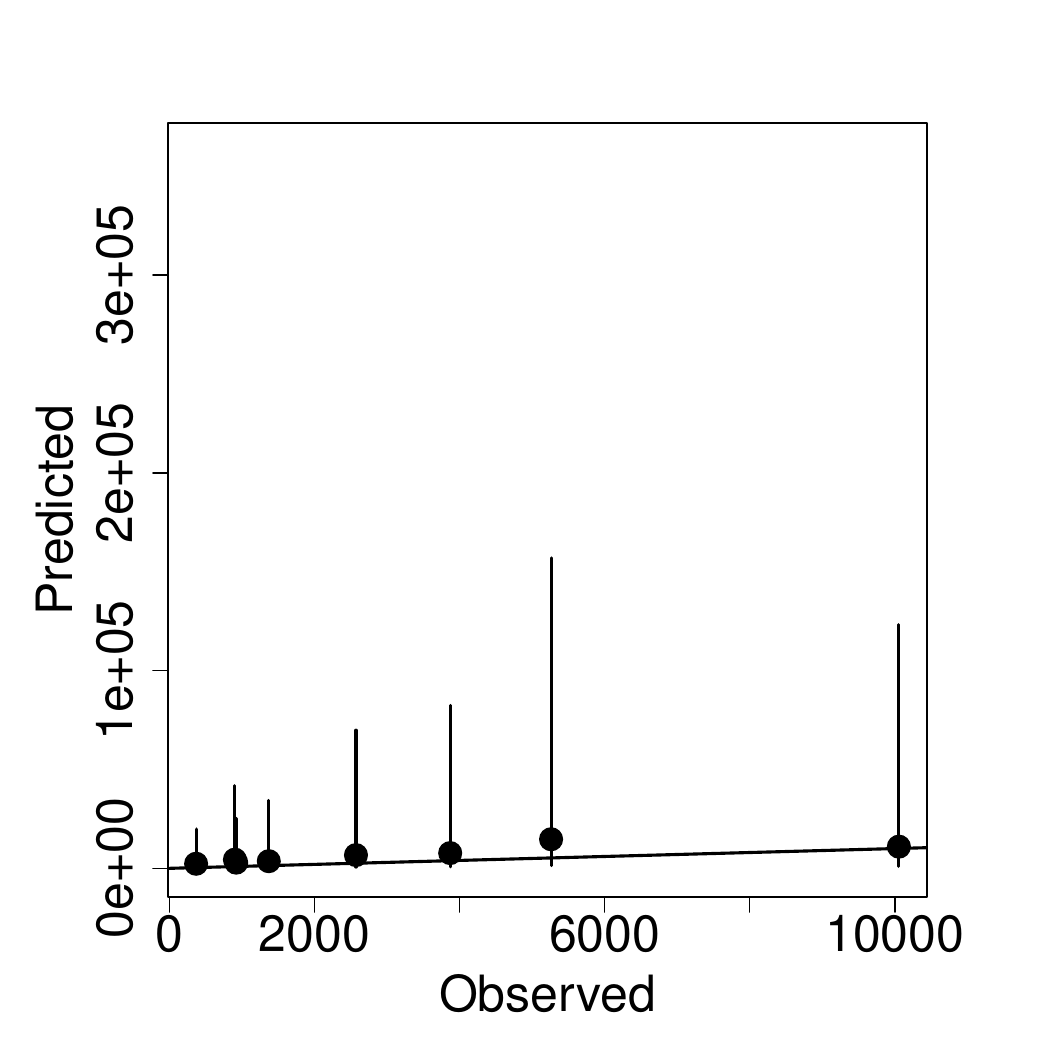} & \includegraphics[width=3.6cm]{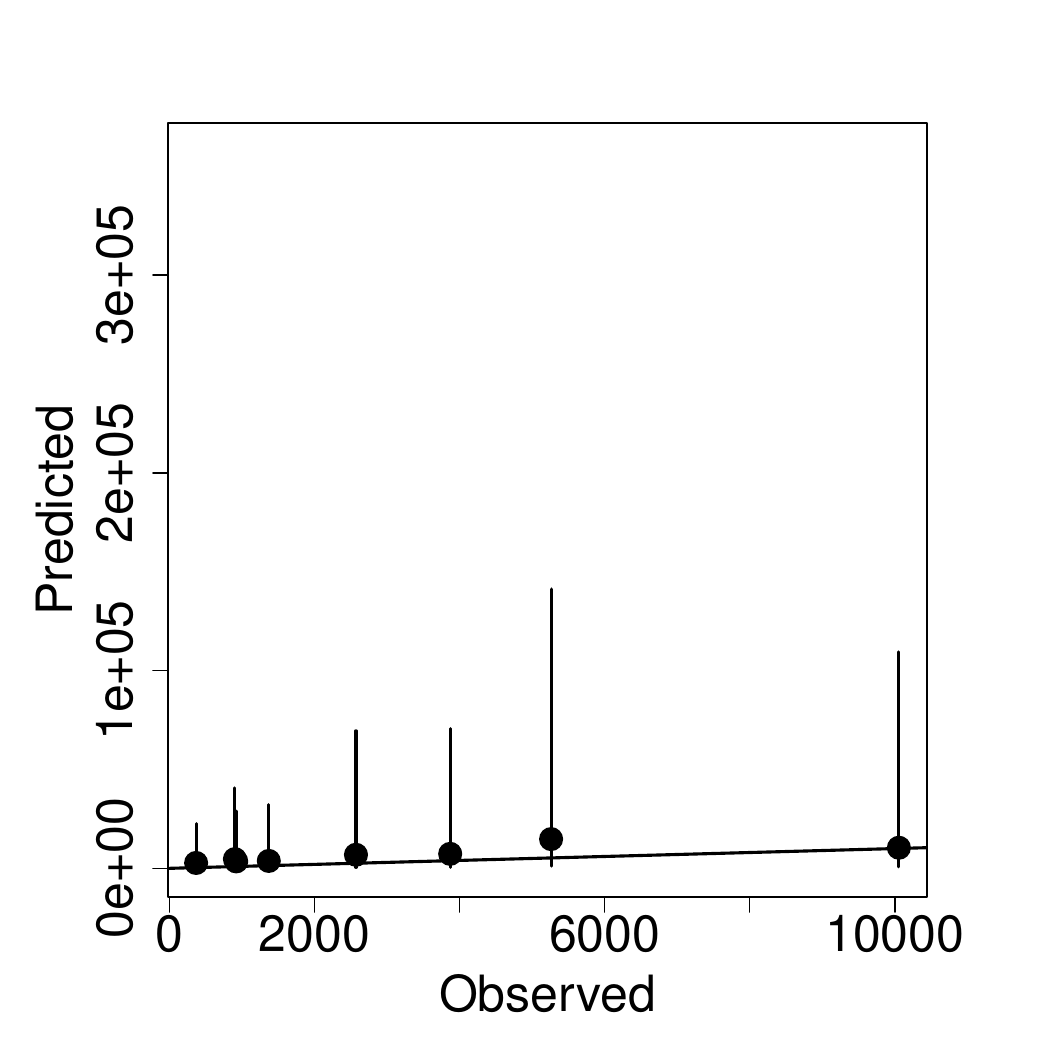} & \includegraphics[width=3.6cm]{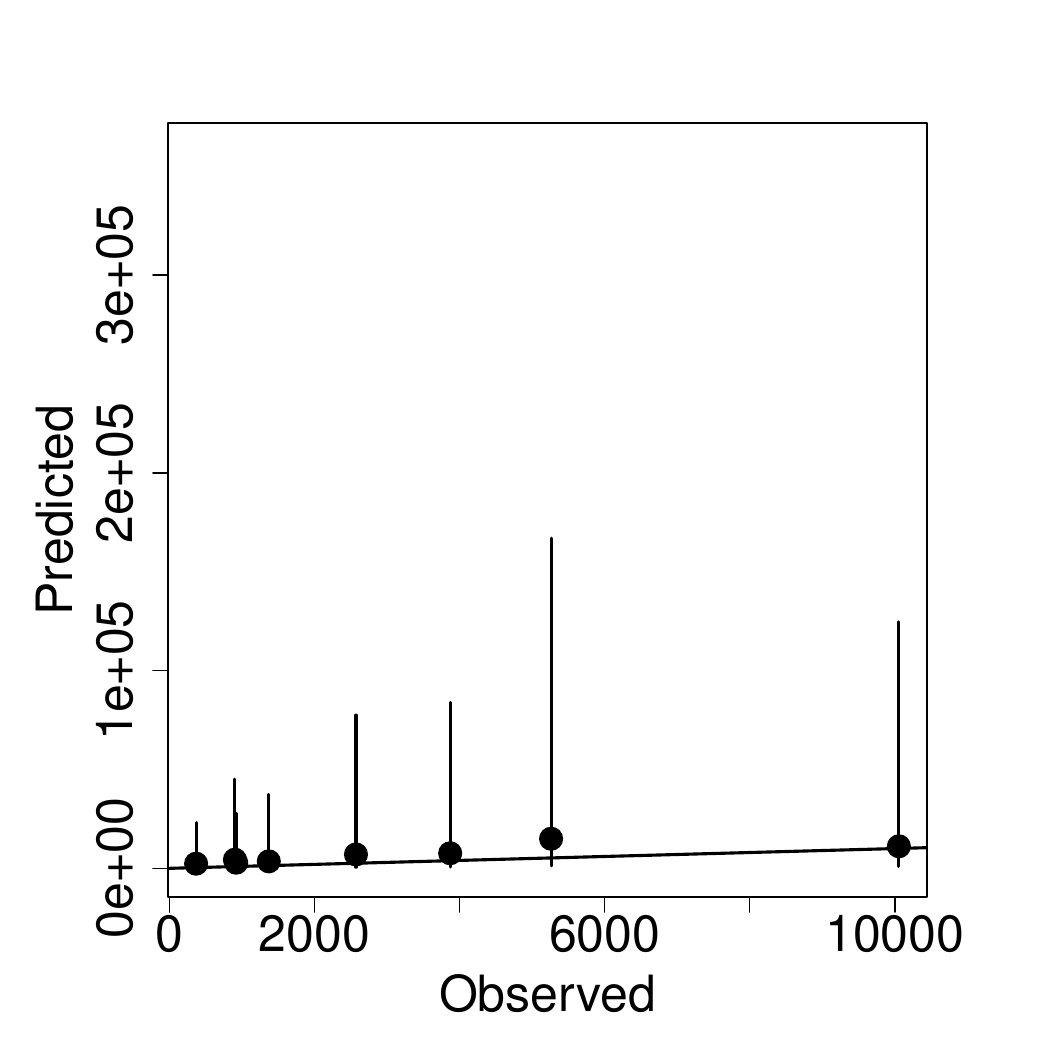}\\
     (e) Skew Normal  & (f) Skew-t & (g) Skew Slash & (h) Skew Variance Gamma\\
    \end{tabular}
    \caption{Case study: Posterior predictive medians (dots) versus the true observed loss reserving for the calendar year 1995 with vertical lines representing the 95\% credible predictive intervals. The diagonal line indicates $y = x$. (a) the Normal model, (b) the Student-t model, (c) the Slash model, (d) the Variance-Gamma model, (e) the Skew-Normal model, (f) the Skew-t model, (g) the Skew Slash model and (h) the Skew Variance-Gamma model.}
    \label{sec32:fig6}
\end{figure}

For predictive purposes, panels of Figures \ref{sec32:fig5} and \ref{sec32:fig6} show the summary of the posterior predictive distribution obtained for the calendar years (1991 and 1995). %that were left out from the inference procedure.
In general, the class of skew models provided interval widths smaller them the non-skewed models, resulting in the prediction of outstanding claim amounts with higher accuracy. The RMSPE measure, the average IS, the CRPS, and the CWI are shown in Table \ref{sec3:tab6}, where as expected, the Gaussian model is the one with the worst predictive performance. As mentioned previously, this model is not adequate to accommodate irregular claims and extreme outliers. Although we are taking into account models that capture thicker tails than Normal, the skewness is not allowed. Notice that models that regard this characteristic perform better than the non-skewed models (see ranking in Table \ref{sec3:tab6}). The measures indicate that the Skew-t has the better performance.

\begin{table}[ht]
\centering
\caption{Case study: Model comparison based on the Width of Credibility Interval (WCI),Root Mean Square Prediction Error (RMSPE) and the Continuous Ranking Predictive Score criteria for the predicted total loss reserve (the lower triangle) under all fitted models.}\label{sec3:tab6}
\begin{tabular}{|l|rrrr|c|}
  \hline
  \textbf{Competing}  & \multicolumn{4}{c|}{\textbf{Measures}} & \multirow{2}{*}{\textbf{Ranking}}\\
  \textbf{Models}     & \multicolumn{1}{c}{average IS} & \multicolumn{1}{c}{average WCI} & \multicolumn{1}{c}{RMSPE} & \multicolumn{1}{c|}{average CRPS} &\\ 
  \hline
  $\mathcal{N}$   & 5.584 &4.515 &1.513 &0.618 &8\\ 
  $\mathcal{ST}$   & 5.405 &4.137 &1.404 &0.578 &5\\ 
  $\mathcal{S}$   & 5.527 &4.354 &1.480 &0.595 &7\\ 
  $\mathcal{VG}$  & 5.451 &4.228 &1.431 &0.587 &6\\ 
  $\mathcal{SN}$  & 5.289 &4.019 &1.378 &0.575 &3\\ 
  $\mathcal{SST}$  & \textbf{5.232} &\textbf{3.743} &\textbf{1.329} &\textbf{0.564} &1\\ 
  $\mathcal{SS}$  & 5.316 &4.012 &1.612 &0.617 &4\\ 
  $\mathcal{SVG}$ & 5.284 &3.856 &1.351 &0.575 &2\\ 
   \hline
\end{tabular}
\end{table}

Table \ref{tab:tab6} presents the posterior summaries derived from the Skew-t model for predicting claim payments, as per comparison measures. To aid in diagnosing the convergence of the chains, we employ the convergence diagnostic (CD) based on the range of -1.96 to +1.96, frequently linked with the 95\% confidence interval for a standard normal distribution. We can observe that both chains appear to display convergence behaviour. The parameter responsible for skewness, $\rho$ is significant and equals -0.909 indicating that the model can capture the asymmetry in the dataset. Besides that, the posterior mean for $\nu$ equals 11.614 shows that the proposed model presents a heavier tail than the Gaussian model, indicating a relatively higher probability of observing extreme values.  

 \begin{table}[ht]
    \caption{Case study: Posterior summary for the Skew-t model.}
    \label{tab:tab6}
    \centering
    \begin{tabular}{|lrrrr|}\hline
         \textbf{Parameter}    & \textbf{Mean} & \textbf{Std. Dev.} & \textbf{IC(95\%)} & \textbf{CD}\\\hline
         $\mu$                 & 8.926         & 0.379 & $\left(8.206,9.679\right)$   &-0.606\\
         $\rho$                & -0.909        & 0.161 & $\left(-0.999,-0.455\right)$ & 0.035\\
         $\sigma^2$            & 0.294         & 0.114 & $\left(0.112,0.556\right)$   &-0.807\\
         $\nu$                 & 11.614        & 4.159 & $\left(4.958,21.064\right)$  &-0.578\\
         $\sigma_{\alpha}^{2}$ & 0.131         & 0.085 & $\left(0.038,0.342\right)$   & 0.776\\
         $\sigma_{\beta}^{2}$  & 0.063         & 0.025 & $\left(0.027,0.122\right)$   &-0.393\\
         $\sigma_{\gamma}^{2}$ & 0.151         & 0.090 & $\left(0.050,0.391\right)$   &-0.110\\\hline
    \end{tabular}
\end{table}

\begin{figure}[H]
    \centering
    \includegraphics[width=12cm]{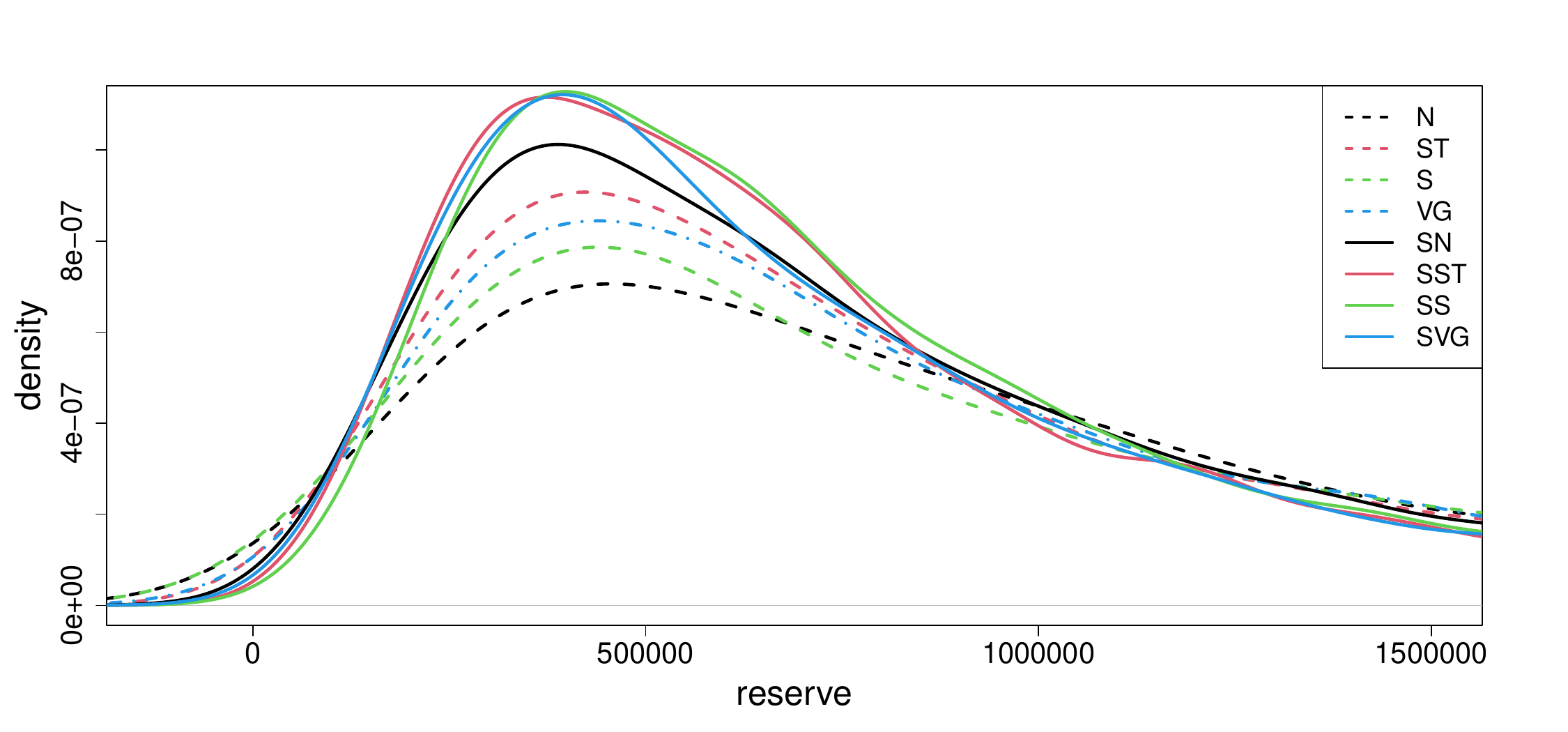}
    \includegraphics[width=12cm]{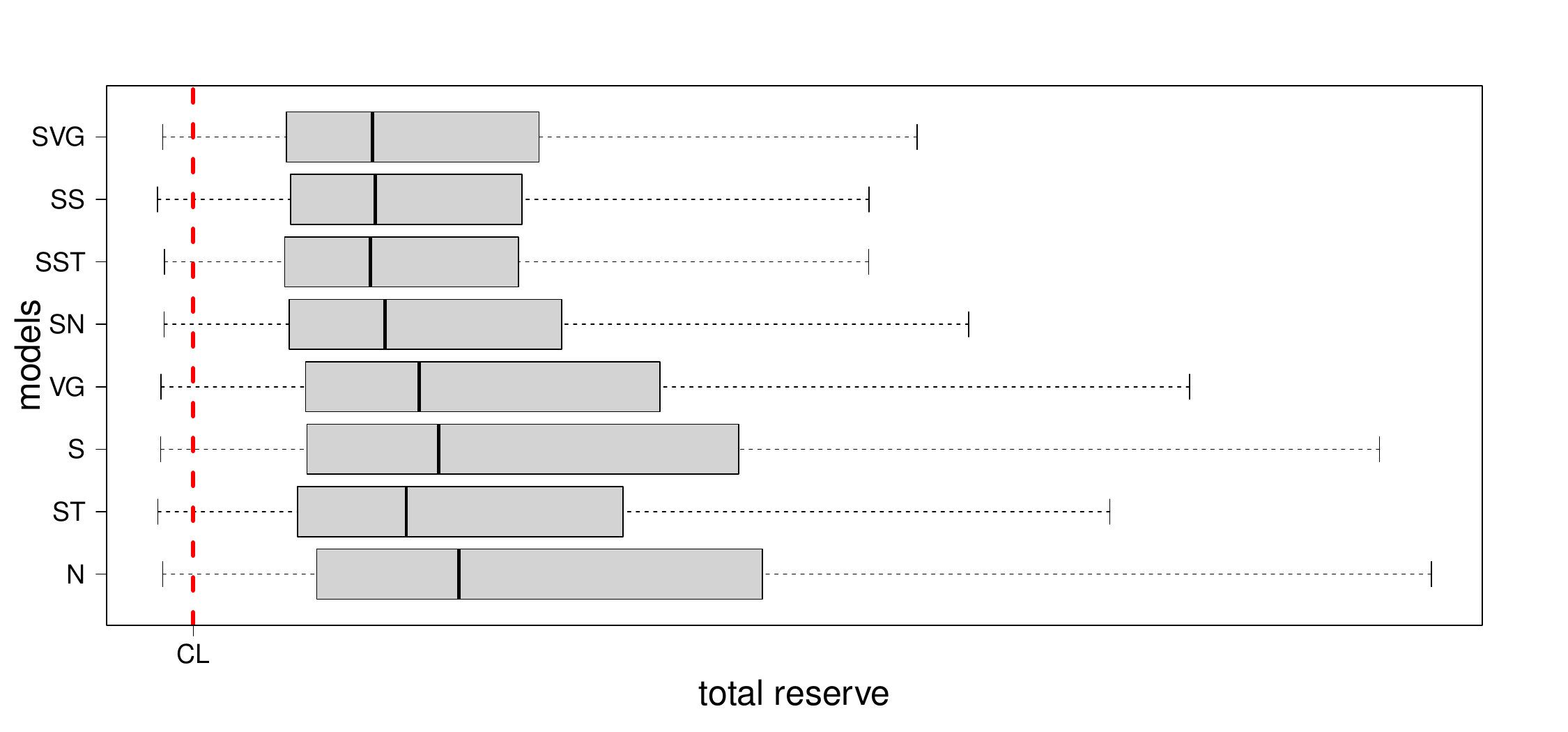}
    \caption{Case study: predictive posterior total loss reserving for all competing models. The dashed red line represents the chain ladder loss reserve.}
    \label{sec32:fig7}
\end{figure}

Figure \ref{sec32:fig7} displays box-plots and the posterior predictive distribution of total loss reserves for non-skewed and skewed models, in comparison to the chain ladder method. The plots indicate that the predictive distribution of total loss reserves for the non-skewed models is wider than for the skewed models. It is worth noting that the validation process excluded the last five years (1991-1995). Note that during this validation, a portion of the true reserve for the triangle $Y_{\nabla}$ is known, with a value of 191,274.00. This finding suggests that the chain ladder method underestimates the total loss reserve, providing a total reserve of 123,776.90. Table \ref{tab:tab9} presents the predictive posterior total reserve for quantiles 20\%, 35\%, 50\%, 65\%, and 80\% for all competing models. It is observed that higher quantiles (above 65\%) result in some overestimation in loss reserving, whereas lower quantiles (below 50\%) exhibit higher accuracy when compared to the chain ladder technique.

\begin{table}[h!]
    \centering
      \caption{Case study: Predictive posterior quantiles of 20\%, 35\%, 50\%, 65\% and 80\% of the total reserve for all competing models.}
    \label{tab:tab9}
    \begin{tabular}{|lccccc|}\hline
        \multirow{2}{*}{\textbf{Models}} & \multicolumn{5}{c|}{\textbf{Predictive Posterior Quantiles}}\\\cline{2-6}
                     & $20\%$ & $35\%$ & $50\%$ & $65\%$ & $80\%$ \\\hline
          $\mathcal{N}$   &415,528.3& 625,046.3& 887,277.5& 1,281,023.6& 2,095,537\\
          $\mathcal{ST}$   &374,646.8& 534,794.5& 735,967.6& 1,018,925.9& 1,609,357\\
          $\mathcal{S}$   &393,488.7& 571,226.8& 829,155.1& 1,236,204.8& 2,009,094\\
         $\mathcal{VG}$   &384,843.8& 562,636.8& 773,439.6& 1,105,735.7& 1,751,480\\
         $\mathcal{SN}$   &354,978.1& 501,559.2& 675,223.2&  927,993.2& 1,370,806\\
         $\mathcal{SST}$   &343,518.2& 479,568.9& 632,237.3&  831,590.2& 1,212,584\\
         $\mathcal{SS}$   &363,925.6& 495,431.3& 647,159.1&  859,286.6& 1,218,528\\
        $\mathcal{SVG}$   &346,078.8& 475,412.1& 638,935.6&  872,867.0& 1,288658\\\hline
    \end{tabular}
\end{table}

\section{Conclusions}\label{sec4}

We have introduced a versatile framework for loss reserving that builds upon the well-established space-state loss reserving modelling. Our framework accommodates heavy tails and skewness in the run-off triangle structure through a scale mixture of skew-normal distributions. The dynamic evolution in the mean model proposed in equation (\ref{sec2:eq10}) incorporates an additional term to capture changes over calendar years, as previously suggested by \cite{Shi2012}. Furthermore, the dynamic evolution in the variance model is capable of accommodating various regimes of variability over time being a crucial aspect in the context of loss-reserving modelling.  As inference is performed under the Bayesian paradigm using MCMC techniques allowing fast computations for inference and prediction.  It takes advantage of the conditionally skew-normal distributions obtained when we condition the distribution of $Z_{ij}$ and $T_{ij}$ on the mixing latent variables. 

We performed a simulation study to investigate the ability of the proposed model to capture asymmetry and fat tails. Our simulated example indicates that the correct model is selected. The non-skew models have worse performance when the data does have the presence of irregular claims and potential outliers and the chain ladder technique fails to produce estimates of loss reserving leading to underestimation. Moreover, it seems that the predictive scoring rules used to compare the models are adequate measures of good predictive performance. Also, we illustrate the performance of our proposed model through a real dataset presented by \cite{Choy2008} and \cite{Jennifer16} as a case study. Our study showed that the skewed class has a better performance in predicting the loss reserving when compared with non-skewed models. 

We conclude that considering the class of the scale mixture of skew-normal distribution produces robust inferences and predictions for the total reserve. Therefore, an extension of this work involves contemplating a multivariate loss reserving model, as in \cite{Goudarzi2018}, but now featuring a class of skew-heavy-tailed distributions.

\appendix

\renewcommand{\thesection}{\Alph{section}}
\renewcommand{\thetable}{\Alph{section}.\arabic{table}}
\renewcommand{\thefigure}{\Alph{section}.\arabic{figure}}
\setcounter{equation}{0}
\renewcommand{\theequation}{\Alph{section}.\arabic{equation}}

\setcounter{section}{0}
\setcounter{figure}{0}
\setcounter{table}{0}

\newpage
\section{Proofs of properties of the proposed model}\label{apA}

In this appendix, we prove the results shown in Section \ref{sec2} and provide de kurtosis and skewness of the proposed model. For this, consider the model given by equation (\ref{sec2:eq1}). We are interested in centers on the first four moments to compute skewness and kurtosis. Remember that the variable $\varepsilon_{ij}$ is a zero mean Skew-Normal distribution, with $\sigma^2$ and $\kappa$ skewness parameter, with $\rho= \frac{\kappa}{\sqrt{1+ \kappa^2}}$. Then,

$$E(\varepsilon_{ij}) = \rho \sqrt{\frac{2}{\pi}}, \hspace{.2cm}  Var(\varepsilon_{ij}) = E\left[ \left( \varepsilon_{ij} - E(\varepsilon_{ij}) \right)^2 \right] = 1 - \frac{2}{\pi}\rho^2$$
$$E(\varepsilon_{ij}^2)= 1, \hspace{0.3cm} E(\varepsilon_{ij}^3)= \sqrt{\frac{2}{\pi}}\left( 3\rho - \rho^3\right), \hspace{0.3cm} E(\varepsilon_{ij}^4)=3.$$

The mean and variance of $Z_{ij}$ as mentioned in Section \ref{sec2} is given by

$$ 
 E(Z_{ij})=  E\left( \mu_{ij} + \sigma \frac{\varepsilon_{ij}}{\lambda_{ij}}\right)  =\mu_{ij} + \sigma \rho\sqrt{\frac{2}{\pi}} E\left(\lambda_{ij}^{-1/2}\right),
$$
\noindent and

\begin{eqnarray} \nonumber
Var(Z_{ij}) &=& E\left[ \left(Z_{ij} - E(Z_{ij}) \right)^2\right]  \\ \nonumber 
&=& \sigma^2 \left( E\left ( \lambda_{ij}^{-1} \varepsilon_{ij}^2\right) - E^2\left( \lambda_{ij}^{-1/2} \varepsilon_{ij} \right)\right)\\ \nonumber
&=& \sigma^2 \left( E\left( \lambda_{ij}^{-1}\right) E\left(\varepsilon_{ij}^2\right) - E^2\left( \lambda_{ij}^{-1/2} \varepsilon_{ij} \right) \right) \\ \nonumber
&=& \sigma^2 \left(E\left( \lambda_{ij}^{-1}\right) - E^2\left( \lambda_{ij}^{-1/2}\right)E^2\left(  \varepsilon_{ij} \right) \right) \\ \nonumber
&=& \sigma^2\left( E\left(\lambda_{ij}^{-1}\right) - \rho^2\left(\frac{2}{\pi}\right)E^2\left(\lambda_{ij}^{-1/2}\right)\right).  \nonumber  
\end{eqnarray}

The expression of the skewness, unconditional on the mixing parameter, is given by

\begin{eqnarray} \nonumber
    Skew\left( Z_{ij} \right) &=& \frac{E\left[ \left(Z_{ij} - E(Z_{ij}) \right)^3\right]}{\left[Var(Z_{ij})\right]^{3/2}}, \\ \nonumber
\end{eqnarray}

where the third central moment $\eta_{3}= E\left[ \left(Z_{ij} - E(Z_{ij}) \right)^3\right]$ is

\begin{align}
     \eta_{3}  = & E\left[ \left( \mu_{ij} + \sigma \varepsilon_{ij} \lambda_{ij}^{-1/2} - \mu_{ij} - \sigma \rho \sqrt{\frac{2}{\pi}} E\left(\lambda^{-1/2}\right)\right)^3 \right] \notag\\
       = & \thinspace\sigma^3  E\left[ \left( \varepsilon_{ij} \lambda_{ij}^{-1/2} - \rho \sqrt{\frac{2}{\pi}} E\left(\lambda^{-1/2}\right)\right)^3 \right] \notag \\ 
     = & \thinspace\sigma^3  E\left[ \left( \varepsilon_{ij} \lambda_{ij}^{-1/2} - \rho \sqrt{\frac{2}{\pi}} E\left(\lambda^{-1/2}\right)\right)^2 \left( \varepsilon_{ij} \lambda_{ij}^{-1/2} - \rho \sqrt{\frac{2}{\pi}} E\left(\lambda^{-1/2}\right)\right) \right] \notag \\
      = & \thinspace\sigma^3 \left[ E(\varepsilon_{ij}^3)E(\lambda_{ij}^{-3/2}) - 3\rho\sqrt{\frac{2}{\pi}}E(\varepsilon_{ij}^2)E(\lambda_{ij}^{-1})E\left[E(\lambda_{ij}^{-1/2})\right] + \frac{4}{\pi}\rho^2 E(\varepsilon_{ij})E(\lambda_{ij}^{-1/2})E\left[E(\lambda_{ij}^{-1/2})^2\right]   \right. \notag \\ 
      + & \left.   \frac{2}{\pi}\rho^2 E(\varepsilon_{ij})E(\lambda_{ij}^{-1/2})E\left[E(\lambda_{ij}^{-1/2})^2\right]  -\left(\frac{2}{\pi}\right)^{3/2}\rho^3 E\left[E(\lambda_{ij}^{-1/2})^3\right] \right]  \notag \\ 
      = & \thinspace \sigma^3 \left[ \sqrt{\frac{2}{\pi}}\left(3\rho- \rho^3\right)E(\lambda_{ij}^{-3/2}) - 3\rho\sqrt{\frac{2}{\pi}}E(\lambda_{ij}^{-1})E\left[E(\lambda_{ij}^{-1/2})\right] + \frac{4}{\pi}\rho^3 \sqrt{\frac{2}{\pi}} E(\lambda_{ij}^{-1/2})E\left[E(\lambda_{ij}^{-1/2})^2\right]   \right. \notag \\ 
      + & \left.   \left(\frac{2}{\pi}\right){ 3/2}\rho^3E(\lambda_{ij}^{-1/2})E\left[E(\lambda_{ij}^{-1/2})^2\right]  -\left(\frac{2}{\pi}\right)^{3/2}\rho^3 E\left[E(\lambda_{ij}^{-1/2})^3\right] \right].  \notag \\ 
\end{align}

Notice that $\lambda_{ij} \rightarrow 1$, then $\eta_3= \sigma^3 \left[ \sqrt{\frac{2}{\pi}}\rho^3\left(\frac{4-\pi}{\pi}\right)\right]$ 
represents a third central moment of a Skew-Normal. \\

The expression of the kurtosis, unconditional on the mixing parameter, is given by

$$Kurt(Z_{ij})= \frac{E\left[ \left(Z_{ij} - E(Z_{ij}) \right)^4\right]}{\left[Var(Z_{ij})\right]^{2}},  $$

where the fourth central moment $\eta_{4}= E\left[ \left(Z_{ij} - E(Z_{ij}) \right)^4\right]$ is

\begin{align}
     \eta_{4}  = & E\left[ \left( \mu_{ij} + \sigma \varepsilon_{ij} \lambda_{ij}^{-1/2} - \mu_{ij} - \sigma \rho \sqrt{\frac{2}{\pi}} E\left(\lambda^{-1/2}\right)\right)^4 \right] \notag\\
       = & \thinspace\sigma^4  E\left[ \left( \varepsilon_{ij} \lambda_{ij}^{-1/2} - \rho \sqrt{\frac{2}{\pi}} E\left(\lambda^{-1/2}\right)\right)^4 \right] \notag \\ 
     = & \thinspace\sigma^4  E\left[ \left( \varepsilon_{ij} \lambda_{ij}^{-1/2} - \rho \sqrt{\frac{2}{\pi}} E\left(\lambda^{-1/2}\right)\right)^2 \left( \varepsilon_{ij} \lambda_{ij}^{-1/2} - \rho \sqrt{\frac{2}{\pi}} E\left(\lambda^{-1/2}\right)\right)^2 \right] \notag \\
      = & \thinspace\sigma^4 \left[ E(\varepsilon_{ij}^4)E(\lambda_{ij}^{-2}) - 4\rho\sqrt{\frac{2}{\pi}}E(\varepsilon_{ij}^3)E(\lambda_{ij}^{-3/2})E\left[E(\lambda_{ij}^{-1/2})\right] + \frac{12}{\pi}\rho^2 E(\varepsilon_{ij}^2)E(\lambda_{ij}^{-1})E\left[E(\lambda_{ij}^{-1/2})\right]   \right. \notag \\ 
      - & \left.   4\rho^3\left(\frac{2}{\pi}\right)^{3/2} E(\varepsilon_{ij})E(\lambda_{ij}^{-1/2})E\left[E(\lambda_{ij}^{-1/2})^3\right]  +\rho^4\frac{4}{\pi^2} E\left[E(\lambda_{ij}^{-1/2})^4\right] \right]  \notag \\ 
      = & \thinspace \sigma^4 \left[ 3 E(\lambda_{ij}^{-2}) - 4\rho\sqrt{\frac{2}{\pi}}\sqrt{\frac{2}{\pi}}(3\rho - \rho^3)E(\lambda_{ij}^{-3/2})E\left[E(\lambda_{ij}^{-1/2})\right] + \frac{12}{\pi}\rho^2 E(\lambda_{ij}^{-1})E\left[E(\lambda_{ij}^{-1/2})\right]   \right. \notag \\ 
      - & \left.   4\rho^3\left(\frac{2}{\pi}\right)^{ 3/2}\rho \sqrt{\frac{2}{\pi}}E(\lambda_{ij}^{-1/2})E\left[E(\lambda_{ij}^{-1/2})^3\right]  - \rho^4\left(\frac{4}{\pi^2}\right) E\left[E(\lambda_{ij}^{-1/2})^4\right] \right]  \notag \\ 
       = & \thinspace \sigma^4 \left[ 3 E(\lambda_{ij}^{-2}) + \frac{12}{\pi}\rho^2 E(\lambda_{ij}^{-1})E\left[E(\lambda_{ij}^{-1/2})\right] - \frac{16}{\pi^2}\rho^4
       E(\lambda_{ij}^{-1/2})E\left[E(\lambda_{ij}^{-1/2})^3\right] \right. \notag \\ 
       -& \left. \left( \frac{24}{\pi}\rho^2 - \frac{8}{\pi}\rho^4\right)E(\lambda_{ij}^{-3/2})E\left[E(\lambda_{ij}^{-1/2})\right]   + \frac{4}{\pi^2}\rho^4 E\left[E(\lambda_{ij}^{-1/2})^4\right] \right]. \notag \\ 
\end{align}

Notice that $\lambda_{ij} \rightarrow 1$, then $\eta_4= 3 - \frac{12}{\pi}\rho^2 + \left(\frac{8\pi - 12}{\pi^2}\right)\rho^4 $ represents a fourth central moment a Skew-Normal. The skewness and kurtosis for the Skew-Gaussian model are

$$Skew(Z_{ij}) = \frac{\eta_3}{\eta_2 ^{3/2}} = \frac{\sqrt{2}(4 - \pi)\rho^3}{(\pi - 2\rho^2)^{3/2}} $$
and
$$Kurt(Z_{ij})= \frac{\eta_4}{\eta_2^2} = 3 + \frac{8(\pi - 3)\rho^4}{(\pi - 2\rho^2)^2}.$$ 

For the class of heavy tails the skewness and kurtosis can be computed considering the expected value for $\lambda_{ij}$ shown in Table \ref{apA:tab1}. 

\begin{table}[H]
\begin{center}
\caption{ Expected value for $\lambda_{ij}$ considering skew heavy tail models.}\label{apA:tab1}
\begin{tabular}{|l|c c c|}
\hline
     &  $\mathcal{SST}$ & $\mathcal{SS}$ & $\mathcal{SVG}$\\
     \hline
 $E(\lambda_{ij}^{-1/2})$    & $\left[{\Gamma(\frac{\nu-1}{2})}/{\Gamma(\frac{\nu}{2})}\right]\left(\frac{\nu}{2}\right)^{1/2}$ &  $ \frac{\nu}{\nu-1/2} $   & $\left[{\Gamma(\frac{\nu+1}{2})}/{\Gamma(\frac{\nu}{2})}\right]\left(\frac{\nu}{2}\right)^{-1/2}$  \\
  $E(\lambda_{ij}^{-3/2})$ & $\left[{\Gamma(\frac{\nu-3}{2})}/{\Gamma(\frac{\nu}{2})}\right]\left(\frac{\nu}{2}\right)^{3/2}$ &  $\frac{\nu}{\nu-3/2}$ & $\left[{\Gamma(\frac{\nu+3}{2})}/{\Gamma(\frac{\nu}{2})}\right]\left(\frac{\nu}{2}\right)^{-3/2}$\\
  $E(\lambda_{ij}^{-1})$ & $ \frac{\nu}{2}/(\frac{\nu}{2}-1)$ &$ \frac{\nu}{\nu-1}$ & $1$ \\
  $E(\lambda_{ij}^{-2})$  &  $\left[{\Gamma(\frac{\nu-4}{2})}/{\Gamma(\frac{\nu}{2})}\right]\left(\frac{\nu}{2}\right)^{2}$&  $\frac{\nu}{\nu-2}$& $\left[{\Gamma(\frac{\nu+4}{2})}/{\Gamma(\frac{\nu}{2})}\right]\left(\frac{\nu}{2}\right)^{-2}$  \\
  \hline
\end{tabular}
\end{center}
\end{table}

\section{Posterior computations}\label{apB}

The prior distributions considered for the parameters, the complete conditional
distributions, and proposal densities considered in the MCMC algorithm are detailed as
follows. %The joint posterior distribution is written as equation (\ref{eq11}).

\subsection{Full conditional of \texorpdfstring{$\mu$}{mu}}

Under the specified Normal prior distribution, the full conditional of $\mu$ is given by
\begin{eqnarray}\nonumber
  p(\mu \mid \cdot) &\propto& \exp\left\{-\frac{\mu^2}{2s_{\mu}^2}-\sum_{i=1}^{n}\sum_{j=1}^{n-i+1}\frac{\lambda_{ij}}{2\sigma^2}\left[\frac{\left(z_{ij}-\mu-\alpha_i-\beta_{ij}-\gamma_{t}-\rho T_{ij}\right)^2}{1-\rho^2}\right]\right\}
\end{eqnarray}
that is,
\begin{eqnarray}\nonumber
\mu \mid \cdot \sim \mathcal{N}\left(m,s^2\right)
\end{eqnarray}
where $m=s^2\left[\sum_{i=1}^{n}\sum_{j=1}^{n-i+1}\frac{\lambda_{ij}\left(z_{ij}-\alpha_i-\beta_{ij}-\gamma_{t}-\rho T_{ij}\right)}{\sigma^2\left(1-\rho^2\right)}\right]$ and $s^2=\left[\frac{1}{\sigma^2\left(1-\rho^2\right)}\sum_{i=1}^{n}\sum_{j=1}^{n-i+1}\lambda_{ij}+\frac{1}{s_{\mu}^2}\right]^{-1}$.

\subsection{Full conditional of \texorpdfstring{$\sigma^2$}{sig2}}

Under the specified Inverse Gamma prior distribution, the full conditional of $\sigma^2$ is given by
\begin{eqnarray}\nonumber
  p(\sigma^2 \mid \cdot) &\propto& \left(\frac{1}{\sigma^{2}}\right)^{a_{\sigma}+1+\frac{n(n+1)}{2}}\exp\left\{-\frac{b_{\sigma}}{\sigma^2}-\sum_{i=1}^{n}\sum_{j=1}^{n-i+1}\frac{\lambda_{ij}}{2\sigma^2}\left[\frac{\left(z_{ij}-\mu-\alpha_i-\beta_{ij}-\gamma_{t}-\rho T_{ij}\right)^2}{1-\rho^2}+z_{ij}^2\right]\right\}
\end{eqnarray}
that is,
\begin{eqnarray}\nonumber
  \sigma^2 \mid \cdot \sim \mathcal{IG}\left(a,b\right)
\end{eqnarray}
where $a=a_{\sigma}+\frac{n(n+1)}{2}$ and $b=b_{\sigma}+\frac{1}{2}\sum_{i=1}^{n}\sum_{j=1}^{n-i+1}\lambda_{ij}\left[\frac{\left(z_{ij}-\mu-\alpha_i-\beta_{ij}-\gamma_{i+j-1}-\rho Z_{ij}\right)^2}{\left(1-\rho^2\right)}+T_{ij}^2\right]$.

\subsection{Full conditional of \texorpdfstring{$\rho$}{ro}}

Under the specified beta prior distribution, the full conditional of $\rho$ is given by
\begin{eqnarray}\nonumber
  p(\rho \mid \cdot) &\propto& (1+\rho)^{c-1-\frac{n(n+1)}{4}}(1-\rho)^{d-1-\frac{n(n+1)}{4}}\\
                     &\times& \exp\left\{-\sum_{i=1}^{n}\sum_{j=1}^{n-i+1}\frac{\lambda_{ij}}{2\sigma^2}\left[\frac{\left(z_{ij}-\mu-\alpha_i-\beta_{ij}-\gamma_{t}-\rho T_{ij}\right)^2}{1-\rho^2}\right]\right\}\boldsymbol{I}_{(-1,1)}(\rho).\nonumber
\end{eqnarray}

Note that $p(\rho \mid \cdot)$ does not have a known distribution, so we obtained samples using the Adaptive Metropolis-Hastings algorithm. Using the transformation of $\rho$ defined as $\zeta=\frac{1}{2}\log\left(\frac{1+\rho}{1-\rho}\right)=\mbox{atanh}(\rho)$, so that $\zeta$ has support in $\mathbb{R}$. Then, the density function of $\zeta$ is given by
\begin{eqnarray}\nonumber
  p\left(\zeta \mid \cdot \right) &\propto& p(\rho(\zeta) \mid  \cdot) \times \mbox{sech}^2(\zeta).
\end{eqnarray}

%In the Adaptive Metropolis algorithm with global adaptive scaling \citep{Andrieu2008} assuming the actual value of $\zeta$ is $\zeta_i$ and we will update its value so that next iteration we had $\zeta_{i+1}$, we follow the steps below:
For this, we considered the global adaptive scaling seen in (\cite{Andrieu2008}), assuming that the actual value of $\zeta$ is $\zeta_i$ and then we will update $\zeta$ so the in following iteration we have $\zeta_{i+1}$. We follow the steps below:
\begin{enumerate}
    \item Initialise $\zeta_0$, $\mu_0$, $\kappa_0$ and $\sigma_0^2$.
    \item At iteration $i+1$, given $\zeta_{i}$, $\mu_i$, $\sigma_{i}^2$ and $\kappa_i$, sample $\zeta_{i+1}\sim\mathcal{N}\left(\zeta_i,\kappa_i\sigma_i^2\right)$ and accept it with probability $\alpha\left(\zeta_{i},\zeta_{i+1}\right)$, otherwise $\zeta_{i+1}=\zeta_{i}$ and update:
    \begin{eqnarray}\nonumber
      \gamma_{i+1}&=&(i+1)^{-a}\nonumber\\
      \log(\kappa_{i+1})&=&\log(\kappa_i)+\gamma_{i+1}\left[\alpha\left(\zeta_{i},\zeta_{i+1}\right)-\alpha_{*}\right]\nonumber\\
      \mu_{i+1}&=&\mu_{i}+\gamma_{i+1}(\zeta_{i+1}-\mu_i)\nonumber\\
      \sigma_{i+1}^{2}&=&\sigma_{i}^{2}+\gamma_{i+1}\left[(\zeta_{i+1}-\mu)^2-\sigma_{i}^{2}\right]\nonumber
    \end{eqnarray}
\end{enumerate}

Once $\zeta_{i+1}$ has been obtained, compute $\rho=\mbox{tanh}(\zeta_{i+1})$. We set $\mu_0=0$, $\sigma_0^2=1$, $\kappa_0=2.38$, $a=0.8$ and $\alpha_{*}=0.234$.

\subsection{Full conditional of \texorpdfstring{$\sigma_{\alpha}^2$}{siga}}

Under the specified Inverse Gamma prior distribution, the full conditional of $\sigma_{\alpha}^2$ is given by
\begin{eqnarray}\nonumber
  p(\sigma_{\alpha}^2 \mid \cdot)\propto\left(\frac{1}{\sigma_{\alpha}^2}\right)^{a_{\alpha}+\frac{n-1}{2}}\exp\left\{-\frac{b_{\alpha}}{\sigma_{\alpha}^2}-\frac{1}{2\sigma_{\alpha}^2}\sum_{i=2}^{n}\left(\alpha_{i}-\alpha_{i-1}\right)^2\right\},
\end{eqnarray}
that is,
\begin{eqnarray}\nonumber
  \sigma_{\alpha}^2 \mid \cdot \sim \mathcal{IG}\left(a,b\right),
\end{eqnarray}
where $a=a_{\alpha}+\frac{n-1}{2}$ and $b=b_{\alpha}+\frac{1}{2}\sum_{i=2}^{n}\left(\alpha_i-\alpha_{i-1}\right)^2$.

\subsection{Full conditional of \texorpdfstring{$\sigma_{\beta}^2$}{sigb}}

Under the specified Inverse Gamma prior distribution, the full conditional of $\sigma_{\beta}^2$ is given by
\begin{eqnarray}\nonumber
  p(\sigma_{\beta}^2 \mid \dots)\propto\left(\frac{1}{\sigma_{\beta}^2}\right)^{a_{\beta}+1+\frac{n(n-1)}{4}}\exp\left\{-\frac{b_{\beta}}{\sigma_{\beta}^2}-\frac{1}{2\sigma_{\beta}^2}\sum_{i=2}^{n}\sum_{j=1}^{n-i+1}\left(\beta_{ij}-\beta_{i-1,j}\right)^2\right\},
\end{eqnarray}
that is,
\begin{eqnarray}\nonumber
  \sigma_{\beta}^2 \mid \dots \sim \mathcal{IG}\left(a,b\right),
\end{eqnarray}
where $a=a_{\beta}+\frac{n(n-1)}{4}$ and $b=b_{\beta}+\frac{1}{2}\sum_{i=2}^{n}\sum_{j=1}^{n-i+1}\left(\beta_{ij}-\beta_{i-1,j}\right)^2$.

\subsection{Full conditional of \texorpdfstring{$\sigma_{\gamma}^2$}{sigg}}

Under the specified Inverse Gamma prior distribution, the full conditional of $\sigma_{\gamma}^2$ is given by
\begin{eqnarray}\nonumber
  p(\sigma_{\gamma}^2 \mid \cdot)\propto\left(\frac{1}{\sigma_{\gamma}^2}\right)^{a_{\gamma}+1+\frac{n-1}{2}}\exp\left\{-\frac{b_{\gamma}}{\sigma_{\gamma}^2}-\frac{1}{2\sigma_{\gamma}^2}\sum_{t=2}^{n}\left(\gamma_{t}-\gamma_{t-1}\right)^2\right\},
\end{eqnarray}
that is,
\begin{eqnarray}\nonumber
  \sigma_{\gamma}^2 \mid \cdot \sim \mathcal{IG}\left(a,b\right),
\end{eqnarray}
where $a=a_{\gamma}+\frac{n-1}{2}$ and $b=b_{\gamma}+\frac{1}{2}\sum_{t=2}^{n}\left(\gamma_t-\gamma_{t-1}\right)^2$.

\subsection{Full conditional of \texorpdfstring{$\lambda_{ij}$}{lamb} and \texorpdfstring{$\nu$}{nu}}

\begin{itemize}
  \item {\bf Student t case:}
\end{itemize}

As $\lambda_{ij}\sim\mathcal{G}(\nu/2,\nu/2)$, the full conditional of $\lambda_{ij}$ is given by
\begin{eqnarray}\nonumber
  p(\lambda_{ij} \mid \cdot) &\propto& \lambda_{ij}^{\frac{\nu}{2}}\exp\left\{-\left(\frac{\nu}{2}+\frac{1}{2\sigma^2}\left[\frac{\left(z_{ij}-\mu-\alpha_i-\beta_{ij}-\gamma_{t}-\rho T_{ij}\right)^2}{1-\rho^2}+T_{ij}^2\right]\right)\lambda_{ij}\right\},
\end{eqnarray}
that is,
\begin{eqnarray}\nonumber
  \lambda_{ij} \mid \cdot \sim \mathcal{G}(a,b),
\end{eqnarray}
where $a=\frac{\nu}{2}+1$ and $b=\frac{\nu}{2}+\frac{1}{2\sigma^2}\left[\frac{\left(z_{ij}-\mu-\alpha_i-\beta_{ij}-\gamma_{t}-\rho T_{ij}\right)^2}{1-\rho^2}+T_{ij}^2\right]$.

Under the specified Gamma prior distribution, the full conditional of $\nu$ is given by
\begin{eqnarray}\nonumber
  p(\nu \mid \cdot) &\propto& \nu^{a_t-1}\left[\frac{\left(\nu/2\right)^{\frac{\nu}{2}}}{\Gamma(\nu/2)}\right]^{n(n+1)/2}\left(\prod_{i=1}^{n}\prod_{j=1}^{n-i+1}\lambda_{ij}\right)^{\frac{\nu}{2}}\exp\left\{-\left[b_t+\frac{1}{2}\sum_{i=1}^{n}\sum_{j=1}^{n-i+1}\lambda_{ij}\right]\nu\right\},
\end{eqnarray}

Note that $p(\nu \mid \cdot)$ does not have a known distribution, so we obtained samples using the Adaptive Metropolis-Hastings algorithm. Using the transformation of $\nu$ defined as $\xi=\log\left(\nu\right)$, so that $\xi$ has support in $\mathbb{R}$. Then the density function of $\zeta$ is given by
\begin{eqnarray}\nonumber
  p\left(\xi \mid \cdot \right) &\propto& p(\nu(\xi) \mid \cdot)\times\exp(\xi)
\end{eqnarray}

Assume that the actual value of $\xi$ is $\xi_i$ and then we will update $\xi$ so the in following iteration we have $\xi_{i+1}$ as:
%In the Adaptive Metropolis algorithm with global adaptive scaling assuming the actual value of $\xi$ is $\xi_i$ and we will update its value so that next iteration we had $\xi_{i+1}$, we follow the steps below:
\begin{enumerate}
    \item Initialise $\xi_0$, $\mu_0$, $\kappa_0$ and $\sigma_0^2$.
    \item At iteration $i+1$, given $\xi_{i}$, $\mu_i$, $\sigma_{i}^2$ and $\kappa_i$, sample $\xi_{i+1}\sim\mathcal{N}\left(\xi_i,\kappa_i\sigma_i^2\right)$ and accept it with probability $\alpha\left(\xi_{i},\xi_{i+1}\right)$, otherwise $\xi_{i+1}=\xi_{i}$ and update:
    \begin{eqnarray}\nonumber
      \gamma_{i+1}&=&(i+1)^{-a}\nonumber\\
      \log(\kappa_{i+1})&=&\log(\kappa_i)+\gamma_{i+1}\left[\alpha\left(\xi_{i},\xi_{i+1}\right)-\alpha_{*}\right]\nonumber\\
      \mu_{i+1}&=&\mu_{i}+\gamma_{i+1}(\xi_{i+1}-\mu_i)\nonumber\\
      \sigma_{i+1}^{2}&=&\sigma_{i}^{2}+\gamma_{i+1}\left[(\xi_{i+1}-\mu)^2-\sigma_{i}^{2}\right]\nonumber
    \end{eqnarray}
\end{enumerate}

Once $\xi_{i+1}$ has been obtained, compute $\nu=\exp(\xi_{i+1})$. We set $\mu_0=0$, $\sigma_0^2=1$, $\kappa_0=2.38$, $a=0.8$ and $\alpha_{*}=0.234$.

\begin{itemize}
  \item {\bf Slash case:}
\end{itemize}

As $\lambda_{ij}\sim\mathcal{B}e(\nu,1)$, the full conditional of $\lambda_{ij}$ is given by
\begin{eqnarray}\nonumber
  p(\lambda_{ij} \mid \cdot) &\propto& \lambda_{ij}^{\nu}\exp\left\{-\left(\frac{\nu}{2}+\frac{1}{2\sigma^2}\left[\frac{\left(z_{ij}-\mu-\alpha_i-\beta_{ij}-\gamma_{t}-\rho T_{ij}\right)^2}{1-\rho^2}+T_{ij}^2\right]\right)\lambda_{ij}\right\}\boldsymbol{I}_{(0,1)}(\lambda_{ij})
\end{eqnarray}
that is,
\begin{eqnarray}\nonumber
  \lambda_{ij} \mid \cdot \sim \mathcal{G}_{(0,1)}(a,b)
\end{eqnarray}
where $\mathcal{G}_{(0,1)}(\cdot,\cdot)$ is the right truncated gamma distribution, $a=\nu+1$ and $b=\frac{\nu}{2}+\frac{1}{2\sigma^2}\left[\frac{\left(z_{ij}-\mu-\alpha_i-\beta_{ij}-\gamma_{t}-\rho T_{ij}\right)^2}{1-\rho^2}+T_{ij}^2\right]$.

Under the specified gamma prior distribution, the full conditional of $\nu$ is given by
\begin{eqnarray}\nonumber
  p(\nu \mid \dots) &\propto& \nu^{a_{s}+\frac{n(n+1)}{2}-1}\exp\left\{-\left(b_{s}-\sum_{i=1}^{n}\sum_{j=1}^{n-i+1}\log(\lambda_{ij})\right)\nu\right\}\boldsymbol{I}_{(1,\infty)}(\nu)
\end{eqnarray}
that is,
\begin{eqnarray}\nonumber
  \nu \mid \dots \sim \mathcal{G}_{(1,\infty)}(a,b)
\end{eqnarray}
where $\mathcal{G}_{(1,\infty)}(\cdot,\cdot)$ is the right truncated Gamma distribution, $a=a_{s}+\frac{n(n+1)}{2}$ and $b=b_{s}-\sum_{i=1}^{n}\sum_{j=1}^{n-i+1}\log(\lambda_{ij})$.

\begin{itemize}
  \item {\bf Variance Gamma case:}
\end{itemize}

As $\lambda_{ij}\sim\mathcal{IG}(\nu/2,\nu/2)$, the full conditional of $\lambda_{ij}$ is given by
\begin{eqnarray}\nonumber
  p(\lambda_{ij} \mid \cdot) &\propto& \lambda_{ij}^{-\frac{\nu}{2}}\exp\left\{-\frac{1}{2}\left[\lambda_{ij}\left(\frac{T_{ij}^2}{\sigma^2}+\frac{\left(y_{ij}-\mu-\alpha_i-\beta_{ij}-\gamma_{t}-\rho T_{ij}\right)^2}{\sigma^2\left(1-\rho^2\right)}\right)+\frac{\nu}{\lambda_{ij}}\right]\right\}
\end{eqnarray}
that is,
\begin{eqnarray}\nonumber
  \lambda_{ij} \mid \cdot \sim \mathcal{GIG}\left(\omega,\chi,\psi\right)
\end{eqnarray}
where $\mathcal{GIG}(\cdot,\cdot,\cdot)$ is the generalized Inverse Gaussian distribution, $\omega=1-\frac{\nu}{2}$, $\chi=\nu$ and $\psi=\frac{T_{ij}^2}{\sigma^2}+\frac{\left(y_{ij}-\mu-\alpha_i-\beta_{ij}-\gamma_{i+j-1}-\rho T_{ij}\right)^2}{\sigma^2\left(1-\rho^2\right)}$. \\

Under the specified gamma prior distribution, the full conditional of $\nu$ is given by
\begin{eqnarray}\nonumber
  p(\nu \mid \cdot) &\propto& \nu^{a_{vg}-1}\left[\frac{\left(\nu/2\right)^{\frac{\nu}{2}}}{\Gamma(\nu/2)}\right]^{n(n+1)/2}\left(\prod_{i=1}^{n}\prod_{j=1}^{n-i+1}\lambda_{ij}\right)^{-\frac{\nu}{2}}\exp\left\{-\left[b_{vg}+\frac{1}{2}\sum_{i=1}^{n}\sum_{j=1}^{n-i+1}\frac{1}{\lambda_{ij}}\right]\nu\right\}
\end{eqnarray}

Note that $p(\nu \mid \dots)$ does not have a known distribution, the Adaptive Metropolis-Hastings is considered. %so we obtained samples using the Adaptative Metropolis Hastigns algorithm. 
Under a transformation of $\nu$ defined as $\xi=\log\left(\nu\right)$, so that $\xi$ has support in $\mathbb{R}$. Then the density function of $\xi$ is given by
\begin{eqnarray}\nonumber
  p\left(\xi| \cdot \right) &\propto& p(\nu(\xi) \mid \dots)\cdot\exp(\xi)
\end{eqnarray}
%In the Adaptive Metropolis algorithm with global adaptive scaling assuming the actual value of $\xi$ is $\xi_i$ and we will update its value so that next iteration we had $\xi_{i+1}$, 
We follow the steps below:
\begin{enumerate}
    \item Initialise $\xi_0$, $\mu_0$, $\kappa_0$ and $\sigma_0^2$.
    \item At iteration $i+1$, given $\xi_{i}$, $\mu_i$, $\sigma_{i}^2$ and $\kappa_i$, sample $\xi_{i+1}\sim\mathcal{N}\left(\xi_i,\kappa_i\sigma_i^2\right)$ and accept it with probability $\alpha\left(\xi_{i},\xi_{i+1}\right)$, otherwise $\xi_{i+1}=\xi_{i}$ and update:
    \begin{eqnarray}\nonumber
      \gamma_{i+1}&=&(i+1)^{-a}\nonumber\\
      \log(\kappa_{i+1})&=&\log(\kappa_i)+\gamma_{i+1}\left[\alpha\left(\xi_{i},\xi_{i+1}\right)-\alpha_{*}\right]\nonumber\\
      \mu_{i+1}&=&\mu_{i}+\gamma_{i+1}(\xi_{i+1}-\mu_i)\nonumber\\
      \sigma_{i+1}^{2}&=&\sigma_{i}^{2}+\gamma_{i+1}\left[(\xi_{i+1}-\mu)^2-\sigma_{i}^{2}\right]\nonumber
    \end{eqnarray}
\end{enumerate}

Once $\xi_{i+1}$ has been obtained, compute $\nu=\exp(\xi_{i+1})$. We set $\mu_0=0$, $\sigma_0^2=1$, $\kappa_0=2.38$, $a=0.8$ and $\alpha_{*}=0.234$.

\subsection{Full conditional of \texorpdfstring{$T_{ij}$}{tij}}

The full conditional of $T_{ij}$ is given by
\begin{eqnarray}\nonumber
  p\left(T_{ij} \mid \cdot \right) &\propto& \exp\left\{-\lambda_{ij}\frac{\left(T_{ij}-\rho(z_{ij}-\mu-\alpha_i-\beta_{ij}-\gamma_t)\right)^2}{\sigma^2(1-\rho^2)}\right\}\boldsymbol{I}_{0,\infty}(T_{ij})
\end{eqnarray}
that is,
\begin{eqnarray}\nonumber
  T_{ij} \mid \cdot \sim \mathcal{TN}_{[0,\infty)}\left(m,s^2\right)
\end{eqnarray}
where $m=\rho(z_{ij}-\mu-\alpha_i-\beta_{ij}-\gamma_t)$ and $s^2=\frac{\sigma^2(1-\rho^2)}{\lambda_{ij}}$.

\subsection{Full conditional of \texorpdfstring{$\alpha_{i}$}{ai}}

Under the initial condition that $\alpha_1=0$, the full conditional of $\alpha_i$ is given by
\begin{eqnarray}\nonumber
  p(\alpha_i \mid \dots) &\propto& \exp\left\{-\frac{1}{2}\sum_{j=1}^{n-i+1}\frac{\lambda_{ij}\left(z_{ij}-\mu-\alpha_i-\beta_{ij}-\gamma_{t}-\rho T_{ij}\right)^2}{\sigma^2\left(1-\rho^2\right)}\right\}\\
  &\times& \exp\left\{-\frac{1}{2\sigma_{\alpha}^2}\left(\alpha_i-\alpha_{i-1}\right)^2\right\}\exp\left\{-\frac{1}{2\sigma_{\alpha}^2}\left(\alpha_{i+1}-\alpha_{i}\right)^2\boldsymbol{I}_{(i+1<n)}\right\}\nonumber
\end{eqnarray}
that is,
\begin{eqnarray}\nonumber
  \alpha_i \mid \dots \sim \mathcal{N}(m,s^2)
\end{eqnarray}
where $m=s^2\left[\sum_{j=1}^{n-i+1}\frac{\lambda_{ij}\left(z_{ij}-\mu-\beta_{ij}-\gamma_{i+j-1}-\rho T_{ij}\right)}{\sigma^2\left(1-\rho^2\right)}+\frac{\left(\alpha_{i-1}+\alpha_{i+1}\boldsymbol{I}_{(i+1<n)}\right)}{\sigma_{\alpha}^2}\right]$ and \\
$s^2=\left[\sum_{j=1}^{n-i+1}\frac{\lambda_{ij}}{\sigma^2\left(1-\rho^2\right)}+\frac{1+\boldsymbol{I}_{(i+1<n)}}{\sigma_{\alpha}^2}\right]^{-1}$.

\subsection{Full conditional of \texorpdfstring{$\beta_{ij}$}{b}}

Under the initial conditions that $\beta_{i1}=0$ and $\beta_{1j}=0$ for all i and j, the full conditional of $\beta_{ij}$ is given by
\begin{eqnarray}\nonumber
  p(\beta_{ij} \mid \cdot) &\propto& \exp\left\{-\frac{1}{2}\frac{\lambda_{ij}\left(z_{ij}-\mu-\alpha_i-\beta_{ij}-\gamma_{t}-\rho T_{ij}\right)^2}{\sigma^2\left(1-\rho^2\right)}\right\}\\
  &\times& \exp\left\{-\frac{1}{2\sigma_{\beta}^2}\left(\beta_{ij}-\beta_{i-1,j}\right)^2\right\}\exp\left\{-\frac{1}{2\sigma_{\beta}^2}\left(\beta_{i+1,j}-\beta_{ij}\right)^2\boldsymbol{I}_{(i+1<n)}\right\}\nonumber
\end{eqnarray}
that is,
\begin{eqnarray}\nonumber
  \beta_{ij} \mid \cdot \sim \mathcal{N}(m,s^2)
\end{eqnarray}
where $m=s^2\left[\frac{\lambda_{ij}\left(z_{ij}-\mu-\alpha_i-\gamma_{t}-\rho T_{ij}\right)}{\sigma^2\left(1-\rho^2\right)}+\frac{\left(\beta_{i-1,j}+\beta_{i+1,j}\boldsymbol{I}_{(i+1<n)}\right)}{s_{\beta}^2}\right]$ and $s^2=\left[\frac{\lambda_{ij}}{\sigma^2\left(1-\rho^2\right)}+\frac{1+\boldsymbol{I}_{(i+1<n)}}{\sigma_{\beta}^2}\right]^{-1}$.

\subsection{Full conditional of \texorpdfstring{$\gamma_{t}$}{g}}

Under the initial condition that $\gamma_1=0$, the full conditional of $\gamma_t$ is given by
\begin{eqnarray}\nonumber
  p(\gamma_t \mid \cdot) &\propto& \exp\left\{-\frac{1}{2}\sum_{j=1}^{t}\frac{\lambda_{t-j+1,j}\left(z_{t-j+1,j}-\mu-\alpha_{t-j+1}-\beta_{t-j+1,j}-\gamma_{t}-\rho T_{t-j+1,k}\right)^2}{\sigma^2\left(1-\rho^2\right)}\right\}\\
  &\times& \exp\left\{-\frac{1}{2\sigma_{\gamma}^2}\left(\gamma_{t}-\gamma_{t-1}\right)^2\right\}\exp\left\{-\frac{1}{2\sigma_{\gamma}^2}\left(\gamma_{t+1}-\gamma_{t}\right)^2\boldsymbol{I}_{(t+1<n)}\right\}\nonumber
\end{eqnarray}
that is,
\begin{eqnarray}\nonumber
  \gamma_t \mid \cdot \sim \mathcal{N}(m,s^2)
\end{eqnarray}
where $m=s^2\left[\sum_{j=1}^{t}\frac{\lambda_{t-j+1,j}\left(z_{t-j+1,j}-\mu-\alpha_{t-j+1}-\beta_{t-j+1,j}-\rho T_{t-j+1,j}\right)}{\sigma^2\left(1-\rho^2\right)}+\frac{\left(\gamma_{t-1}+\gamma_{t+1}\boldsymbol{I}_{(t+1<n)}\right)}{\sigma_{\gamma}^2}\right]$ and $s^2=\left[\sum_{j=1}^{t}\frac{\lambda_{t-j+1,j}}{\sigma^2\left(1-\rho^2\right)}+\frac{1+\boldsymbol{I}_{(t+1<n)}}{\sigma_{\gamma}^2}\right]^{-1}$.

\section{Model Comparison Criteria}\label{apC}
We regard the root mean square prediction error and scoring rules in a Bayesian context as measures for comparing models based on their
posterior predictive distribution.  In particular, we consider the 
Interval Score, the Width of Credibility Interval and the Continuous Ranked Probability Score.
\subsection{Root Mean Square Prediction Error}
The root mean squared prediction error measures the expected squared distance between the predictive value and the true observed value (validation observation).
\begin{equation}\label{eq:mspe}
RMSPE(z, z^{pred}) = \sqrt{\frac{1}{n_0}\sum_{i=1}^{n_0}\frac{1}{n_0}\sum_{j=1}^{n_0} \left(z_{ij} - z_{ij}^{pred}\right)^2},
\end{equation}
\noindent where $z$ is the validation observation, $z^{pred}$ the predictive value and $n_0$ the length of $z$.
\subsection{Interval Score} 
Interval forecast is a crucial special case of quantile prediction \citep{GneitRaf07}. It compares the predictive credibility interval with the true observed value (validation observation), and it considers the uncertainty in the predictions such that the model is penalised if an interval is too narrow and misses the true value. The Interval Score (IS) is given by 
\begin{equation}\label{eq:IS}
{\rm IS}(u,l;z) = (u-l) + \frac{2}{\psi}(l - z)I_{\left[z< l\right]}  + \frac{2}{\psi}( z - u)I_{\left[z> u\right]},
\end{equation}
\noindent where $l$ and $u$ represent for the forecaster quoted $\frac{\psi}{2}$ and $1-\frac{\psi}{2}$ quantiles based on the predictive distribution and $z$ is the validation observation. If $\psi=0.05$ the resulting interval has 95\% credibility.

\subsection{Width of Credibility Interval}
The Width of Credibility Interval (WCI) is defined based on credible intervals and is given by
\begin{equation}\label{eq:WCI}
WCI(u,l) = (u-l),
\end{equation}
\noindent where $l$ and $u$ represent the lower and upper intervals, respectively shown in the Interval Score measure.
\subsection{Continuous Ranked Probability Score}
This measure generalizes the absolute error which it reduce if $F$ is a deterministic forecast, that is, a point measure. Thus, the CRPS provides a direct way to compare deterministic and probabilistic forecasts \cite{GneitRaf07}. The Continuous Ranked Probability Score is given by
\begin{equation}\label{eq:CRPS}
CRPS(F, z) = E_{F}|z^{pred} - {\tilde{z}}^{pred}|  - \frac{1}{2}E_{F}| z^{pred} - z|, 
\end{equation}
\noindent where $z^{pred}$ and $\tilde{z}^{pred}$ are independent replicas of the predictive distribution, $z$ the validation observation and $F$ the predictive distribution of the model.

%%%%%%5 Apendice D
%%%%%%%%%%%%%%%%%%%%%%%%%%%%%%%%%%%%%%%%%%%%%%%%%%%%%%%%%%%%%%%%%%%%%%%%%%%%%%%%%%%
\begin{landscape}
\section{Run-off triangle data}\label{apD}
\begin{table}[H]
    \centering
        \caption{Dataset1: The amount of claims paid to the insureds of an insurance product during the period from 1978 to 1995.}
    \label{tabapC:1}
    \adjustbox{max width=\linewidth}{
    \begin{tabular}{c|ccccccccccccc:ccccc}
    \hline
  accident   &   \multicolumn{18}{c}{development year $j$} \\
  year $i$ &  1&2&3&4&5&6&7&8&9&10&11&12&13&14&15&16&17&18 \\
 \hline
1978&3323&8332&9572&10172&7631&3855&3252&4433&2188&333&199&692&311  &{\bf 0}&{\bf 405}&{\bf 293}&{\bf 76}&{\bf 14} \\ 
1979&3785&10342&8330&7849&2839&3577&1404&1721&1065&156&35&259&{\cellcolor{lightgray}{\bf 250}}&{\bf 420}&{\bf 6}&{\bf 1}&{\bf 0}&  \\
1980&4677&9989&8745&10228&8572&5787&3855&1445&1612&626&1172&{\cellcolor{lightgray}{\bf 589}}&{\cellcolor{lightgray}{\bf 438}}&{\bf 473}&{\bf 370}&{\bf 31}& &  \\
1981&5288&8089&12839&11828&7560&6383&4118&3016&1575&1985&{\cellcolor{lightgray}{\bf 2645}}&{\cellcolor{lightgray}{\bf 266}}&{\cellcolor{lightgray}{\bf 38}}&{\bf 45}&{\bf 115}& & &  \\
1982&2294&9869&10242&13808&8775&5419&2424&1597&4149&{\cellcolor{lightgray}{\bf 1296}}&{\cellcolor{lightgray}{\bf 917}}&{\cellcolor{lightgray}{\bf 295}}&{\cellcolor{lightgray}{\bf 428}}&{\bf 359}& & & &  \\
1983&3600&7514&8247&9327&8584&4245&4096&3216&{\cellcolor{lightgray}{\bf 2014}}&{\cellcolor{lightgray}{\bf 593}}&{\cellcolor{lightgray}{\bf 1188}}&{\cellcolor{lightgray}{\bf 691}}&{\cellcolor{lightgray}{\bf 368}}& & & & &  \\
1984&3642&7394&9838&9733&6377&4884&11920&{\cellcolor{lightgray}{\bf 4188}}&{\cellcolor{lightgray}{\bf 4492}}&{\cellcolor{lightgray}{\bf 1760}}&{\cellcolor{lightgray}{\bf 944}}&{\cellcolor{lightgray}{\bf 921}}&{\cellcolor{gray}} & & & & & \\
1985&2463&5033&6980&7722&6702&7834&{\cellcolor{lightgray}{\bf 5579}}&{\cellcolor{lightgray}{\bf 3622}}&{\cellcolor{lightgray}{\bf 1300}}&{\cellcolor{lightgray}{\bf 3069}}&{\cellcolor{lightgray}{\bf 1370}}&  {\cellcolor{gray}} & {\cellcolor{gray}}& & \\
1986&2267&5959&6175&7051&8102&{\cellcolor{lightgray}{\bf 6339}}&{\cellcolor{lightgray}{\bf 6978}}&{\cellcolor{lightgray}{\bf 4396}}&{\cellcolor{lightgray}{\bf 3107}}&{\cellcolor{lightgray}{\bf 903}}&{\cellcolor{gray}} & {\cellcolor{gray}}& {\cellcolor{gray}}& & & \\
1987&2009&3700&5298&6885&{\cellcolor{lightgray}{\bf6477}}&{\cellcolor{lightgray}{\bf7570}}&{\cellcolor{lightgray}{\bf5855}}&{\cellcolor{lightgray}{\bf5751}}&{\cellcolor{lightgray}{\bf 3871}}&  {\cellcolor{gray}} & {\cellcolor{gray}}& {\cellcolor{gray}}& {\cellcolor{gray}}& & & \\
1988&1860&5282&3640&{\cellcolor{lightgray}{\bf 7538}}&{\cellcolor{lightgray}{\bf 5157}}&{\cellcolor{lightgray}{\bf 5766}}&{\cellcolor{lightgray}{\bf6862}}&{\cellcolor{lightgray}{\bf 2572}} & {\cellcolor{gray}} &  {\cellcolor{gray}}& {\cellcolor{gray}}& {\cellcolor{gray}}& {\cellcolor{gray}}&& & & \\
1989&2331&3517&{\cellcolor{lightgray}{\bf 5310}}&{\cellcolor{lightgray}{\bf 6066}}&{\cellcolor{lightgray}{\bf 10149}}&{\cellcolor{lightgray}{\bf9265}}&{\cellcolor{lightgray}{\bf 5262}}&  {\cellcolor{gray}} &  {\cellcolor{gray}}& {\cellcolor{gray}}& {\cellcolor{gray}}& {\cellcolor{gray}}& {\cellcolor{gray}}&& & & & \\
1990&2314&{\cellcolor{lightgray}{\bf 4487}}&{\cellcolor{lightgray}{\bf 4112}}&{\cellcolor{lightgray}{\bf 7000}}&{\cellcolor{lightgray}{\bf 11163}}&{\cellcolor{lightgray}{\bf 10057}}& {\cellcolor{gray}} &  {\cellcolor{gray}}& {\cellcolor{gray}}& {\cellcolor{gray}}& {\cellcolor{gray}}& {\cellcolor{gray}}& {\cellcolor{gray}}&& & & & \\ \cdashline{1-19}
1991&{\bf 2607}&{\bf 3952}&{\bf 8228}&{\bf7895}&{\bf 9317}&  & & & & & & &    & & & & & \\
1992&{\bf 2595}&{\bf 5403}&{\bf6579}&{\bf15546}& & & & & & & & & & & & & & \\
1993&{\bf 3155}&{\bf 4974}&{\bf7961}& & & & & & & & & & & & & & & \\
1994& {\bf 2626}&{\bf 5704}& & & & & & & & & & & & & & & & \\
1995&{\bf 2827}& & & & & & & & & & & & & & & & & \\
\hline
    \end{tabular}
    }
\end{table}

\end{landscape}
\newpage
\bibliographystyle{apalike}
\bibliography{references}

%\appendix
\end{document}